\documentclass[useAMS,usenatbib,twocolumn]{mnras}
\usepackage{natbib}
\usepackage{graphicx}
\usepackage{xcolor}
\usepackage{times}
\usepackage{rotate}
\usepackage{color}
\usepackage{hyperref}
\usepackage{amsmath,bm}
\usepackage{url} 
\usepackage{mathtools}
\usepackage{graphicx}
\usepackage{amssymb}
\usepackage{subfig}

\usepackage{eso-pic}

\AddToShipoutPictureBG*{%
  \AtPageUpperLeft{%
    \hspace{0.75\paperwidth}%
    \raisebox{-3.5\baselineskip}{%
      \makebox[0pt][l]{\textnormal{DES-2021-0640}}
}}}%

\AddToShipoutPictureBG*{%
  \AtPageUpperLeft{%
    \hspace{0.75\paperwidth}%
    \raisebox{-4.5\baselineskip}{%
      \makebox[0pt][l]{\textnormal{FERMILAB-PUB-21-123-AE}}
}}}%

\newcommand{\eg}{{\it e.g.,}}

\newcommand{\ifib}{\ensuremath{i_{\rm fib2}}}
\newcommand{\ic}{\ensuremath{i_{\rm cmod}}}
\newcommand{\lm}{\ensuremath{\log_{10}(M_*)}}
\newcommand{\photoz}{photo-$z$~}
\newcommand{\photozs}{photo-$z$s~}

\newcommand{\eric}[1]{\textcolor{black}{#1}}

\newcommand{\new}[1]{\textcolor{black}{#1}}

\newcommand{\mmnew}[1]{\textcolor{black}{#1}}
\newcommand{\kidsnew}[1]{\textcolor{black}{#1}}
\newcommand{\report}[1]{\textcolor{black}{#1}}

\definecolor{purple}{RGB}{76, 0,153}
\newcommand{\ch}[1]{{\color{black}{#1}}}

  \hypersetup{draft}
 
\begin{document}
  

\title[Lensing Without Borders]{Lensing Without Borders. I. A Blind Comparison of the Amplitude of Galaxy-Galaxy Lensing Between Independent Imaging Surveys}




\author[Leauthaud and Amon et al.]{
\parbox{\textwidth}{
\Large
A.~Leauthaud,$^{1}$\footnote{alexie@ucsc.edu}
A.~Amon,$^{2}$\footnote{amon2018@stanford.edu}
S.~Singh,$^{3,4}$
D.~Gruen,$^{2,5}$
J.~U.~Lange,$^{1,2}$
S.~Huang,$^{6}$
N.~C.~Robertson,$^{7}$
T.~N.~Varga,$^{8,5}$
Y.~Luo,$^{1}$
C.~Heymans,$^{9,10}$
H.~Hildebrandt,$^{10}$
C.~Blake,$^{11}$
% ...
}
\vspace{0.4cm}
\\
\\
Affiliations are listed at the end of the paper.
\vspace{0.1cm}
}

\maketitle
\label{firstpage}

 
\begin{abstract} Lensing Without Borders is a cross-survey collaboration created to assess the consistency of galaxy-galaxy lensing signals ($\Delta\Sigma$) across different data-sets and to carry out end-to-end tests of systematic errors. We perform a blind comparison of the amplitude of $\Delta\Sigma$ using lens samples from BOSS and six independent lensing surveys. We find good agreement between empirically estimated and reported systematic errors which agree to better than 2.3$\sigma$ in four lens bins and three radial ranges. For lenses with $z_{\rm L}>0.43$ and considering statistical errors, we detect a 3-4$\sigma$ correlation between lensing amplitude and survey depth. This correlation could arise from the increasing impact at higher redshift of unrecognised galaxy blends on shear calibration and imperfections in photometric redshift calibration. At $z_{\rm L}>0.54$ amplitudes may additionally correlate with foreground stellar density. The amplitude of these trends is within survey-defined systematic error budgets which are designed to include known shear and redshift calibration uncertainty. Using a fully empirical and conservative method, we do not find evidence for large unknown systematics.  Systematic errors greater than 15\% (25\%) ruled out in three lens bins at 68\% (95\%) confidence at $z<0.54$. Differences with respect to predictions based on clustering are observed to be at the 20-30\% level. Our results therefore suggest that lensing systematics alone are unlikely to fully explain the ``lensing is low" effect at $z<0.54$. This analysis demonstrates the power of cross-survey comparisons and provides a promising path for identifying and reducing systematics in future lensing analyses. 	
\end{abstract}

\begin{keywords}
cosmology: observations -- gravitational lensing -- large-scale structure of Universe
\end{keywords}
 

 
 



\section{Introduction}

The pursuit to constrain the equation of state of dark energy has motivated a number of imaging weak lensing surveys. A number of the surveys are now complete such as the Sloan Digital Sky Survey \citep[SDSS,][]{Gunn:1998}, the Canada France Hawaii Telescope (CFHT) Lensing Survey \citep[CFHTLenS\footnote{https://www.cfhtlens.org}, ][]{Heymans:2012}, the Deep Lens Survey \cite[DLS, ][]{Jee:2013}, the Red-sequence Cluster Lensing Survey \citep[RCSLenS, ][]{Hildebrandt2016}, and the CFHT Survey of Stripe 82 \citep[CS82, ][]{Leauthaud:2017aa}. Analysis of a number of weak lensing surveys are ongoing including the Dark Energy Survey \citep[DES\footnote{https://www.darkenergysurvey.org}, ][]{The-Dark-Energy-Survey-Collaboration:2015}, the Kilo Degree Survey \citep[KiDS\footnote{http://kids.strw.leidenuniv.nl}, ][]{Kuijken:2015}, and the Hyper Suprime Cam survey \citep[HSC\footnote{https://hsc.mtk.nao.ac.jp/ssp}, ][]{Aihara:2018aa}. Looking forward, a number of Stage 4 surveys will also be carried out within the next decade including the Legacy Survey of Space and Time \citep[LSST\footnote{https://www.lsst.org},][]{LSST-Science-Collaboration:2009}, the \textit{Euclid}\footnote{https://sci.esa.int/web/euclid} mission \cite[][]{Laureijs:2011}, and the \textit{Roman}\footnote{https://roman.gsfc.nasa.gov} mission \citep[][]{WFIRST}.

 
 
 As the statistical precision of these surveys has grown,  intriguing differences with respect to the best fit model from the \textit{Planck} experiment have begun to emerge. Assuming a standard 6-parameter $\Lambda$CDM model, recent cosmic shear measurements \citep[][]{Hikage:2019aa, Asgari2020, desy3} appear to prefer slightly lower values for the $S_8$ cosmological parameter compared to the best fit \textit{Planck} cosmology \citep{planck2020}. Another such difference is the ``lensing is low" effect. This is the observation that the lensing amplitude around luminous red galaxies is lower than predicted by their clustering in a \textit{Planck} cosmology \citep[][]{Cacciato:2013, Leauthaud:2017aa, Lange2019, Singh:2020}. Measurements of the $E_{\rm G}$ statistic \citep[e.g.][]{Blake2016,Amon2018, Singh:2020, Blake2020} and joint cosmic shear and BOSS clustering analyses (see for example \citealt[][]{Heymans2021} and references therein) draw similar conclusions.

Using data from the SDSS main survey, \citet[][]{Cacciato:2013} studied clustering and galaxy-galaxy lensing measurements (hereafter ``gg-lensing"). While their constraints on $S_8$ were consistent with  WMAP \citep[Wilkinson Microwave Anisotropy Probe,][]{Dunkley:2009} at the time of publication, their  results correspond to a lower value of $S_8$ compared to \textit{Planck}. Also using the SDSS main sample, \citet[][]{Mandelbaum:2013} obtained cosmological constraints using large scale measurements of lensing and clustering. However, due to the limited volume of the main sample and the radial scale cuts employed, their constraints have relatively large errors and are consistent with both WMAP and \textit{Planck}.  Using the larger CMASS sample from the Baryon Oscillation Spectroscopic survey \citep[BOSS,][]{Alam:2017aa} and lensing data from a combination of CFHTLenS and CS82, \citet[][]{Leauthaud:2017aa} showed that the observed lensing signal around CMASS LRGs is lower than predicted from the clustering. Specifically, they found $\Delta\Sigma_{\rm predicted}/\Delta\Sigma_{\rm obs} \sim 1.2-1.3$ (20-30\% level differences in the lensing amplitude) where $\Delta\Sigma_{\rm predicted}$ is the signal predicted from a variety of galaxy halo models applied to the clustering. \citet[][]{Lange2019} confirmed and extended these results to a wider range in redshift using CFHTLenS data. \citet[][]{Lange2019} also showed the effect to be  relatively independent of galaxy stellar mass (albeit with lower signal-to-noise due to smaller sample sizes when dividing by galaxy mass). \citet[][]{Singh:2020} extended the results of \citet[][]{Mandelbaum:2013} and studied the lensing and clustering of the BOSS LOWZ sample using lensing from SDSS as well as \textit{Planck} CMB lensing. Using only the large scale signal, they constrain the $S_8$ parameter to be $\sim$15\%  lower than predicted by \textit{Planck} at the 2-4$\sigma$ level. Their CMB lensing analysis prefers a 10\% (1$\sigma$) lower value of $S_8$ but with lower significance due to the larger errors from CMB lensing. Finally, \citet[][]{Lange2020} showed the ``lensing is low" effect to be independent of both halo mass ($M_{\rm halo}>10^{13.3}$ h$^{-1} $M$\odot$ ) and radial scale ($r<$ 60 h$^{-1}$ Mpc). 

Taken together, these results could offer tantalising hints of physics beyond $\Lambda$CDM. However, lensing measurements are notoriously difficult and understanding (and controlling for) systematic errors is one of the most challenging aspects for any lensing analysis. The weak lensing community is acutely aware of the need to quantify and mitigate systematic errors and has been actively engaged in reducing systematic errors. There have been a number of community efforts to combat systematics, such as the Shear TEsting Programme (STEP) \citep{Heymans2006, Massey2007}, the GRavitational lEnsing Accuracy Testing (GREAT) challenges \citep{Bridle2009, Mandelbaum2014}, and the PHoto-z Accuracy Testing (PHAT) program \citep[][]{Hildebrandt2010}. As underscored by existing efforts on this front, two key challenges are: 1) the accurate measurement of the lensing shear from galaxy shapes (in the presence of noise, the Point Spread Function, and galaxy blends), and 2) the determination of photometric redshifts (or redshift distributions), for source galaxies. While systematic errors from shape measurements and redshifts have in the past been sub-dominant compared to statistical errors, the increase in statistical precision afforded by larger survey areas means that even greater attention must be paid to systematic errors. Of particular concern is the possibility that the data may be affected by an unknown systematic that has yet to be quantified.

Systematic errors may be categorized into three types: the ``known knowns", the ``known unknowns", and the ``unknown unknowns". The ``known knowns" are effects already accounted for in systematic error budgets. The `known unknowns" are effects that are currently being studied and will be incorporated into future systematic error budgets. The ``unknown unknowns" are effects that have not been thought about and may not accounted for. If the differences with respect to \textit{Planck} (after including the known systematic errors) continue to increase in significance, then the question of ``unknown unknowns" will become of considerable interest.

Lensing Without Borders (hereafter LWB) is an inter-survey collaboration, exploiting the overlap on the sky of existing lensing surveys with the BOSS spectroscopic survey to perform direct and empirically motivated tests for systematic effects in measurements of  gg-lensing, following the methodology in  \citet[][]{Amon2018}. LWB has two goals: 1) to empirically search for systematic trends that could be used to reduce systematic floors, and 2) to empirically test if large ``unknown unknowns" systematic effects are present in the data.

The premise underlying LWB is that the gg-lensing signal around BOSS galaxies measures $\Delta\Sigma$, the excess differential surface mass density, a physical quantity. As such the measured $\Delta\Sigma$ values for BOSS galaxies should agree, independently of the lensing data employed (modulo sample variance and inhomogeneity in the lens sample). BOSS provides spectroscopic redshifts for lenses which enables a more accurate measurement of $\Delta\Sigma$. We perform a blind comparison of the amplitude of the $\Delta\Sigma$ signal using four lens samples from BOSS and using the sources catalogues and methodologies from six distinct lensing surveys (SDSS, CS82, CFHTLenS, DES, HSC, KiDS). As shown in \citet[][]{Luis_Bernal_2018}, constraints on systematic errors improve when considering a large number of independent measurements, even if some measurements are more uncertain than others.

LWB provides an empirical end-to-end test of systematics in gg-lensing that is sensitive to both the shear calibration of the data, the redshift estimation, as well as the methodology for computing $\Delta\Sigma$. The framework developed here also provides a first handle on determining the origin of amplitude offsets (shear calibration, redshift estimation, methodology), however, future work will focus more specifically on developing methodologies for disentangling such effects. 

In the radial range of consideration in this paper ($r<10$ Mpc), statistical constraints on the amplitude of the gg-lensing signal vary from $\sigma_{\rm amp}\sim 0.04-0.1$ depending on the survey at hand (these numbers will depend on which lensing data set is being used, the cuts made on the lens sample, and the radial range under consideration). Here, $\sigma_{\rm amp}$ is the statistical error on the ratio  $\Delta\Sigma_{\rm predicted}/\Delta\Sigma_{\rm obs}$ where $\Delta\Sigma_{\rm predicted}$ is the predicted signal based on galaxy clustering (which should be the same for all surveys). With reported tensions between lensing and clustering in a \textit{Planck} cosmology being at the 10-30\% level \citep[][]{Leauthaud:2017aa, Lange2019, Singh:2020, Lange2020}, the tests proposed here will be able to check for large unknown systematics that could lead to such differences. However, our tests rely on the assumption that all of the lensing surveys are independent, have been analysed independently, and are not subject to confirmation bias. 

The goal of this paper is to provide the first direct and empirically motivated test of the consistency of the galaxy-galaxy lensing amplitude across lensing surveys and to develop a framework for such comparisons. While the precision of the tests in this paper is limited by the existing overlap between various lensing surveys and BOSS, the LWB methodologies developed in this paper will become more powerful both as the overlap between lensing surveys increases, as well as the overlap between lensing surveys and spectroscopic surveys such as DESI \citep[Dark Energy Spectroscopic Instrument, ][]{DESI2016}.

Our methodology is outlined in Section \ref{Methodology}. Section \ref{specz} describes the foreground lens sample and Section \ref{wldata} gives a brief description of the weak lensing data used in this paper. The various methodologies used to compute $\Delta\Sigma$ are described in Section \ref{computeds}. Section \ref{homogenity} presents tests on the homogeneity of the BOSS samples. Our results are presented in Section \ref{Results} and discussed in Section \ref{discussion}. Section \ref{conclusions} presents a summary and our conclusions. We use a flat $\Lambda$CDM cosmology with $\Omega_{\rm m}=0.3$, $H_0=70$ km s$^{-1}$ Mpc$^{-1}$. We assume physical coordinates to compute $\Delta\Sigma$\footnote{see Appendix C in \citet{Dvornik2018}}.

\section{General Methodology for Galaxy Galaxy Lensing}\label{ggl}

Here we describe in general terms how to convert tangential shear into $\Delta\Sigma$. The full details, including team specific approaches, are presented in Section \ref{computeds}.

\subsection{From $\gamma$ to $\Delta\Sigma$}

The shear signal induced by a given foreground mass distribution on a
background source galaxy will depend on the transverse proper distance
between the lens and the source and on the redshift configuration of
the lens-source system. A lens with a projected surface mass density,
$\Sigma(r)$, will create a shear that is proportional to the {\em
  surface mass density contrast}, $\Delta\Sigma(r)$:

\begin{equation}
  \Delta \Sigma(r)\equiv\overline{\Sigma}(< r)-\overline{\Sigma}(r)=\Sigma_{\rm c}\times\gamma_{\rm t}(r).
\label{dsigma}
\end{equation}

Here, $\overline{\Sigma}(< r)$ is the mean surface density within
proper radius $r$, $\overline{\Sigma}(r)$ is the azimuthally averaged
surface density at radius $r$
\citep[\eg][]{Miralda-Escude:1991,Wilson:2001}, and $\gamma_{\rm t}$ is the
tangentially projected shear. The geometry of the lens-source system
intervenes through the critical surface mass
  density $\Sigma_\mathrm{c}$:

\begin{equation}\label{eq:sigmacrit1}
\Sigma_\mathrm{c} = \frac{c^2}{4\pi G} \frac{D_{\rm A}(z_{\rm s})}{D_{\rm A}(z_{\rm L}) D_{\rm A}(z_{\rm L},z_{\rm s})},
\end{equation}

\noindent where $D_{\rm A}(z_{\rm L})$ and $D_{\rm A}(z_{\rm s})$ are angular diameter distances to the lens and source, and $D_{\rm A}(z_{\rm L},z_{\rm s})$ is the angular diameter distance between the lens
and source. When the redshifts (or redshift distribution) of source galaxies are known, each
estimate of $\gamma_{\rm t}$ can be directly converted to an estimate of
$\Delta\Sigma(r)$.

To measure $\Delta\Sigma(r)$ with high signal-to-noise, the lensing
signal must be stacked over many foreground lenses and background
sources. In order to optimise the signal-to-noise of this stacking process, an inverse variance weighting scheme is commonly employed when
$\Delta\Sigma_{ij}$ is summed over many lens-source pairs. Each
lens-source pair is attributed a weight $w_{ij}$ that is often (but not always) the estimated variance of the shear measurement. The excess projected
surface mass density is the weighted sum over all lens-source pairs:

\begin{equation}
  \Delta\Sigma =  
  {\sum_{j=1}^{N_{\rm Lens}} \sum_{i=1}^{N_{\rm Source}} w_{ij} \times \gamma_{{\rm},ij}\times \Sigma_{{\rm c},ij}
    \over \sum_{j=1}^{N_{\rm Lens}} \sum_{i=1}^{N_{\rm Source}}w_{ij}} ~.
\label{ds_equation}
\end{equation}

To remove systematic bias and obtain the optimal covariance \citep{Mandelbaum:2005a,Singh:2017}, it has become common to subtract the measurement around random points (see Section \ref{computeds}.). In addition, sometimes additional weights may be applied to the lens sample (e.g. see Section \ref{bossdata}).

\subsection{Correction terms due to imperfect knowledge of source redshifts}\label{photozcorrections}

To compute $\Delta\Sigma$ we must select background source galaxies. However, source galaxies typically only have photometric redshifts. These redshifts may be biased and the source selection may be imperfect. A number of correction terms are applied to $\Delta\Sigma$ estimates to account for such effects. These are:

\begin{enumerate}
    \item The boost factor. A “boost correction factor” is sometimes applied in order to account for
the dilution of the signal by physically associated sources 
\citep[\eg][]{Kneib:2003aa,Sheldon:2004, Hirata:2004aa,
  Mandelbaum:2006}. This correction factor is usually computed by
comparing the weighted number density of source galaxies for the lens
sample to the weighted number density of source galaxies around random
points. However, the validity of boost correction factors is debated \citep[e.g.][]{Melchior:2015aa,Applegate:2014aa,Simet:2015, Leauthaud:2017aa}.
    \item The dilution factor. The ``background'' sample may
contain a number of galaxies that are actually in the foreground
($z_{\rm s}<z_{\rm L}$). Because foreground galaxies are
unlensed, the inclusion of these galaxies will cause $\Delta\Sigma$ to
be underestimated.
    \item The $f_{\rm bias}$ correction factor. $\Delta\Sigma$ estimates can be biased due to imperfect calibration of photo-$z$'s. Furthermore, even with perfectly calibrated point source photo-$z$'s, $\Delta\Sigma$ can be biased because of the non linear response of  $\Delta\Sigma$ to source redshifts (via the $\Sigma_{\rm c}$ factor). Instead of using a point source estimate, some teams prefer to integrate over a source redshift probability distribution function (PDF). However, this integration will only be accurate if the full shape of the PDF is well calibrated. In other terms, \emph{an unbiased mean $P(z)$ does not guarantee an unbiased $\Delta\Sigma$}. For these reasons, a correction factor called $f_{\rm bias}$ is sometimes applied (see Section \ref{fbiassection}). This correction factor is computed using a representative sample of galaxies with spectroscopic redshifts. Often, $f_{\rm bias}$ is written in a way that also corrects for the dilution factor.
\end{enumerate}

\section{Methodology}\label{Methodology}

\subsection{General approach and limitations}

In this paper, we use weak lensing data from CS82, CFHTLenS, HSC, KiDS, SDSS, and DES. A description of these data are given in Section \ref{wldata}. 

\begin{figure*}
\centering
\begin{tabular}{@{}c@{}}
    \includegraphics[width=16cm]{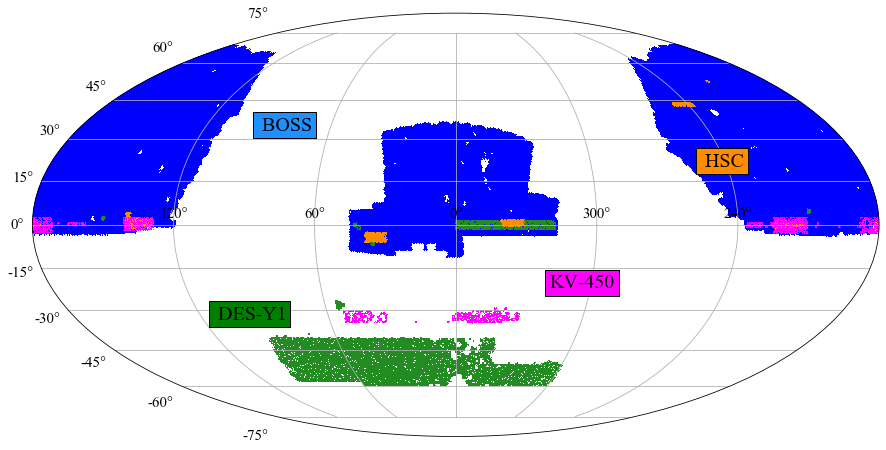}
 \end{tabular}

  \vspace{\floatsep}
  
\begin{tabular}{@{}c@{}}
   \includegraphics[width=16cm]{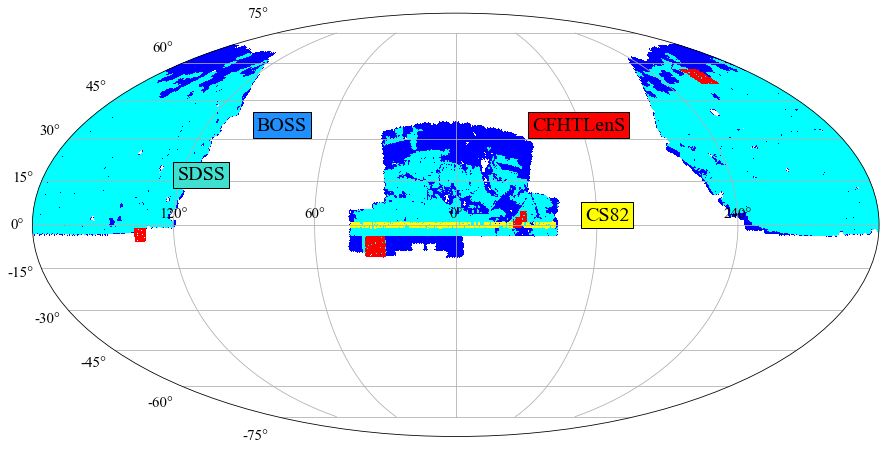}
  \end{tabular}
\caption{Footprints of a subset of  the weak lensing surveys and their overlap with the BOSS survey. BOSS is shown in dark blue, SDSS weak lensing in cyan, HSC in orange, CFHTLenS in red, CS82 in yellow, DES-Y1 in green, and KV-450 (KiDS)  in magenta.}
\label{fig:map}
\end{figure*}

For the lens sample, we select a common set of lenses from BOSS (see Section \ref{specz}). Figure \ref{fig:map} displays the footprints of different surveys considered in this paper. Table \ref{overlaptable} gives the overlap between BOSS and various lensing surveys\footnote{\eric{Binary masks with nside=2048 were used.}}. Currently, apart from the SDSS lensing catalogue, the overlap between BOSS and existing lensing surveys is typically of order 100-200 deg$^2$, however, this overlap will rapidly expand over the next few years to reach of order $\sim$1000 deg$^2$.

\begin{table*}
  \caption{Overlap between current weak lensing surveys and the spectroscopic BOSS, reported in deg$^2$. These area values are just rough estimates and should not be used for computations.}
\begin{tabular}{@{}lcccccccc}
\hline
 & SDSS lensing & HSC Y1 & DES Y1 & KV-450 & CFHTLenS & CS82 \\
\hline
BOSS & 8359 & 166 & 160 &  204 &  118 & 144 \\
SDSS lensing & - & 160 & 134 &  196 &  108 & 130 \\
HSC Y1 & 160 & - & 26 &  68 & 32 & 11 \\
DES Y1 & 134 & 26 & - &  0 &  20 & 67 \\
KV-450 & 196 & 68 & 0 &  -  & 3 & 0 \\
CFHTLenS & 108 & 32 & 20 &  3 & - & 7 \\
CS82 & 130 & 11 & 67 &  0 & 7 & - \\
\hline
\end{tabular}
\label{overlaptable}
\end{table*}

One of the main assumptions behind our methodology is that the lens sample selects a homogeneous sample of dark matter halos across the BOSS footprint. However, there may be inhomogeneity in the BOSS lens sample. This is tested in Section \ref{homogenity} and Section \ref{postblindhomogtest}. There are two other caveats to our analysis. First, we do not account for cross-covariance between surveys (overlap areas are modest and are quoted in Table \ref{fig:map}). Section \ref{jointcovariance} outlines a methodology for accounting for cross-covariance when the overlap between survey footprints increases. Ignoring this cross-covariance means that our systematic errors may be overestimated (and that our main conclusions are conservative). Second, our tests rely on the assumption that all of the lensing surveys are independent, and have been analysed independently. However, there may be systematic errors
that are common between different lensing surveys (e.g. a common redshift calibration sample, such as COSMOS-30 and/or similar shear measurement methods) which cannot be tested here.

\subsection{Computation of $\Delta\Sigma$}

Prior to computing $\Delta\Sigma$ we agreed that all teams would compute the signal under the following set of assumptions:

\begin{enumerate}
\item A fixed fiducial cosmology (as given in Section 1).
\item A fixed radial binning scheme. We use 10 logarithmically spaced bins from 0.05 Mpc to 15 Mpc.
\item Physical transverse distances are used for the computation of $\Delta\Sigma$.
\item Data points are compared at the mean $r$ value of the bin (see justification below), where $r$ is a physical transverse radius. This value is the same for all surveys.
\item The lens and random files provided to each team correspond to the intersection between the BOSS footprint and the footprint of each shear catalogue.
\item Our fiducial test uses systematic weights that are applied to lenses to ensure that the spatial variations of the lenses follow those of the randoms (see Section \ref{bossdata}). 
\item We also perform an additional test for the CMASS sample in which we measure the lensing signal without systematic weights.
\end{enumerate}

The effective value of $r$ within bins depends on the scaling of the underlying signal ($\Delta\Sigma$) which is same for all the surveys. It also depends on the weighting imposed by the survey window (or the distribution of source galaxies), which can be different for different surveys. On small scales, the effects of survey masks are expected to be small, in which case the mean value of $r$ within the bins, $\overline{r}=(r_{\rm high}-r_{\rm low})/2$, is close to the effective value for the measured $\Delta\Sigma$ \citep[see equation D3 and figure D2 in][]{Singh:2020}\footnote{The impact of binning in the context of cosmic shear measured in angular bins has been discussed in other work \citep[e.g.,][]{Krause:2017, Troxel:2018,Asgari:2019} but conclusions from these papers are not directly relevant to the case of gg-lensing which measures the signal in physical bins and over a much narrower redshift range.}. In the mock tests performed by \citet[][]{Singh:2020} for SDSS, binning effects with $\overline{r}$ were $<$ 1\% at $r<$ 60 Mph/$h$, smaller than the 10-30\% differences of concern for this paper.

There are also a number of other choices required for a $\Delta\Sigma$ calculation. The following aspects were intentionally not discussed and were not homogenised amongst teams:

\begin{enumerate}
\item How to write the estimator for $\Delta\Sigma$.
\item How to use the redshift information for each source.
\item How (and if) to compute and apply boost factors.
\item How (and if) to compute and apply dilution factors.
\item How (and if) to apply any further correction factors for photo-$z$ biases.
\item Computation of the covariance matrix.
\end{enumerate}

Each team was responsible for the computation of $\Delta\Sigma$. Teams were asked to perform all tests deemed necessary before unblinding. Section \ref{computeds} provides the specific details on how each team computed $\Delta\Sigma$. 

\subsection{Blinding strategy}\label{blinding}

We agreed that each team would compute $\Delta\Sigma$ independently. In the blinded phase, each team applied a multiplicative scale dependent offset to their $\Delta\Sigma$ values. We opted for a scale dependent offset so that no guesses could be made as to which scales were in better agreement. Each team randomly drew two numbers $\alpha$ and $\beta$ with values between [0.80, 1.2] and then multiplied their $\Delta\Sigma$ values by a radially dependent factor $f(r)$:

\begin{equation}
f(r)=\frac{1}{9}\left[ (\beta - \alpha)r + 10\alpha-\beta\right],
\end{equation}

\noindent where $r$ is expressed in physical Mpc. This blinding strategy results in a radial dependent offset between signals at the 20\% level. All figures were made with this blinding strategy during the blinded phase. 

There are already $\Delta\Sigma$ values published for CMASS and LOWZ. Hence, our tests could not be made 100\% blind. But to make tests as blind as possible, we imposed redshift cuts on the BOSS samples so that it was not possible to compare directly with other published values. These are described in Section \ref{lenssamples}.

\subsection{Aspects of tests agreed to before analysis}

This section describes aspects of the tests that were decided upon before the analysis was conducted.

\begin{enumerate}
\item Small scales were distinguished from large scales when comparing signals. This is because smaller scales are subject to boost factor correction uncertainties, whereas large scales will be more affected by error estimates (correlated shape noise and sample variance). The scales R$_1$=[0.05,1] Mpc and R$_2$=[1,15] Mpc were analysed both separately and jointly. The motivation for these scales is based on the idea that boost correction factors should mainly only affect $\Delta\Sigma$ below 1 Mpc. 
\item Data from each survey were fit with a model in which only the overall amplitude was allowed to vary (see next section). This is because the current errors on gg-lensing do not provide good constraints on slope variations \footnote{A slope variation would result when a measured $\Delta\Sigma$ does not have the same shape as $\Delta\Sigma_{\rm HOD}$.}. Hence, we only tested for amplitude shifts. The radial ranges $R_1$ and $R_2$ were fit both separately and jointly. The resulting amplitudes are noted $A_1$ (for the R$_1$ range), $A_2$ (for the R$_2$ range), and $A$ (for the full range).
\item Amplitudes were compared across different surveys.
\item A set of post-unblinding tests was also defined and is described further in Section \ref{Results}. It was agreed to use 3$\sigma$ as a threshold for determining trends to be significant.
\item \mmnew{It was agreed to not comment on any survey being deemed either ``high" or ``low". Doing so would amount to sigma-clipping and would introduce confirmation bias into the results by lowering the estimated value of $\sigma_{\rm sys}$}.
\item \mmnew{Monte Carlo tests were used to show that given the number of bins, the errors, and the number of surveys used, there is $\sim$6\% probability of having one survey appear either ``high" or ``low" across all lens bins. It was therefore agreed to not comment on this aspect and we also strongly encourage readers not to do so}.
\end{enumerate}

\subsection{Amplitude fitting}\label{afit}

Our goal is to detect differences between the amplitudes of the $\Delta\Sigma$ signal, as measured by different surveys. One common, yet fairly insensitive way is a direct $\chi^2$ test between the data points. Given knowledge about the shape of $\Delta\Sigma(r)$, and its covariance, a more stringent test can be done based on a \emph{matched filter}\footnote{Fitting the amplitude of a model to the data can be thought of as an optimal linear combination the data vector, yielding one number of interest. We then perform tests, such as $\chi^2$, using this one number.}. Here we opt to use the latter because we are primarily interested in comparing the amplitudes of measurements from different surveys.

For a data vector $d$, with covariance matrix $\mathsf{C}$, a linear combination $A$ can be written as 
$A=w^{\mathsf{T}}d=\sum_i w_i d_i$ with a weight vector $w$. The variance of this linear 
combination is $\sigma_{\rm A}^2=w^\mathsf{T}\mathsf{C}w=\sum_{ij}w_i w_j C_{ij}$. When the true shape of the noiseless signal (i.e. the expectation value of $d$) is known as $t$, one 
can show that the linear combination of $d$ with the highest possible signal-to-noise ratio is given by the 
matched filter amplitude $A$ with weights $w\propto\mathsf{C}^{-1}t$.

In our case, $d$ is the difference between the $\Delta\Sigma$ data vectors measured by two surveys. To define a matched filter, we need to know both the true shape and the covariance matrix of $d$. For the first ingredient, we expect that a potential non-zero $d$ is primarily due to multiplicative errors, e.g. arising from shear or redshift calibration errors. That is, $d$ has a radial shape close to that of $\Delta\Sigma$ itself. \report{For the true profile assumed for our matched filter we adopt $\Delta\Sigma$ as predicted by a Halo Occupation Distribution (HOD) analysis of the CMASS clustering signal from \citet[][]{Leauthaud:2017aa}, hereafter noted $\Delta\Sigma_{\rm HOD}$. This model was obtained by fitting a standard HOD model to the two-point clustering of the CMASS sample and then by population a dark matter simulation with this HOD and predicting $\Delta\Sigma(r)$.  The redshift range over which the clustering was measured (full CMASS sample) is different from the redshift ranges of the lens samples used in this paper. We should thus not expect the lensing amplitude here to match the prediction from clustering, but the general shape of $\Delta\Sigma_{\rm HOD}$ for BOSS samples does not vary strongly with redshift \citep[e.g.,][]{Leauthaud:2017aa}, and so this model is good enough for our purpose}.

The second ingredient for the matched filter is the covariance matrix of $d$. We assume that any pair of surveys 1 and 2, who have measured $\Delta\Sigma$ with covariances $\mathsf{C}_1$ and $\mathsf{C}_2$, are uncorrelated, such that the covariance of $d$ is simply $(\mathsf{C}_1+\mathsf{C}_2)$. We have verified empirically that the optimal filter defined this way for any pair of surveys is not too different from the filter assuming $\mathsf{C}\propto\sum_j \mathsf{C}_j$, where the sum runs over all surveys $j$. We will use the latter in order to be able to compare the amplitudes of all surveys on the same footing. 

In summary, for each survey $j$, and one of three radial ranges (small, large, and all radii), we will determine a matched amplitude
\begin{equation}
A_j = \frac{w^{\mathsf{T}}\Delta\Sigma_j}{w^{\mathsf{T}}\Delta\Sigma_{\rm HOD}},
\end{equation}
and its uncertainty,
\begin{equation}
\sigma_j^2=w^\mathsf{T} \mathsf{C}_j w,
\end{equation}
where,
\begin{equation}
w = \left(\sum_j \mathsf{C}_j\right)^{-1}\Delta\Sigma_{\rm HOD} \; .
\end{equation}

Note that because the operations are linear, the difference between two amplitudes is the same as the amplitude of the difference between the two corresponding data vectors (for which this matched filter was derived). In line with our focus on inter-comparing lensing surveys, our figures will report $A-\overline{A}$ only, which is not sensitive to an amplitude difference between the lensing signal and the clustering-based prediction. Here, $\overline{A}$ is the mean amplitude averaged over the lensing surveys.

The validity of our tests do not rely on the model having the correct shape - it remains a test on a linear combination of the data that should be zero in the absence of biases. The sensitivity of the test, however, does depend on $\Delta\Sigma_{\rm HOD}$. Had we used the matched filter amplitude for each individual survey, i.e. with $\bm{w}_j = \mathsf{C}_j^{-1}\Delta\Sigma_{\rm HOD}$, then this would not be the case: each survey would weight the signal differently as a function of radius, and an offset between $\Delta\Sigma_{\rm HOD}$ and the correct model could manifest as a non-zero difference in amplitudes for mutually consistently calibrated surveys with differently structured covariance matrices.

\subsection{Searching for trends caused by correlated systematic errors}\label{trendmethod}

One of the key goals of this paper is to investigate if correlations exist between measured lensing amplitudes and survey properties (e.g., $n_{\rm star}$, survey depth) that should, in principle, have no impact on $\Delta\Sigma$ (Section \ref{postblindhomogtest} and Section \ref{meansourcez}). If found, such correlations could provide important clues as to the origin of systematic errors. These could be both known or unknown systematic errors. We seek to pin-point trends caused systematic errors. \report{For this, we both use the reported statistical errors, as well as the sum in quadrature of statistical and systematic errors to conduct these tests}. 

\begin{figure*}
\centering
\includegraphics[width=15cm]{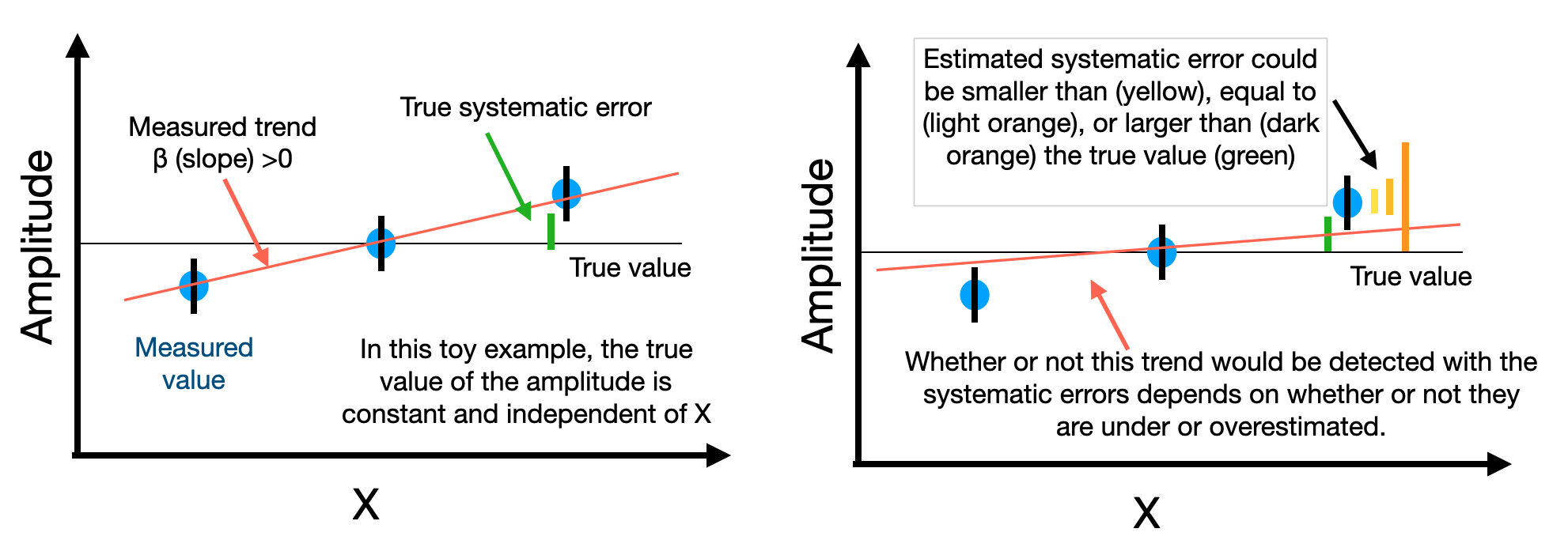}
\caption{Illustrative example of the detection of a trend originating from systematic effects. Left: amplitude of the lensing signal versus parameter X. In this toy example, the true value of the amplitude is constant with X. However, a systematic error that correlates with X causes a trend (here with slope $\beta>0$) in the relationship between amplitude and slope. The true systematic error (green horizontal line) is the rms spread between the measured data points (blue) and the horizontal line. Right: the estimated systematic error ($\sigma_{\rm sys, est}$) could be smaller, equal to, or larger than the true value ($\sigma_{\rm sys, true}$). Indeed, the true level of systematic error is rarely known. If $\sigma_{\rm sys, est}<\sigma_{\rm sys, true}$, the trend \report{may} be detected. If  $\sigma_{\rm sys, est} \geq \sigma_{\rm sys, true}$, the trend \report{may no longer} be detected. Trends may or may not be detected if the estimated systematic error is summed in quadrature with the statistical error. The use of the statistical error bar (left) will yield a detection of trends if they are present in the data, \report{but the significance of the trend could be overestimated.} }
\label{fig:staterrors}
\end{figure*}

The left hand side of Figure \ref{fig:staterrors} illustrates an example in which a systematic error correlates with a given parameter X (e.g., redshift, $n_{\rm star}$, etc. ) and causes a trend in the lensing amplitude versus X (in this example, the measured trend is detected with a positive slope $\beta>0$). The errors in the left hand figure are the statistical errors on the measurements. The green line indicates the true level of systematic error in these data (the rms deviation between the horizontal line and the blue data points). 

The right hand side of Figure \ref{fig:staterrors} now considers the addition of the estimated systematic errors. Systematic errors are educated guesses and may underestimate or overestimate the true value. \report{For example, current lensing surveys rarely report estimates of the \emph{error on the systematic error}}. If the estimated systematic error underestimates the true value, then the trend with $\beta>0$ \report{may} still be detected. If the estimated systematic error is equal or larger than the true value, then the trend \report{may} no longer be detected. Whether or not a trend would be detected will depend on how close the estimated systematic error is to the true value \report{and how many data points are available}. \report{Thus, using the sum in quadrature of the statistical and the estimated systematic errors may not provide any insight into sources of systematic error}.

In the case of a single dominant systematic error that correlates with parameter X, the statistical errors \report{will increase the probability of detecting the trend}, as illustrated on the left hand side of Figure \ref{fig:staterrors}. However, the picture will be more complicated if multiple kinds of systematic error with distinct physical origins are present in the data. In this case, the correct errors to use would be the sum in quadrature of the statistical error and the true values of those systematic errors that do not correlate with the parameter under investigation (e.g., parameter X in Figure \ref{fig:staterrors}). However, systematic errors are not known at this level of detail (and the true values are not usually known). \report{Because we are working in the regime of systematic uncertainties, where the true errors are not exactly known, there is no perfect way of carrying out these tests. The use of statistical errors will enhance the probability of the detecting trends if they are present in the data, but the significance of these trends could be overestimated, especially if multiple different kinds of systematic error are present in the data. In this paper, we will carry out tests both using statistical errors, as well as the sum in quadrature of statistical and systematic errors, keeping in mind the advantages and disadvantages of both choices.}

\subsection{Estimate of global systematic error}\label{sigmasysmethod}

A second key goal in the paper is to use the measured spread between the amplitudes of $\Delta\Sigma$ as an empirical and end-to-end estimate of systematic errors. This global estimate will be noted $\sigma_{\rm sys}$ and is computed as follows. We first compute the reduced  $\chi^2$ between amplitudes (measured following the methodology described in Section \ref{afit}). When the reduced $\chi^2$ of the data, $\chi^2_{\nu}$, is greater than 1, we report the value of $\sigma_{\rm sys}$ that yields $\chi^2_{\nu}=1$. We assume that each amplitude data point is drawn from a normal distribution with $\sigma^2=\sigma_{\rm stat}^2+\sigma_{\rm sys}^2$ where $\sigma_{\rm stat}$ is the error on the amplitude for each survey. We also derive 68\% and 95\% confidence intervals on $\sigma_{\rm sys}$. For this, we consider the expected probability distribution for $\chi^2$ with degrees of freedom $\nu=n-1$ where $n$ is the sample size ($n=6$ for LOWZ and $n=5$ for CMASS). We find the range of $\sigma_{\rm sys}$ values that produces a $\chi^2$ that is within the central 68\% and 95\% of the distribution. Monte Carlo tests were used to validate this methodology. 

Because we use the spread between the data points as a means to estimate the overall systematic error, the number we quote should be thought of as an ensemble estimate over all of the surveys under consideration. \mmnew{Monte Carlo tests were used to show that the $\sigma_{\rm sys}$ value that we estimate is roughly equal to the mean systematic error among surveys}.
 
 Our empirical estimate is a multiplicative bias on the amplitude of $\Delta\Sigma$. More specifically, if we consider a variable $S_{\rm sys}$ that is drawn from a Gaussian of width $\sigma_{\rm sys}$ and unknown mean, the relation between the true value $\Delta\Sigma_{T}$ and the measured value $\Delta\Sigma_{M}$ is:

\begin{equation}\label{syserrequation}
    \Delta\Sigma_{M}(r)=(1+S_{\rm sys})\times \Delta\Sigma_{T}(r)\; ,
\end{equation}

\noindent where $S_{\rm sys}$ is independent of $r$ and where $S_{\rm sys}$ can take on a different value for each survey. The mean of $S_{\rm sys}$ is unknown because we cannot use the methods here to determine the absolute value of $\Delta\Sigma$. For example, we would not be able to detect a systematic bias if this bias were common to all of the lensing surveys and had a similar impact on $\Delta\Sigma$.

\subsection{Effective redshift weighting of lens samples}

Different surveys apply a different effective weight to the lens sample \citep[e.g.,][]{Nakajima:2012aa, Mandelbaum:2013, Simet:2016, Leauthaud:2017aa}. However, amplitude variations in $\Delta\Sigma$ across the CMASS redshift range have been found to be small \citep[][]{Leauthaud:2017aa, Blake2020}. We also use narrow redshift bins for our lens samples in order to mitigate this effect. This topic is discussed further in Section \ref{effectivelensz}.

\section{Foreground Lens Data}\label{specz}

\subsection{BOSS survey}\label{bossdata}

BOSS is a spectroscopic survey of 1.5 million galaxies over 10,000
deg$^2$ that was conducted as part of the SDSS-III program
\citep[][]{Eisenstein:2011} on the 2.5 m aperture Sloan Foundation
Telescope at Apache Point Observatory \citep[][]{Gunn:1998,
  Gunn:2006}. A general overview of the BOSS survey can be found in
\citet[]{Dawson:2013}, the BOSS spectrographs are described in
\citet[]{Smee:2013}, and the BOSS pipeline is described in
\citet[]{Bolton:2012}. BOSS galaxies were selected from Data Release 8
\citep[DR8,][]{Aihara:2011} {\it ugriz} imaging
\citep[][]{Fukugita:1996} using a series of color-magnitude cuts. 

The BOSS selection uses the following set of colours:

\begin{eqnarray}
c_\parallel &=& 0.7(g_{\rm mod}-r_{\rm mod}) + 1.2[(r_{\rm mod}-i_{\rm
  mod}) - 0.18] \\
 c_\perp &=& (r_{\rm mod}-i_{\rm mod}) - (g_{\rm mod}-r_{\rm mod})/4 - 0.18\\
d_\perp &=& (r_{\rm mod}-i_{\rm mod})-(g_{\rm mod}-r_{\rm mod})/8.0
\end{eqnarray} 

The subscript ``mod'' denotes model magnitudes, which are derived by
adopting the better fitting luminosity profile between a de
Vaucouleurs and an exponential luminosity profile in the $r$-band
\citep[][]{Stoughton:2002}. All magnitudes are corrected for Galactic extinction using the dust maps of \citet{Schlegel:1998}.


BOSS targeted two primary galaxy samples: the LOWZ sample at
$0.15<z<0.43$ and the CMASS sample at $0.43<z<0.7$. The LOWZ sample is
an extension of the SDSS I/II Luminous Red Galaxy (LRG) sample
\citep[][]{Eisenstein:2001} to fainter magnitudes and is defined
according to the following selection criteria:

\begin{eqnarray}
|c_\perp| &<& 0.2\label{lowz_cperp}\\
r_{\rm cmod} &<& 13.6 + c_\parallel/0.3\label{lowz_sliding}\\
16 < r_{\rm cmod} &<& 19.6\label{lowz_flux_limit}\\
r_{\rm psf} - r_{\rm cmod} &>& 0.3\label{star_gal}
\end{eqnarray}

Here, PSF magnitudes are denoted with the
subscript ``psf''. The subscript ``cmod'' denotes composite
model magnitudes, which are calculated from the best-fitting linear
combination of a de Vaucouleurs and an exponential luminosity profile
\citep[][]{Abazajian:2004}. Equation \ref{lowz_cperp} sets the colour boundaries of the sample;
equation \ref{lowz_sliding} is a sliding magnitude cut which selects
the brightest galaxies at each redshift; equation
\ref{lowz_flux_limit} corresponds to the bright and faint limits and
equation \ref{star_gal} is to separate galaxies from stars. In a
similar fashion to the SDSS I/II Luminous Red Galaxy (LRG) sample, the
LOWZ selection primarily selects red galaxies \citep[][]{Reid:2016}. 


The CMASS sample targets galaxies at higher redshifts with a surface
density of roughly 120 deg$^{-2}$. CMASS targets are selected from
SDSS DR8 imaging according to the following cuts:

\begin{eqnarray}\label{eq:cmass}
|d_\perp| &>& 0.55\label{cmass_dperp}\\
i_{\rm cmod} &<& 19.86 +1.6(d_\perp-0.8)\label{cmass_sliding}\\
17.5 &<& i_{\rm cmod} < 19.9\label{cmass_flux_limit}\\
r_{\rm mod} - i_{\rm mod} &<& 2\\
i_{\rm fib2}&<&21.5
\end{eqnarray}

\noindent where $i_{\rm fib2}$ is the estimated $i$-band magnitude in a
2\arcsec\ aperture diameter assuming 2\arcsec\ seeing. Star-galaxy
separation on CMASS targets is performed via:

\begin{eqnarray}
i_{\rm psf} - i_{\rm mod} &>&0.2+0.2(20.0-i_{\rm mod})\\
z_{\rm psf} - z_{\rm mod} &>&9.125-0.46z_{\rm mod}.
\end{eqnarray}

In this paper, we use catalogues from Data Release 12 \citep[DR12][]{Alam:2015}. We use the large-scale structure catalogues described in \citet[]{Reid:2016} that were generated via the {\sc mksample} code and that can be found at \url{https://data.sdss.org/sas/dr12/boss/lss/}\footnote{Exact file names are galaxy\_DR12v5\_CMASS\_North.fits.gz and so on and so forth.}. 

These large-scale structure catalogues include information about the BOSS selection function, survey masks, imaging quality masks, as well as weights to correct for various selection effects. In this paper, we will be concerned with understanding if inhomogeneities in the BOSS samples may lead to variations in the mean halo mass of the sample across different regions. We will return to this topic in Section \ref{homogenity}.

Veto masks are applied to the LSS catalogues \citep[]{Reid:2016}. These masks reject regions where BOSS galaxies cannot be observed. Among other things, these masks impose a cut that rejects areas of the survey that are too close to bright stars (the bright star mask), that have non photometric imaging conditions, where the seeing is poor, and with high extinction. 



In the early phase of the survey, an incorrect star-galaxy separation scheme was used for LOWZ. We do not use any LOWZ galaxies in regions where this happened (chunks 2-6 corresponding to the LOWZE2 and LOWZE3 samples). As a result, the areas covered by CMASS and LOWZ are different. See Appendix A in \citet[][]{Reid:2016}.

In DR12, a ``combined" sample was also created. We do not use the combined sample here. The reason for this is because the CMASS sample is more subject to observational effects (seeing, stellar density). We wish to study the impact of these effects on the lensing signal in isolation from the LOWZ sample. Also, we do not wish to use the LOWZE2 and E3 samples which are in the combined sample.

The BOSS LSS catalogues include various weights designed to minimise the impact of artificial observational effects that can impact estimates of the true galaxy over-density field. A full description of these weights is given in \citet[]{Reid:2016}. We briefly summarise the weights here:

\begin{itemize}
\item $w_{\rm cp}$: accounts for galaxies that did not obtain redshifts due to fibre collisions by up-weighting the nearest galaxy from the same target class.
\item $w_{\rm noz}$: weighting scheme designed to deal with galaxies for which the spectroscopic pipeline failed to obtain a redshift.
\end{itemize}

Taking these two weights together, the overall redshift weight is $w_{\rm z}=w_{\rm cp}+w_{\rm noz}-1$. In addition, there is also a set of weights that are designed to correct for variations in the CMASS samples with stellar density and seeing. Because the LOWZ sample is brighter than CMASS, it does not require these extra weights. The angular systematic weights for CMASS are: 

\begin{itemize}
\item $w_{\rm star}$: a weight to account for variations in the CMASS number density with stellar density. $w_{\rm star}(n_{\rm s},i_{\rm fib2})=(A_{i\rm{fib2}}+B_{i\rm{fib2}}n_{\rm s})^{-1}$. Variations in the number density with stellar density were found to correlate with galaxy surface brightness, in particular, the $i_{\rm{fib2}}$ magnitude. As the stellar density increases, on average, galaxies with lower magnitudes in a 2$\arcsec$ fibre are lost from the sample.
\item $w_{\rm see}$: the seeing based weight. There is a correlation between the number density and local seeing, due to star galaxy separation. For CMASS, the effect is such that in poor seeing conditions, the number density decreases because compact galaxies are classified as stars and are removed from the sample.
\end{itemize}

The total angular systematic weight for each galaxy is $w_{\rm systot}=w_{\rm star}w_{\rm see}$. Finally, the total weight for CMASS is constructed as $w_{\rm tot}=w_{\rm systot}(w_{\rm cp}+w_{\rm noz}-1)$\footnote{There are also the so-called ``FKP" weights ($w_{\rm FKP}$) based on Feldman, Kaiser, and Peacock 1994. These are weights that are designed maximise the signal to noise of 3D clustering statistics, not to correct for systematic effects, and are not relevant for the present study.}. 

The BOSS systematic weights were designed to up-weight galaxies to create a sample with constant number density. We apply the systematic weights to our lens samples so that the spatial distribution of the randoms follow that of the lens sample. However, applying the BOSS weights will not guarantee a sample with fixed halo mass across the survey -- indeed selection effects could lead to spatial inhomogeneity in the mean halo mass across the survey. In Section \ref{homogenity} we explore the impact of the inhomogeneity of the BOSS samples on $\Delta\Sigma$. We also design a set of post-unblinding tests that can be found in Section \ref{postblindhomogtest}.

\subsection{Lens samples}\label{lenssamples}

We use four distinct lens samples. Two are based on LOWZ and two are based on CMASS. Specifically, the samples we use are:

\begin{itemize}
\item L1: LOWZ sample with $0.15<z<0.31$
\item L2: LOWZ sample with  $0.31<z<0.43$
\item C1: CMASS sample with $0.43<z<0.54$
\item C2: CMASS sample with $0.54<z<0.7$
\end{itemize}

These redshift cuts are designed to  ensure that the signals cannot be compared with any other published values. Fine redshift bins were also desirable in order to minimise differences in the mean effective redshift across surveys (see Section \ref{effectivelensz}). 

We apply $w_{\rm tot}$ to the lens samples to ensure that the distribution of the randoms follows the variations in the lens samples. We also further test how our results vary if $w_{\rm tot}$ is not applied. 

In BOSS, redshift dependent effects are taken into account with the systematic weights. For example, the $w_{\rm star}$ weight includes a magnitude dependence via  $i_{\rm fib2}$ which accounts for redshift dependent variations in the number density.  Previous analyses of BOSS data have binned by redshift, most notable is the final DR12 cosmological analysis  which had arbitrary redshift binning across the combined LOWZ and CMASS samples \citep[][]{Alam:2017aa}.

Each lensing team has provided a {\sc healpix} mask \citep[][]{gorski2005} corresponding to the footprint of their shear catalogue. The BOSS lens and random catalogues are masked by each of the survey {\sc healpix} masks before computing $\Delta\Sigma$.

\section{Weak Lensing Data}\label{wldata}

This section provides brief descriptions on the various lensing data sets used in this paper. Readers are referred to the original survey and shear catalogue papers for the full details. The footprints of each of the lensing surveys involved in this collaboration are shown in Figure \ref{fig:map} together with the footprint of BOSS. These lensing surveys differ in terms of their location on the sky, coverage area, data quality, depth, and number of source galaxies. Beyond that, their analyses differ in shear and redshift calibration techniques. These differences are summarised in Table \ref{wlsurveytable}. 


\begin{table*}
  \centering
  \caption{Overview of lensing surveys used in this paper and methodologies used to compute $\Delta\Sigma$. First section: general properties of weak lensing surveys. We quote the survey area in deg$^2$ (after masking out bright stars and other artefacts), the characteristic seeing (FWHM), the photometric bands available for photometric redshift estimation, the median unweighted redshifts of the source distribution, and the effective weighted galaxy number density (see Equation 1 in \citealt{Heymans:2012}) after photo-$z$ quality cuts measured in galaxies per square arc-minutes.  Second section:  method used to compute photometric redshifts, calibration samples used to ensure unbiased redshifts ($z$-reference sample), and choices regarding whether the mean redshift or the full $p(z)$ distribution was calibrated to be unbiased ($z$-calibration type). Third section: choices for the selection of background galaxies (also see Section \ref{computeds}). All surveys use a galaxy-by-galaxy $z_{\rm phot}$ point estimate to select background galaxies. Fourth section: choices regarding the computation of $\Delta\Sigma$. This includes the redshift adopted for the computation of $\Sigma_{\rm c}$ (Equation \ref{eq:sigmacrit1}). Here choices differ with respect to the use of a point source estimate or the $p(z)$. As detailed in Section \ref{photozcorrections}, integrating over the $p(z)$ does not guarantee an unbiased estimate of $\Delta\Sigma$ unless $z$-calibration type is also of type $p(z)$. Finally, we also specify choices regarding the boost factor correction, $f_{\rm bias}$, and the dilution factor. For $f_{\rm bias}$, KiDS is marked with a star symbol to indicate that the method employed should be equivalent to an $f_{\rm bias}$ and dilution correction, but the methodology used is different (see Section \ref{kidsds}).}
\begin{tabular}{@{}lcccccc}
\hline
& SDSS & HSC-Y1 & CS82 & CFHTLenS & KiDS-VIKING-450 & DES-Y1 \\
\hline
Area [deg$^2$]& 9243 & 137 & 129.2 & 126 & 341 & 1321 \\
FWHM [$\arcsec$] & 1.2 & 0.58 & 0.6 & 0.6 & 0.66 & 0.96 \\ 
filters &  \textit{ugriz} & \textit{grizY} & \textit{ugriz} & \textit{ugriz} & \textit{ugriZYJHK$_{\rm s}$
} & \textit{griz}  \\ 
$z_{\rm med}$  & 0.39 & 0.80 & 0.57 & 0.7  & 0.67 & 0.59 \\
$n_{\rm eff}$ & 1.18 & 21.8 & 4.5 & 15.1 & 6.93 & 6.3 \\
\hline
$z$-name & {\sc zebra} & {\sc frankenz} & \textsc{bpz} & \textsc{bpz} & \textsc{bpz}$+$DIR & \textsc{bpz} \\
$z$-method & SED & Machine learning & SED & SED & kNN & SED \\
$z$-reference sample & SPECZ & SPECZ+COSMOS30 & SPECZ & none & SPECZ & COSMOS30 \\
$z$-calibration type & none & full $p(z)$ shape & mean $z_{\rm phot}$ & none  & full $n_{\rm s}(z)$ shape & mean of $p(z)$ \\
\hline
$z$-usage in source selection & Point estimate & Point estimate & Point estimate & Point estimate & Point estimate & Point estimate \\
Source selection cut 1 & $z_{\rm s} > z_{\rm L}$ & $z_{\rm s} > z_{\rm L} + 0.1$ & $z_{\rm s} > z_{\rm L} + 0.1$ & $z_{\rm s} > z_{\rm L} + 0.1$ & $z_{\rm s} > z_{\rm L} + 0.1$ & $z_{\rm s} > z_{\rm L} + 0.1$ \\
Source selection cut 2 & none & $z_{\rm s}> z_{\rm L} + \sigma_{68}$ & $z_{\rm s}> z_{\rm L} + \sigma_{95}/2.0$ & none & $0.1 < z_{\rm s} \leq 1.2$ & none \\
\hline
$\Sigma_{\rm c}$ computation & Point estimate & Point estimate & Point estimate & $p(z_{\rm s})$ & $n(z_{\rm s})$ & $p(z_{\rm s})$  \\
Boost factor correction & yes & no & no & no & no & yes \\
dilution correction & yes & yes & yes & no & yes$^*$ & yes \\
$f_{\rm bias}$ & yes & yes & yes & no & yes$^*$ & yes \\
\hline
\end{tabular}
\label{wlsurveytable}
\end{table*}

\subsection{SDSS}

The SDSS survey \citep{York:2000} imaged approximately 9000 square degrees of the sky.
		  We use the shape catalogue provided by \cite{Reyes:2012aa} which is based on the re-gaussianization technique developed by \cite{Hirata:2003aa}. Briefly, the algorithm uses
		  adaptive moments to measure the PSF-convolved 
		  galaxy shapes and then corrects for the PSF
		  using the adaptive moments of the measured PSF, while also accounting for the non-gaussianity of both PSF and the galaxy light profiles. The shear calibration factor ($1+m\sim1.04\pm0.02$) is derived using simulations performed by \cite{Mandelbaum:2012,Mandelbaum2018}.

Photometric redshift estimates for 
		 source galaxies were obtained by 
		 \cite{Nakajima:2012aa}, using the template fitting method {\sc zebra} 
		 \citep{Feldman2006} on $ugriz$ SDSS DR8 photometry. 
		 Following \cite{Nakajima:2012aa}, a representative spectroscopic sample is used to estimate and correct for the bias in $\Delta\Sigma$
		 caused by imperfect photometric redshifts.

\subsection{HSC}

The Wide layer of the Hyper Suprime-Cam Subaru Strategic Program aims to cover 1,400~deg$^2$ of the sky in $grizy$ using the Hyper Suprime-Cam \citep{Miyazaki:2018, Komiyama:2018} Subaru 8.2m telescope. The survey design is described in \citet[][]{Aihara:2018}, the HSC analysis pipeline is described in \citet[][]{Bosch:2018aa}, and validation tests of the pipeline photometry are described in \citet[][]{Huang:2018ac}. 

In this paper, we use the shear catalogue associated with the first data release \cite[DR1,][]{Aihara:2018aa}. This catalogue covers an area of 136.9 deg$^2$ split into six fields (see Figure \ref{fig:map}) and has a  mean $i$-band seeing of 0.58$\arcsec$ and a 5$\sigma$ point-source depth of $i\sim$26. We refer the reader to \citet[][]{Mandelbaum:2018ab} for details regarding the first year shear catalogue. Only a brief description is given here. For HSC Y1, galaxy shapes are estimated on the coadded $i$-band images using a moments-based shape measurement method and the re-Gaussianization PSF correction method \citep[][]{Hirata:2003aa}. The shear calibration is described in \citet[][]{Mandelbaum:2018aa}. The HSC Y1 shear catalogue uses a conservative source galaxy selection including a magnitude cut of $i<24.5$. The unweighted and weighted source number densities are 24.6 and 21.8 arcmin$^{-2}$, respectively. 

A variety of photometric redshifts have been computed for the HSC Y1 catalogue \citep[][]{Tanaka:2018aa}. Here we use the {\sc frankenz} photo-$z$'s described in \citet[][]{Speagle:2019aa}, which uses a hybrid method that combines Bayesian inference with machine learning. In brief, {\sc frankenz} derives photo-$z$'s for each object by computing a posterior-weighted average of the redshift distributions of its nearest photometric neighbours in the training set, taking into account observational uncertainties. 
The S16A HSC photo-$z$'s were trained on a catalogue of $\sim$300k sources including a combination
of spectroscopic, grism, prism, and many-band photometric redshifts covering a wide redshift,
colour, and magnitude range. Using the \texttt{best} photo-$z$ value from \citet[][]{Speagle:2019aa}, the source distribution in this paper has a mean redshift of $z_{\rm s}=0.95$ and a median of $z_{\rm s}=0.8$. A series of tests validating our galaxy-galaxy lensing measurements using {\sc frankenz} photo-z's can be found in \citet[][]{Speagle:2019aa}. 

\subsection{CS82}

The CS82 survey is 160 degrees$^2$ (before masking cuts are applied) of imaging data along the SDSS Stripe 82 region. We briefly summarise the key features of the CS82 weak lensing catalogue and refer the reader to  \citet[][]{Leauthaud:2017aa} for further details. CS82 is  is built from 173 MegaCam \citep[][]{Boulade:2003} $i$'-band images taken under excellent seeing conditions
(median seeing is 0.6\arcsec). The limiting magnitude of the survey is
$i$'$\sim$24.1. The images were processed based on the procedures presented in
\citet[][]{Erben:2009} and shear catalogues were constructed using the
same weak lensing pipeline developed by the CFHTLenS collaboration
using the \emph{lens}fit Bayesian shape measurement method
\citep[][]{Miller:2013}. A series of quality cuts is applied to construct the CS82 source catalogue (see \citealt[][]{Leauthaud:2017aa} for details). Shear calibration was performed using the same methodology as CFHTLenS.

Photo-$z$'s were computed from SDSS {\it ugriz} imaging by \citet[][]{Bundy:2015} using the Bayesian
photometric redshift software {\sc bpz} \citep[][]{Benitez:2000,
  Coe:2006}. The peak of the posterior distribution given by {\sc
  bpz}, $z_{\rm B}$, is used for sources redshifts, and a fiducial \photoz quality cut of {\sc odds}$>0.5$ is applied to reduce the catastrophic outlier rate. The CS82
survey overlaps with a number of spectroscopic surveys. Among these, the DEEP2
\citep[][]{Newman:2013a} catalogue spans the magnitude range of the CS82 and was the most useful in terms of assessing the photometric redshifts. A representative spectroscopic sample was used to estimate and correct for the bias caused by photometric redshifts \citep[][]{Leauthaud:2017aa}. After applying photo-$z$ quality cuts, the CS82 source catalogues corresponds to an effective weighted galaxy number density\footnote{Here we use
  $n_{\rm eff}$ as defined by Equation 1 in \citet[]{Heymans:2012}} of $n_{\rm eff}=4.5$ galaxies arcmin$^{-2}$.

\subsection{CFHTLenS}
CFHTLenS analysed 172 square degrees of imaging data from the wide component of the CFHT Legacy survey ($ugriz$ imaging to a 5$\sigma$ point source limiting magnitude of $i_{\rm AB}=25.5$). The observing strategy reserved the best seeing (seeing $< 0.8"$) conditions for the lensing $i$-band filter, the primary object detection filter, and follow-up with the other bands in the poorer seeing conditions.

The data reduction for CFHTLenS was conducted with the {\sc theli} pipeline \citep{Schirmer2004,Erben2005} following the procedures outlined in \citet{Erben:2013}. The dataset shares a similar data processing pipeline to KiDS, where the shape measurement of galaxies was conducted using the \emph{lens}fit model fitting code \citep{Miller:2013}. Shear multiplicative bias terms were characterised as a function of the signal-to-noise ratio and galaxy size using image simulations, thereby allowing for the calculation of the multiplicative bias term for an arbitrary selection of galaxies. 

Photometric redshifts, $z_{\rm B}$, were estimated using the Bayesian photometric redshift algorithm \citep[{\sc bpz},][]{Benitez:2000} and $ugriz$-band data. A probability distribution of true redshifts was estimated from the sum of the uncalibrated {\sc bpz} redshift probability distributions. As such the CFHTLenS analysis represents a snapshot of our best understanding of photometric redshift accuracy in 2012 \citep{Hildebrandt:2012}. This approach has since been demonstrated to carry systematic error \citep{Choi:2015}. Current weak lensing surveys focus on optimal methods to calibrate their photometric redshift distributions \citep[e.g.,][]{Tanaka:2018aa,Hildebrandt2016,Hoyle2018,Speagle:2019aa,Wright2020cfht,Buchs:2019}. 

For cosmic shear, \citet{Choi:2015} found the largest bias in the mean redshift of the source sample to be 0.04. This corresponds to a shift of $0.6 \sigma$ in the cosmological constraints for cosmic shear. However, the response of galaxy-galaxy lensing to redshift errors is different and the \citet{Choi:2015} results cannot be directly translated into errors on $\Delta\Sigma$. Instead, here we evaluate the impact of this photo-$z$ bias on $\Delta\Sigma$ and include this in the reported CFHTLenS systematic error budget (Section \ref{cfhtlensdsmeasure}).

\subsection{KiDS}
The KiDS survey \citep{Kuijken:2015} will span 1350 deg$^2$ on completion, in two patches of the sky with the \textit{ugri} optical filters, as well as forced-aperture photometry on 5 infrared bands from the overlapping VISTA Kilo-degree Infrared Galaxy (VIKING) survey \citep{Edge2013}, yielding the first well-matched wide and deep optical and infrared survey for cosmology and more accurate photometric redshifts.  It uses the wide-field camera, OmegaCAM, at the VLT Survey Telescope at ESO Paranal Observatory, optimally designed for lensing with high-quality optics and seeing conditions in the detection \textit{r}-band filter with a median of $ <0.7 \arcsec$. 

This paper uses 450 deg$^2$ of KiDS-VIKING 9-band imaging data
\citep[KV-450,][]{Wright2018}. With an effective, unmasked area of 360 deg$^2$, this dataset has an effective number density of $n_{\rm eff} = 6.93$ galaxies arcmin$^{-2}$. Galaxy shapes were measured from the $r$-band data using a self-calibrating version of \emph{lens}fit \citep{Miller:2013, fenech-conti/etal:2016}. A weight, $w_{\rm s}$, is also assigned based on the quality of the shape measurement. Utilising a large suite of image simulations, the multiplicative shear bias was deemed to be at the percent level for the entire KiDS ensemble \citep[][]{Kannawadi2019}. 

The redshift distribution for KiDS galaxies was determined via four different approaches, which were shown to produce consistent results in a cosmic shear analysis \citep{Hildebrandt2020}. The preferred method of that analysis, used here, is the `weighted direct calibration' (DIR) method, which exploits an overlap with deep spectroscopic fields. Following the work of \cite{Lima:2008}, the spectroscopic galaxies are re-weighted in 9-band colour space to obtain a true redshift distribution. A sample of KiDS galaxies is selected using their associated $z_{\rm B}$ value, estimated from the nine-band photometry as the peak of the redshift posterior output by {\sc bpz} \citep{Benitez:2000}. The resulting redshift distribution is well-calibrated in the range $0.1 < z_{\rm B} \leq 1.2$ (see \citealt{Wright2020cfht} for a detailed mock catalogue analysis that quantified the accuracy of the DIR method for a KV-450 like survey). 

\subsection{DES}

The DES survey conducted its first year of survey operation (Y1) between August 31, 2013 and February 9, 2014 \citep{DrlicaWagner2018} from the  4-meter Blanco Telescope and the Dark Energy Camera \citep{Flaugher2015}.
DES Y1 covers 
two non-contiguous areas near the southern galactic cap: The ``SPT'' area (1321 deg$^2$), which overlaps the footprint of the South Pole Telescope Sunyaev-Zel'dovich Survey \citep{Carlstrom2011}, and the ``S82'' area (116 deg$^2$), which overlaps the Stripe-82 deep field of the Sloan Digital Sky Survey \citep[SDSS;][]{Annis2014}.  Each area  within these footprints was revisited three to four times to reach sufficient photometric depth in the four $griz$ DES bands. In this paper, we only use the S82 area which overlaps with BOSS. 

For the DES Y1 data two independent shape catalogues were created: \textsc{metacalibration} \citep{Sheldon2017, Huff2017}, and \textsc{im3shape} \citep{Zuntz2013} both of which were found suitable for cosmological analyses. In the present study we only consider the \textsc{metacalibration} shape catalogue as it provides the larger surface source density of 6.28 arcmin$^{-2}$ over the full Y1 footprint. The \textsc{metacalibration} approach, instead of relying on calibrating shear bias from image simulations, makes use of the actual observed galaxy images to de-bias shear estimates, estimating a response $\mathsf{R}$ of measured ellipticity to shear. \textsc{metacalibration} also provides a photometric catalogue derived from its internal galaxy model fits. 

Photometric redshifts for the Y1 source catalogue were initially estimated using the \textsc{bpz} algorithm, and the mean redshift of the resulting sample of galaxies calibrated by matching to galaxies with high-quality photometric redshifts in COSMOS \citep{Laigle:2016aa} by magnitude, colour and size \citep{Hoyle2018}, and by cross-correlation with a photometric luminous red galaxy sample \citep{GattiVielzeuf2018,Davis2018}. To properly account for selection effects, the photometric redshifts were calculated with two different input photometries, one using the fiducial DES Y1 GOLD photometry catalogue \citep[][for $n(z)$ estimation]{DrlicaWagner2018}, and one using the \textsc{metacalibration} derived photometry catalogues (for selection and weighting of galaxies). The performance of the redshift estimates have been validated and \cite{McClintock2019} quantified the COSMOS-derived bias correction for $\Delta \Sigma$.

\section{Computation of $\Delta \Sigma$}\label{computeds}

This section describes how each team computed $\Delta\Sigma$. This section provides a snap-shot picture of each different team's approach to the computation of $\Delta\Sigma$ (also see Table \ref{wlsurveytable}). For the full details on the methodology, and tests regarding the validity of each computation, the reader is referred to survey specific papers. See Section \ref{ggl} for an introduction to terminology and for the definition of $\Delta\Sigma$.

\subsection{Computation of $\Delta\Sigma$ and notation}
\label{sec:notation}

Here we define common notation used in the computation of $\Delta\Sigma$. We then give the details of each team's specific computation.

\subsubsection{Redshifts and critical surface mass density}

Lenses have spectroscopic redshifts and their redshifts are noted $z_{\rm L}$. For source galaxies, redshift probability distributions are denoted as $p(z_{\rm s})$, point source estimates of redshifts are denoted $z_{\rm s}$, and an ensemble redshift distribution is denoted $n(z_{\rm s})$. Photometric redshifts are a noisy and, in some cases, a biased estimate of the true source redshift. For this reason boost, dilution and $f_{\rm bias}$ corrections are sometimes required when computing the critical surface density. 

Teams employ three different approaches for the computation of the critical surface mass density. First,  the critical surface mass density can be computed for each lens-source pair and with a source point source estimate following Equation \ref{eq:sigmacrit1}. This is the methodology employed by SDSS, HSC, CS82, and DES. 

Second, the critical surface density may be computed for a lens-source pair but using a $p(z)$. Here the inverse critical surface density is estimated:

\begin{equation}
\centering
\Sigma^{\rm inv}_{\rm c,pz}(z_{\rm L},z_{\rm s}) = \frac{4\pi G D_{\rm A}(z_{\rm L})}{c^2} \int^{\infty}_{z_{\rm L}} dz_{\rm s} \, {p}(z_{\rm s})  \frac{D_{\rm A}(z_{\rm L},z_{\rm s})}{D_{\rm A}(z_{\rm s})} \, ,
\label{sigma_crit_pz}
\end{equation}
and the critical surface density is then:
\begin{equation}
\centering
\Sigma_{\rm c,pz} = 1/\Sigma^{\rm inv}_{\rm c,pz}.
\end{equation}

If the per-source photometric redshift probability distributions are an accurate representation of the statistical and systematic redshift error, then this approach removes the necessity for a dilution or $f_{\rm bias}$ correction, when $p(z_s)$ is normalised as $\int_{0}^{\infty}p(z_s) dz_{\mathrm{s}}=1$.  As shown, for example in \citet[]{Hildebrandt2020}, however, the posterior redshift probability distribution functions estimated by {\sc BPZ}, are inherently biased.   As such, this approach is not recommended, but we include it nevertheless as this was the methodology originally employed by CFHTLenS in \citet{Ford:2015}, where additionally the $p(z_s)$ was normalised as $\int_{z_\mathrm{L}}^{\infty}p(z_s) dz_{\mathrm{s}}=1$ such that the dilution factor was unaccounted for.

Third, the critical surface mass density may also be computed for each lens galaxy with redshift $z_{\rm L}$ but for the ensemble source population (after lens-source separation cuts). In this case, the effective inverse critical surface mass density is noted $\overline{\Sigma}^{\rm inv}_{\rm c, nz}$ and is computed following:

\begin{equation}
\label{eqn:sigcritinv}
\centering
\overline{\Sigma}^{\rm inv}_{\rm c,nz}(z_{\rm L},{n}(z_{\rm s})) = \frac{4\pi G D_{\rm A}(z_{\rm L})}{c^2} \int^{\infty}_{z_{\rm L}} dz_{\rm s} \, {n}(z_{\rm s})  \frac{D_{\rm A}(z_{\rm L},z_{\rm s})}{D_{\rm A}(z_{\rm s})},
\end{equation}

\noindent where $\int^{\infty}_{0} n(z_{\rm z})=1$. The effective surface density is then:

\begin{equation}
\centering
\overline{\Sigma}_{\rm c,nz} = 1/\overline{\Sigma}^{\rm inv}_{\rm c,nz}.
\label{kidssigmacrit}
\end{equation}
\ch{If the ensemble redshift distribution estimate is an accurate and unbiased measurement of the true ensemble distribution (for example through calibration with an external spectroscopic sample) then both the dilution and $f_{\rm bias}$ correction are automatically accounted for with this approach} . 
This is the methodology employed by KiDS.

Testing of the equivalence between these different approaches is warranted and will be carried out using mock simulations in the DESI lensing mock challenge (Lange et al in prep).

\subsubsection{Weighting schemes}

An inverse variance weight is applied to lens-source pairs and is noted:

\begin{equation}
w_{\rm Ls} = \frac{\Sigma_{\rm c}^{-2}}{\sigma^2_{\rm e} + \sigma^2_{\rm rms}}
\equiv \frac{\Sigma_{\rm c}^{-2}}{\sigma_{\rm s}^2},
\label{eq:weight}
\end{equation}

\noindent where $\sigma_{\rm s}$ is the total shape noise, $\sigma_{\rm rms}$ is the intrinsic shape dispersion per component, 
and $\sigma_{e}$ is the per-component shape measurement error. For shape catalogues that use \emph{lens}fit, the \emph{lens}fit weight is $w_{\rm lf}^{-1} \sim \sigma_{\rm e}^2+\sigma_{\rm rms}^2$ and $w_{\rm lf}$ is used for weighting (note that in the notation used here, $w_{\rm Ls}$ includes the $\Sigma_{\rm c}^{-2}$ term whereas $w_{\rm lf}$ is the \emph{lens}fit approximation to the total shape noise).

DES uses a different weight, firstly because they choose to normalise the individual source's contribution to shear rather than in units of $\Sigma$, and secondly because they do not weight by the inverse shape noise variance of the individual source. The equation for the DES weight applied to each source's shape is  
\begin{equation}
w_{\rm Ls}^{\gamma} = \Sigma_{{\rm c,MCAL}}^{-1}\left(z_{{\rm L}}, z_{{\rm s, mean}}^{{\rm MCAL}}\right)\,
\label{eq:weightdes}
\end{equation}

\noindent where MCAL indicates a metacalibration redshift (see Section \ref{desds}). This weight can be thought of as a weight on $\gamma$, hence the $\Sigma_{\rm c}^{-1}$ term instead of the $\Sigma_{\rm c}^{-2}$ term used in equation \ref{eq:weight}. 

\subsubsection{f$_{\rm bias}$ correction factor}\label{fbiassection}

The $f_{\rm bias}$ correction factor accounts for biases that arise when converting $\gamma$ to $\Delta\Sigma$ using sources with photometric redshifts (see Section \ref{photozcorrections} and a more detailed derivation in Appendix \ref{fbiasappendix}). This term is computed using a representative sample of galaxies (hereafter called the ``calibration catalogue") following:

\begin{equation}
    f_{\rm bias}^{-1} = \frac{\sum_{\rm Ls}  w_{\rm calib, s} \ \sigma_s^2 \ \Sigma_{\rm c, Ls, P}^{-1} \ \Sigma_{\rm c, Ls, T}^{-1}}{\sum_{\rm Ls} w_{\rm calib, s} \ \sigma_s^2 \ \Sigma_{\rm c, Ls, P}^{-2}}\,,
    \label{fbias}
\end{equation}

\noindent where $\sigma_{\rm s}$ is shape noise of calibration sources, $\Delta\Sigma_{\rm crit, Ls, P}$ represents the (possibly biased) value of $\Delta\Sigma_{\rm crit}$ measured with \photozs, $\Delta\Sigma_{\rm crit, Ls, T}$ represents the true value of $\Delta\Sigma_{\rm crit}$, and the sum is performed over all possible pairs of lenses and sources from the calibration catalogue. The calibration weight, $w_{\rm calib}$ may account for: a) the sample variance of the calibration sample or b) colour differences between the overall source sample and the calibration sample. The form of $f_{\rm bias}$  written here includes the dilution effect by sources that scatter above $z_{\rm L}$ but which are actually located at lower redshifts than $z_{\rm L}$. Equation \ref{fbias} is written in terms of $f_{\rm bias}^{-1}$ because of the dilution factor and to avoid issues in the computation of $\Delta\Sigma_{\rm T}$ when $z_s<z_{\rm L}$ (resulting in an ill defined $\Sigma_{\rm c}$ term). The relation between $\Delta\Sigma_{\rm P}$ and $\Delta\Sigma_{\rm T}$ is:

\begin{equation}
\Delta\Sigma_{\rm T} = f_{\rm bias} \Delta\Sigma_{\rm P}. 
\end{equation}

\noindent DES employs a similar equation but without the shape noise weight\footnote{The DES Y1 catalogue does not have shape noise estimates for source galaxies but it is expected that later versions of DES source catalogues will include such estimates.}. Specifically, DES uses $\tilde{f}$ defined as:

\begin{equation}
    \tilde{f}_{\rm bias}^{-1} = \frac{\sum_{\rm Ls} w_{\rm calib, s} \Sigma_{\rm c, Ls, P}^{-1} \Sigma_{\rm c, Ls, T}^{-1}}{\sum_{\rm Ls} w_{\rm calib, s}  \Sigma_{\rm c, Ls, P}^{-2}}\,,
    \label{fbiasdes}
\end{equation}

\subsubsection{Effective Lens Redshift}\label{zeff}

Finally, each survey also computes the effective lens redshift for each of the samples. The effective redshift of each lens sample is

\begin{equation}
z_{\rm eff} = \frac{\sum_{\rm Ls} w_{\rm sys} w_{\rm Ls} z_{\rm L}}{\sum_{\rm Ls} w_{\rm sys} w_{\rm Ls}},
\end{equation}

\noindent where the sum is taken over all lens source pairs and $w_{\rm sys}$ is the BOSS systematic weight applied to each lens.

\subsection{SDSS}

The methodology of \cite{Singh:2018aa} is used to compute $\Delta\Sigma$. A photo-$z$ point estimate, $z_\text{s}$, is used to select source galaxies behind lenses ($z_\text{s}>z_\text{L}$) as well as to compute the $\Sigma_\text{c}$ factors and to weight each lens-source pair with the weighting scheme given in Equation \ref{eq:weight}. The maximum likelihood redshift is taken as the point source estimate. The representative spectroscopic sample from \citet{Nakajima:2012aa} is used to correct for biases arising from photometric redshifts. These corrections are of order $10\%$ (estimated at $\sim2\%$ accuracy, see also tests in \citealt{Singh:2018aa}) and increase with the effective redshift of the lens sample.

Following \cite{Mandelbaum:2005a}, the measurement around random points is subtracted to remove the additive systematic bias and also to obtain the optimal covariance \citep{Singh:2017}. $\Delta \Sigma$ is computed as a function of physical radius $r$ as:

\begin{equation}
{\Delta\Sigma}(r) = 
\frac{f_{\rm bias}}{2 \mathcal R (1+m)}({\Delta\Sigma}_{\mathrm{L}}(r) - {\Delta\Sigma}_{\mathrm{R}}(r)),
\label{eq:ds1}
\end{equation}
\noindent where ${\Delta\Sigma}_{\mathrm{L}}$ is the stacked signal around lens galaxies, 
${\Delta\Sigma}_{\mathrm{R}}$ is the stacked profile around a much larger number of random positions that share the same redshift distribution as lenses, and 
$f_{\rm bias}\approx1.09 (1.2)$ is the correction for photo-z calibration errors for the L1 and the L2 samples respectively.
This factor corrects both for photo-$z$ bias and the dilution of the signal caused by  sources that are below the lens redshift but get scattered above it due to photo-$z$ error. The $1+m$ term is the correction for the shear multiplicative bias with $1+m\sim0.96$. $\mathcal R$ is the shear responsivity 
factor. The SDSS lensing catalog employs a single $m$ and $\mathcal R$ value for all galaxies, defined at the full shape catalogue level. 

The signal around lens galaxies is computed as:

\begin{equation}
{\Delta\Sigma}_{\rm L}(r) = 
\frac{\Sigma_{\rm Ls}^{r} w_{\rm sys} w_{\rm Ls} \ \epsilon_{\rm t,Ls} 
\Sigma_{\rm c, Ls}}{\Sigma_{\rm Rs}^{r} w_{\rm sys} w_{\rm Ls}},
\label{eq:ds2}
\end{equation}

\noindent where $\sum_{\rm Ls}^{r}$ indicates a sum over all lens-source pairs with separation $r$.  The sum in the denominator is taken over random source pairs ($\Sigma_{\rm Rs}^{r}$) which applies a boost correction which is important at small scales ($r\leq 1$ [Mpc]). 
The signal around random points, ${\Delta\Sigma}_{\mathrm{R}}(r)$, is computed in a similar fashion to Equation \ref{eq:ds2} but the sums are taken over random-source pairs instead of lens-source pairs.

Shear calibration and photo-$z$'s are both estimated to be around the 2\% level \citep[][]{Reyes:2012aa,Nakajima:2012aa}. From  tests using cross correlations, photo-$z$ calibration uncertainty is around 5\%. We therefore quote 5\% as upper limit on the photo-$z$ calibration systematics. Adding these in quadrature yields an estimated $\sim 6\%$ systematic error.

The covariance of the measurements is estimated using jackknife method with 100 approximately equal area regions. The weighted mean redshifts of the L1 and L2 lens samples are $z_L=0.223$ and $z_{\rm L}=0.357$.

\subsection{HSC}\label{hscds}

The methodology described in \citet{Speagle:2019aa} is used to compute $\Delta\Sigma$. The HSC calculation closely follow the SDSS approach with a few differences that are highlighted below. The full details of the calculation, as well as a number of tests validating the robustness of the signals, can be found in \citealt{Speagle:2019aa}. The \texttt{best} photo-$z$ value 
from {\sc frankenz} is used as a point estimate for the photometric redshift for each source galaxy, $z_\text{s}$. The \texttt{medium} photo-$z$ quality cut from \citet[][]{Speagle:2019aa} is applied. This cut requires $\chi_5^2 \leq 6$ and $z_{\rm risk} \leq 0.25$, where $\chi_5^2$ describes the goodness-of-fit using a five-degree $\chi^2$ distribution and $z_{\rm risk}$ is the ``risk'' that the point estimate is incorrect 
as defined in \citet{Tanaka:2018aa}. These photo-$z$ cuts keep about 75\% of all source galaxies. Source-lens separation is performed by requiring $z_{\rm s}>z_{\rm L}+0.1$ and $z_{\rm s}>z_{\rm L}+\sigma_{68}$ where $\sigma_{68}$ is the 1-$\sigma$ confidence limit of the photo-$z$. In a similar fashion to Equation \ref{eq:ds1}, $\Delta \Sigma$ is computed as a function of physical radius $r$ following

\begin{equation}
{\Delta\Sigma}(r) = 
f_{\rm bias}({\Delta\Sigma}_{\mathrm{L}}(r) - {\Delta\Sigma}_{\mathrm{R}}(r)).
\label{eq:randomsubtraction}
\end{equation}

The signal around lens galaxies is computed as:

\begin{equation}
{\Delta\Sigma}_{\rm L}(r) = \frac{1}{2 \mathcal{R}(r) [1+K(r)]}
\frac{\Sigma_{\rm Ls}^{r} w_{\rm sys} w_{\rm Ls} \ \epsilon_t^{(\rm Ls)} 
\Sigma_{\rm c}^{(\rm Ls)}}{\Sigma_{\rm Ls}^{r} w_{\rm sys} w_{\rm Ls}}.
\label{eq:ds3}
\end{equation}

This equation is similar to Equation \eqref{eq:ds2} with three differences. First, the normalisation in the denominator is $\Sigma_{\rm Ls}^{r}$ instead of $\Sigma_{\rm Rs}^{r}$ (summed weights over lens-source pairs instead of random-source pairs) because boost factor corrections are not applied. Secondly, whereas SDSS uses a single value for $\mathcal R$, here we compute:

\begin{equation}
\mathcal{R}(r) = 1 - \frac{\Sigma_{\rm Ls}^{r} w_{\rm sys} w_{\rm Ls}  \ \sigma_{\rm rms, s}^2}{\Sigma_{\rm Ls}^{r} w_{\rm sys} w_{\rm Ls}},
\label{eq:rfactor}
\end{equation}

\noindent This is because in the HSC shape catalogue, $\sigma_{\rm rms}$ depends on galaxy properties like SNR and resolution (also see Equation 23 in \citealt{Speagle:2019aa}). Third, in HSC, the correction for multiplicative bias is $1/[1+K(r)]$ instead of $1+m$. This is because in HSC, each galaxy has an $m$ value (see \citealt{Mandelbaum:2018ab} for details about the calibration of HSC weak lensing catalogue). As described in \citet{Speagle:2019aa}, $K(r)$ is computed following:

\begin{equation}
K(r) = \frac{\Sigma_{\rm Ls}^{R} w_{\rm sys} w_{\rm Ls} \ m_{\rm s}}{\Sigma_{\rm Ls}^{R} w_{\rm sys} w_{\rm Ls}}.
\label{eq:kfactor}
\end{equation}

The signal around random points, ${\Delta\Sigma}_{\rm R}(r)$, is computed in a similar fashion to Equation \ref{eq:ds2} but the sums are taken over random-source pairs instead of lens-source pairs.

The COSMOS many-band catalogue \citep[][]{Laigle:2016aa} is used to compute corrections due to photo-$z$'s biases and dilution effects (the $f_{\rm bias}$ term). For these signals, the values for $f_{\rm bias}$ range between $f_{\rm bias}=1.00$ and $f_{\rm bias}=1.02$. In \citet{Speagle:2019aa}, a number of tests were performed on the robustness of the gg-lensing signal with regards to the photo-z calibration. Each source galaxy has quantities denoted $P_{\rm phot}$ and $F_{\rm phot}$ which indicate what kind of redshift it was primarily trained on (e.g. photo-$z$, spec-$z$, grism-$z$). By computing the gg-lensing signal with various values of $P_{\rm phot}$ and $F_{\rm phot}$,  \citet{Speagle:2019aa} showed that the gg-lensing signals are stable with respect to the origin of the training redshifts. 

To compute the uncertainty of the $\Delta\Sigma$ signal, lens and random samples are grouped into 41 roughly equal-area sub-regions. A $N=10000$ bootstrap re-sampling is used to estimate errors for $\Delta\Sigma$. The weighted mean redshifts of the four lens samples are $z_{\rm L}=0.23$, $z_{\rm L}=0.36$, $z_{\rm L}=0.49$, $z_{\rm L}=0.59$. The code used to compute $\Delta\Sigma$ (\textsc{dsigma}) is publicly available at \url{https://github.com/johannesulf/dsigma}. The systematic error is estimated to be of order 5\% (roughly Gaussian and 1$\sigma$).

\subsection{CS82}
The CS82 lensing signals are computed using the same code as HSC (\textsc{dsigma}). The main difference with \citet[][]{Leauthaud:2017aa} is that here the signal around random points in subtracted. But this does not have a large effect on the results and the derived signals are consistent with those derived in \citet[][]{Leauthaud:2017aa}. Photometric redshifts are derived using $ugriz$ photometry and the {\sc bpz} algorithm. Each source galaxy is assigned a point source redshift corresponding to the $z_{\rm B}$ value from {\sc bpz}. A cut of  $\rm{ODDS}>0.5$ was applied to the source catalogue in order to reduce the number of source galaxies with catastrophic redshift failures. Source background selection is performed by requiring that
$z_{\rm s}>z_{\rm L}+0.1$ and $z_{\rm s}>z_{\rm L}+\sigma_{95}/2.0$ where $\sigma_{95}$ is the 95 per cent confidence limit on the source redshift. \citet[][]{Leauthaud:2017aa} showed that the CMASS lensing signal did not vary when a more stringent lens-source separation scheme was employed. Boost factors were not applied. The $f_{\rm bias}$ term was applied with values ranging from $f_{\rm bias} = 0.98$ to $f_{\rm bias} = 1.03$ using a representative sample of spectroscopic redshifts (reweighed to match the colour and magnitude distribution of the source sample) described in \citet[][]{Leauthaud:2017aa}.

Errors on $\Delta\Sigma$ are computed via jack-knife. Because the same code is used as for HSC (\texttt{dsigma}), all other aspects of the calculation are as given in Section \ref{hscds}. The weighted mean redshifts of the four lens samples are $z=0.227$, $z=0.362$, $z=0.488$, and $z=0.586$. The systematic error is roughly estimated to be $\sim$6\% (roughly Gaussian and 1$\sigma$).

\subsection{CFHTLenS}\label{cfhtlensdsmeasure}


The photometric redshift probability distribution for each galaxy, $p(z_{\rm s})$, is computed from $ugriz$-band photometry using the {\sc bpz} algorithm, as well as a point estimate redshift per galaxy, $z_{\rm s}=z_{\rm B}$ \citep{Hildebrandt:2012}. Galaxies where the peak of their $p(z_{\rm s})$ are in the range $0.15<z_{\rm B}<1.3$ are used. The full redshift probability distribution is used to measure $\Delta \Sigma$ and $z_{\rm B}>z_{\rm L} + 0.1$ is requied. This lens-source separation has been shown to significantly reduce the amplitude of the boost correction \citep[see for example,][]{Amonir}.  The $p(z_{\rm s})$ is used to estimate $\Sigma^{\rm inv}_{\rm c,pz}$ (Equation \ref{sigma_crit_pz}) for each source pair. The weighted stacked $\Delta \Sigma$ is then calculated via

\begin{equation}
	\Delta\Sigma_{\mathrm{L}}(r) = \frac{1}{1+\overline{m}_{\rm s}} \frac{\sum_{\rm Ls} w_{\rm sys} w_{\rm lf} \, \epsilon_{{\rm t}} \, \Sigma_{{\rm c,pz}}^{\rm inv} }{ \sum_{\rm Ls}w_{\rm sys}{w_{\rm lf}} (\Sigma_{{\rm c,pz}}^{\rm inv})^2 }  \, ,
	\label{eq:dscfhtlens}
\end{equation}

The multiplicative bias correction, $\overline{m}_{\rm s}$, is calculated for a given lens sample as

\begin{equation}
    \overline{m}_{\rm s} = \frac{\sum_{\rm s} w_{\rm sys}  w_{\rm lf} \ m_{\rm s}}{\sum_{\rm s} w_{\rm sys} w_{\rm lf}} \, ,
\end{equation}

\noindent where $m_{\rm s}$ is the per galaxy multiplicative bias and $w_{\rm lf}$ is the \emph{lens}fit weight. The difference with regards to equation \ref{eq:kfactor} used by HSC is that this equation uses $w_{\rm lf}$ instead of $w_{\rm Ls}$. The difference between these two quantities is that $w_{\rm Ls}$ includes a $\Sigma_{\rm c}^{-2}$ term.

The signal around random lenses is subtracted from the signal around the lenses,
\begin{equation}
{\Delta\Sigma}(r) = {\Delta\Sigma}_{\mathrm{L}}(r) - {\Delta\Sigma}_{\mathrm{R}}(r) \, .
\end{equation}

Following \citet{Ford:2015}, boost, dilution, and $f_{\rm bias}$ correction factors are not calculated or applied.  The error that is then incurred is accounted for in this analysis with a significant systematic error budget. The methodology of  \citet{Filaments2020} is used to compute a systematic error due to the error in the uncalibrated photometric redshifts, $p(z_{\rm s})$. A photo-$z$ shift of $\delta z_s=0.04$ is used to capture the photo-$z$ bias found by \citet{Choi:2015}. The $p(z_{\rm s})$ is shifted by $\pm\delta z_{\rm s}$ and two new functions $\Sigma^{\rm inv}_{\rm c}(z_{\rm l})^{\pm}±$ are computed. The full measurement and error analysis is repeated using both the $\Sigma^{\rm inv}_{\rm c}(z_{\rm l})^+$ and $\Sigma^{\rm inv}_{\rm c}(z_{\rm l})^-$. The difference, $\Delta\Sigma_{\rm bias}$, is averaged over all scales. This photo-z uncertainty is the main systematic uncertainty for CFHTLenS. This systematic error is estimated to be up to 6\% for the LOWZ lens sample and up to 10\% for CMASS (roughly Gaussian and $1 \sigma$).  After unblinding, a 5\% systematic error on $m$ \citep[][]{Kuijken:2015,Kilbinger:2017} was also included. This increased the systematic errors but did not change any of the main conclusions. The final numbers are reported in Table \ref{syserrtable}.

Statistical errors are computed via bootstrapping over measurements using 1000 patches. The weighted mean redshifts of the four lens samples are $z_{\rm L}=0.23$, $z_{\rm L}=0.36$, $z_{\rm L}=0.49$, $z_{\rm L}=0.60$. 

\subsection{KiDS}\label{kidsds}

The KiDS lensing signal is computed similarly to the methodology outlined in \citet[][]{Dvornik2018} and \citet[][]{Amon2018}. A point estimate of the photometric redshift per galaxy, $z_{\rm B}$, is derived using $ugriZYJHK_s$ photometry and the {\sc bpz} algorithm. This redshift is used to define the source samples and for source-lens separation. The source galaxy sample is first limited to  $0.1 < z_{\rm B} \leq 1.2$. \ch{Then, further source-lens separation cuts are applied to significantly reduce the amplitude of the boost correction. These are defined as $z_{\rm B} > z_{\rm L}+0.1$, following tests in \citet[][]{Amonir}.} 

\ch{The ensemble redshift distribution of the source sample behind each lens, $n(z_{\rm s}|z_{\rm L})$, is estimated using a direct calibration method (DIR) that employs a diverse and representative set of spectroscopic samples \cite[]{Hildebrandt2020}. Specifically DIR calibrated redshift distributions $n(z)$ are determined for a series of photometric redshift $z_{\rm B}$ slices of width 0.1. A critical surface density is then computed (equation~\ref{kidssigmacrit}) for a series of discrete lens values ($z_{\rm i,L}=0.0,0.1,0.2$ etc.) and a composite DIR-calibrated source redshift distribution of `background' galaxies with $z_{\rm B}>z_{\rm i,L}+0.1$. Linear interpolation is then used to compute the critical surface density for each lens in the full KiDS-BOSS sample.   If the DIR-calibration results in an unbiased and accurate representation of the true source redshift distribution then both the dilution and $f_{\rm bias}$ correction are already included with this approach.

The multiplicative shear calibration correction \citep[][]{Kannawadi2019} is estimated for the ensemble source and lens galaxy population.  Extending the method described in \cite{Dvornik2018} to the higher KV-450 redshifts, the shear calibration is estimated for 11 linear source photometric redshift bins between $0.1<z_{\rm B} \leq 1.2$. These corrections are then optimally weighted and stacked following:
\begin{equation}
	\overline{m}=\frac{\sum_{\rm i} w_{\rm i}' m_{\rm s}}{\sum_{\rm i} w_{\rm i}'} \, ,
	\label{eq:biascorr}
\end{equation}
where $w' = w_{\rm s} D(z_{\rm L},z_{\rm s}) / D(z_{\rm s})$. The resulting correction $\overline{m}\approx-0.014$ is independent of the distance $r$ from the lens, and reduces the effects of multiplicative bias to within $\pm 2\%$ \citep[][]{Kannawadi2019}.}

The signal around lens galaxies is computed: 

\begin{equation}
	\Delta\Sigma_{\mathrm{L}}(r) = \frac{1}{1+\overline{m}} \frac{\sum_{\rm Ls} w_{\rm sys}w_{\rm Ls} \, \epsilon_{{\rm t}} \, \overline{\Sigma}_{{\rm c,nz}} }{ \sum_{\rm Ls}w_{\rm sys}{w_{\rm Ls}} }  \, ,
	\label{eq:dskids}
\end{equation}

\noindent where $\overline{\Sigma}_{{\rm c,nz}}$ is given in Equation \ref{kidssigmacrit}. 

The signal around random lenses is subtracted as follows:

\begin{equation}
{\Delta\Sigma}(r) = {\Delta\Sigma}_{\mathrm{L}}(r) - {\Delta\Sigma}_{\mathrm{R}}(r) \, .
\end{equation}
Errors are computed using a bootstrap method using regions of 4 $ {\rm deg}^2$. The weighted mean redshifts of the four lens samples are $z_{\rm L}=0.23$, $z_{\rm L}=0.36$, $z_{\rm L}=0.49$, $z_{\rm L}=0.58$.

Similar to the method employed by CFHTLenS, KiDS computes a contribution to the systematic uncertainty due to the error in the sample's calibrated redshift distribution, $n(z_{\rm s})$, by reporting an additive systematic error. This is determined by propagating $-0.06< \delta z_s < 0.014$, as advised by \citet{Wright2020cfht}.  The difference between the two measurements, $\Delta\Sigma_{\rm sys}$, is averaged over all scales and taken as the systematic error. This systematic error is estimated to be up to 2\% for LOWZ and up to 3\% for CMASS. This is the dominant systematic uncertainty for the KiDS measurements. After unblinding, the systematic error on $m$ as estimated in \citet[][]{Kannawadi2019} was also included. This increased the systematic errors by 1\% but did not change any of the main conclusions. The final numbers are reported in Table \ref{syserrtable}.

\subsection{DES}\label{desds}

The DES lensing signal is computed following the methodology outlined in \cite{McClintock2019} and using the \textsc{metacalibration} weak lensing source galaxy catalogue for DES Y1 \citep{Zuntz2018}.

The \textsc{metacalibration} algorithm \citep{Huff2017,Sheldon2017} provides estimates on the ellipticity $\epsilon$ of galaxies, the response of the ellipticity estimate on shear $\mathsf{R}_\gamma$, and of the ensemble mean ellipticity on shear-dependent selection $\mathsf{R}_\mathrm{sel}$. These are applied in the shear estimator to correct for the bias of the mean ellipticity estimates.

The DES shear response is broken into two terms: $\mathsf{R}^{\rm T}_{\gamma,s}$ is the shear response measured for individual galaxies, averaged over both ellipticity components, and $\langle \mathsf{R}^{\rm T}_\mathrm{sel} \rangle$ is the shear response of the source selection. The latter is a single mean number computed for each source galaxy ensemble. The DES catalogue also contains a multiplicative bias correction term (one number per source catalogue, similar to SDSS).

Two different photometric redshift estimates are used. The first is based on fluxes measured in the \textsc{metacalibration} process. In this case, the redshift used is the mean of the $p(z)$ estimated from the \textsc{metacalibration} photometry and is denoted $z^{\rm MCAL}_{\rm mean}$. The second is a random draw from the $p(z)$ estimated from the Y1 GOLD MOF photometry \citep{DrlicaWagner2018} (hereafter denoted $z^{\rm MOF}_{\rm MC}$). Both are estimated using the \textsc{bpz} algorithm \citep{Hoyle2018}. In order to properly account for the selection response term, the \textsc{metacalibration} redshifts are  used for source selection ($z^{\rm MCAL}_{\rm mean}>z_{\rm L}+0.1$) and for the weight, $w_{\rm Ls}^{\gamma}$. The $z^{\rm MOF}_{\rm MC}$ redshifts, preferable due to the higher quality of the photometric information, are used to convert shear to $\Delta\Sigma$. This sample can be at $z^{\rm MOF}_{\rm{MC}}<z_{\rm L}$ despite the $z^{\rm MCAL}_{\rm mean}$-based source selection selection. That fact that different redshifts are used to weight the signal and to compute $\Delta\Sigma$ requires a modified $\Delta\Sigma$ estimator, described below.

Equation \ref{eq:sigmacrit1}, and point source redshifts, are used to compute the critical surface density. However, $z^{\rm MCAL}_{\rm mean}$ is used for $w_{\rm Ls}^{\gamma}$ and 
$z^{\rm MOF}_{\rm MC}$ is used for $\Sigma_{\rm c,MOF}$.

Photometric redshift estimates and their associated uncertainties are calibrated using the \citet{Laigle:2016aa} COSMOS photometric redshifts and using the algorithms described in \citet{Hoyle2018} and \citet{McClintock2019}. Unlike the DES shear two-point functions \citep{Prat2018,Troxel2018,desy1kp}, the calibration of redshifts for $\Delta\Sigma$ are not refined by the result of the cross-correlation techniques \citep{Gatti2017,Davis2018}\footnote{This analysis uses source galaxies at $z>0.9$ where there is a dearth of spectroscopic galaxies for calibration purposes. See Figure 3 in \citet[][]{GattiVielzeuf2018}}. The consistency of the two \citep{Hoyle2018}, however, is evidence for the validity of the former.

The lensing estimator is given by

\begin{equation}
	\label{eq:updated_deltasigma_estimate}
	\Delta\Sigma(r) = \frac{1}{1+f_{\rm cl}}\frac{1}{\hat{f}_{\rm bias}^{-1}+m} \left(\Delta\Sigma_L(r) - \Delta\Sigma_R(r)\right) ,
\end{equation}	

\noindent with the signal around lens galaxies estimated as

\begin{eqnarray}
    \Delta\Sigma_L(r) & = & \frac{\sum\limits_{\rm{Ls}} w_{\rm sys}\,w_{\rm Ls}^\gamma\, \Sigma_{\rm c, Ls,MOF}^{-1}}{\sum\limits_{\rm{Ls}} w_{\rm sys}\,w_{\rm Ls}^\gamma\, \Sigma_{\rm c, Ls,MOF}^{-1}\, \left(\mathsf{R}^{\rm T}_{\gamma,s}  + \langle \mathsf{R}^{\rm T}_\mathrm{sel} \rangle \right)} \nonumber \\
    & \times & \frac{\sum\limits_{\rm{Ls}}w_{\rm sys}\,w_{\rm Ls}^\gamma\,  \epsilon_{\rm t}^{\rm Ls} }{\sum\limits_{\rm{Ls}} w_{\rm sys}\,w_{\rm Ls}^\gamma\, \Sigma_{\rm c, Ls,MOF}^{-1}} \; .
    \label{eq:mv12}
\end{eqnarray}

\autoref{eq:mv12} is equivalent to Equation 12 in \citet{McClintock2019}. Here, we have ordered the terms for comparison with the estimators used by other surveys. For instance, the correction by the mean response in the first term here is similar to the $1/(1+K)$ term in \autoref{eq:ds3} and the $1/(1+m)$ term in Equation \ref{eq:dskids}. The second term can be interpreted as a weighted mean tangential ellipticity in the nominator, normalised by a weighted mean $\Sigma_{{\rm c}}^{-1}$ estimated from MOF photometry. These use the \textsc{metacalibration}-derived weights of \autoref{eq:weightdes}. 

\autoref{eq:updated_deltasigma_estimate} subtracts the signal around random points and corrects it for systematic errors in photometric redshifts through $\hat{f}_{\rm bias}^{-1}$ and shear through a multiplicative bias correction $m$. Terms proportional to $\hat{f}_{\rm bias}^{-1}\times m$ are neglected. The signal is divided by $(1+f_{\rm cl})$ to apply a boost factor. Here $f_{\rm cl}$ is the fractional contribution from galaxies falsely identified as sources to the weighted mean shear, estimated using $p(z)$ decomposition \citep[][]{Varga2019, Gruen2014}. All correction terms are defined and estimated as in \citet{McClintock2019}\footnote{The impact of systematic weights is expected to be minor on the recovered boost factors, and as a computational simplification were assumed to be unity with respect to the boost factor calculation.}.  The analysis setup used in these calculations is made publicly available in the \textsc{xpipe} package\footnote{\url{https://github.com/vargatn/xpipe}}.

The measurement used here differs from various other DES analyses where systematic uncertainties were incorporated at the model/likelihood level, and their amplitudes varied according to their respective prior. In the present study we apply the correction directly to the data vector, while estimating the corresponding systematic uncertainties for each lens redshift bin and for the inner and outer radial ranges respectively. Shear calibration and photometric redshift systematic errors are estimated using the methodology of  \cite{McClintock2019}. The combined systematic uncertainty is estimated to be 2$\%$ for the three lower redshift bins, and 3$\%$ for the highest redshift bin. When incorporating the covariance of boost factor estimates to the net systematic error budget of the different radial ranges, we find a combined upper limit for the different radial ranges across all lens redshift bins respectively at the level of 2-3\%. 

The weighted mean redshifts of the four lens samples are $z_{\rm L}=0.23$, $z_{\rm L}=0.36$, $z_{\rm L}=0.49$, $z_{\rm L}=0.59$. 

\section{Homogeneity of BOSS samples}\label{homogenity}

The validity of the tests we seek to perform rely on the assumption that BOSS selects a homogeneous sample of foreground galaxies living in similar dark matter halos. This assumption may be invalid if the properties of the CMASS and LOWZ samples (e.g. luminosity, colour, stellar mass) vary spatially (each survey's submitted measurement is performed on a different patch of
the sky and therefore with different samples drawn from BOSS). The BOSS clustering team identified several factors leading to inhomogeneity in the BOSS samples \citep[][]{Ross:2011aa,Ross:2012,Reid:2016,Ross:2017aa}. The goal of this section is to investigate inhomogeneity in the sub-regions probed by each survey footprint.

\subsection{Overall homogeneity}

We first study the overall homogeneity of the CMASS and LOWZ samples. We apply the masks of each of the lensing surveys to the BOSS catalogues to extract distributions of colour, \ic, \ifib, $z$, and \lm~within each of the subregions. For M$^*$, we use the Granada masses\footnote{\url{https://www.sdss.org/dr16/spectro/galaxy_granada/}}
 described in \citet[][]{Ahn:2014}. Figure \ref{fig:homoghist} and Table \ref{photometrictable} demonstrate that the basic properties of the two samples are spatially homogeneous across the regions of interest. A further visual confirmation of this is provided by Figure \ref{fig:a1spatialmaps}.

\begin{figure*}
\centering
\includegraphics[width=\linewidth]{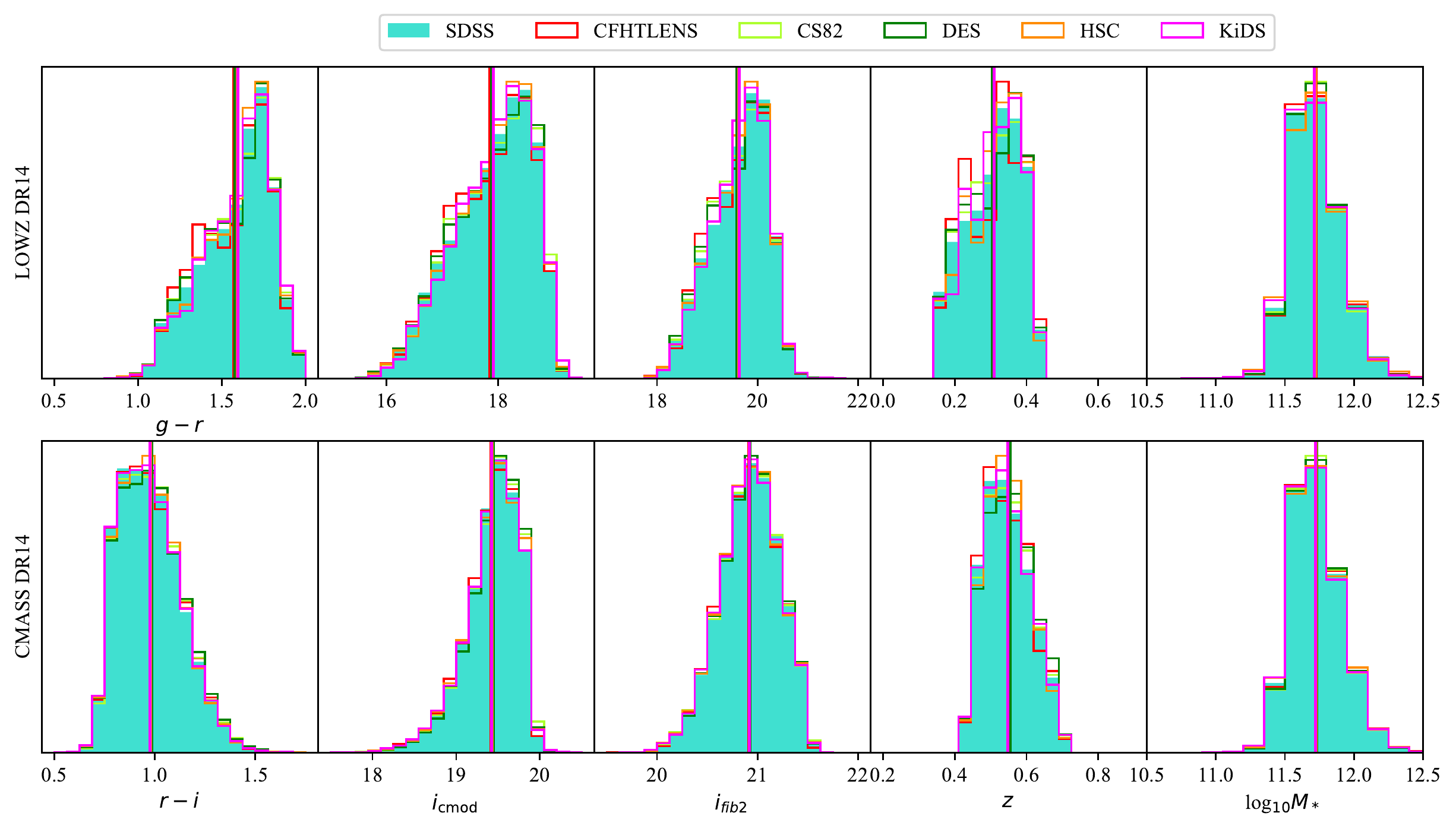}
\caption{Distributions of colour, \ic, \ifib, $z$, and \lm~for LOWZ (upper panels) and CMASS (lower panels) in the regions of overlap with each of the lensing surveys. The DR14 catalogue with photometry tied to Pan-STARRS 1 (PS1) as described in \citet[][]{Finkbeiner:2016} was used for magnitudes. The CMASS selection includes a cut at $i_{\rm fib2}<21.5$ (Equation \ref{eq:cmass}). The reason this sharp cut is not apparent in this Figure is because of scatter between DR14 and the original CMASS targeting catalogue. }
\label{fig:homoghist}
\end{figure*}

\begin{table*}
  \caption{Mean values of photometric quantities for the overall CMASS and LOWZ samples.  This table also provides differences between the mean quantities in each lensing survey footprint and the overall BOSS samples. We consider the colour differences used in the LOWZ (CMASS) selections. We use ($r-i$) for CMASS and ($g-r$) for LOWZ.}
\begin{tabular}{@{}lcccc}
\hline
Mean Values & $i_{\rm cmod}$ & $g-r$ & $r-i$ & $\log_{10}M_*$ \\
\hline
SDSS CMASS  & 19.4186 $\pm$ 0.0004  & - & 0.9780 $\pm$ 0.0002& 11.7264 $\pm$ 0.0002
\\
SDSS LOWZ  & 17.694 $\pm$ 0.001 & 1.5126 $\pm$ 0.0004 & - & 11.7227 $\pm$ 0.0003 \\
\hline
Differences& $\Delta(i_{\rm cmod})$ & $\Delta(g-r)$ & $\Delta(r-i)$ & $\Delta(\log_{10}M_*)$ \\
\hline
& & CMASS &  &  \\
\hline
HSC Y1 & 0.0054 $\pm$ 0.003 & - & 0.0028 $\pm$ 0.001 & 0.001$\pm$0.001\\
DES Y1 & 0.014 $\pm$ 0.003 & - & 0.0044 $\pm$ 0.001 & 0.005$\pm$0.001\\
KiDS & 0.0104 $\pm$ 0.002 & - & -0.0011 $\pm$ 0.001 & -0.004$\pm$0.001\\
CFHTLenS & -0.00036 $\pm$ 0.003 & - & -0.0017 $\pm$ 0.001 & 0.003$\pm$0.002\\
CS82 & 0.020 $\pm$ 0.002 & - & 0.0064 $\pm$ 0.001 & 0.003$\pm$0.001 \\
\hline
& & LOWZ &  &  \\
\hline
HSC Y1 & -0.0033 $\pm$ 0.01  & 0.009$\pm$0.003 &- & -0.006$\pm$0.002 \\
DES Y1& 0.0202 $\pm$ 0.008 &   -0.007$\pm$0.002 &- & -0.004$\pm$0.002 \\
KiDS & 0.0205 $\pm$ 0.009 &  0.011$\pm$0.003 &-& -0.013$\pm$0.002 \\
CFHTLenS & -0.01* $\pm$ 0.01  &  -0.013$\pm$0.004 &-& -0.002$\pm$0.002 \\
CS82 & 0.024 $\pm$ 0.007  & -0.002$\pm$0.002 & - & -0.008 $\pm$0.002 \\
\hline
\end{tabular}
\label{photometrictable}
\end{table*}

\subsection{Galactic hemisphere}\label{galactichemisphere}


SDSS imaging is carried out into two large contiguous areas in the North Galactic Cap (NGC) and the South Galactic Cap (SGC). Figure \ref{fig:map} displays the overlap between the NGC, the SGC, and the lensing surveys. The SDSS lensing catalogues have the most overlap with BOSS and cover most of the NGC and about half of the SGC. CFHTLenS and HSC have fields in both the NGC and the SGC. KiDS only overlaps with BOSS in the NGC. CS82 and DES only overlap with BOSS in the SGC. It is thus important to understand whether or not BOSS samples give rise to the same lensing signals in the NGC and the SGC. 

\citet[]{Schlafly:2010aa} and \citet[]{Schlafly:2011aa} found photometric offsets between the NGC and the SCG. The photometric calibration of the DR13 catalogue (which came after BOSS targeting was complete) was tied to Pan-STARRS 1 (PS1), as described in \citet[][]{Finkbeiner:2016}. This procedure led to new flat fields and zero points in the $g$, $r$, $i$, and $z$ bands, and new flat fields (but not new zero points) in the $u$ band. The updated photometry results in a 0.015 magnitude difference in $c_{\parallel}$ compared to when BOSS targeting was performed.



\citet[]{Ross:2017aa} find a 1\% difference in the number density of CMASS between the NGC and the SGC. Differences for LOWZ are larger:  the projected density of LOWZ is 7.6\% higher in the SGC compared to the NGC. \citet[]{Ross:2012} find these differences to be consistent with the level of color offsets determined by \citet[]{Schlafly:2011aa} and that the sliding cut $c_{\parallel}$ imparts the largest differences. Because of the small color offsets, the North and the South may correspond to slightly different galaxy populations. As a result, the BOSS team treats galaxies in the North and in the South as two separate samples.

Figure A1 in \citet{Alam:2017aa} investigates the differences in the clustering scales between the NGC and the SGC (also see \citealt[][]{Lee2019} for angular clustering $w(\theta)$). The power spectrum of CMASS is consistent for both hemispheres over the scales $0.05<k~[h^{-1} {\rm Mpc}]<0.3$. There is, however, an amplitude shift in the power spectrum $P(k)$ (a 4\% shift in the amplitude of the power spectrum monopole) for LOWZ but it can be explained by taking into account the color shifts between SDSS photometry in the north and south described previously (see Appendix A in \citealt{Alam:2017aa} and \citealt[][]{Lee2019}). 

  
    

Because galaxy number counts are steep \footnote{\report{the number of galaxies in a sample rises steeply as a function of the limiting magnitude of the sample.}}, a small color offset can easily result in variations in number density, but the variations in the galaxy (dark matter halo) selection may still be comparatively small. Using the calculation outlined in Appendix \ref{onep_flux_calibration}, and assuming the \citet[][]{Leauthaud:2012a} stellar-to-halo mass relation, a 0.015 shift in flux corresponds to a halo mass shift of 0.006 dex. This imparts less than a 1\% shift on $\Delta\Sigma$. However, these are only rough estimates based on the impact of magnitude shifts on $M^*$. Without a detailed understanding of the impact of the color offsets on the galaxy (and underlying halo) selection, it is not trivial to translate differences in clustering amplitudes or number densities into differences in $\Delta\Sigma$. We therefore implement additional tests of the impact of North versus South via two alternative methods. First, we estimate the mean shift in $M^*$ in different regions directly. Second, we use lensing from SDSS to perform a direct and empirical test on potential shifts in $\Delta\Sigma$.

We first consider the DR14 catalogue with updated and better calibrated photometry from Pan-STARRS 1 (PS1), as described in \citet[][]{Finkbeiner:2016}. \eric{We take the existing LOWZ and CMASS catalogues and cross-correlate them with the DR14 photometry}\footnote{\eric{\citet[][]{Lee2019} studies how the sample changes with cuts applied to DES photometry, and a similar behavior might be seen in the case of cuts applied to Pan-STARRS photometry.}}. Figure \ref{fig:homoghist} shows that the updated photometry does not have a large impact on distributions in color, \ic, \ifib, $z$, and \lm.  Table \ref{overlaptable} lists differences between the mean values of these quantities and the overall BOSS sample. Differences are small and photometric calibration should therefore not impact $\Delta\Sigma$ (also see Appendix \ref{onep_flux_calibration}).

We now carry out a more direct test of the impact of differences between the North and the South on $\Delta\Sigma$. The SDSS survey is the only lensing survey with enough coverage in both hemispheres to perform a direct test (see Figure \ref{fig:map}). SDSS is too shallow to accurately measure lensing for CMASS, so this test is limited to the LOWZ sample. However, as described above, we expect such effects to be more important for LOWZ than for CMASS. Figure \ref{fig:lowz_NS} shows $\Delta\Sigma$ measurements obtained for the  full LOWZ sample as well as from the North and South regions separately. Errors shown in the figure are obtained using jackknife (68 regions for the North, 32 regions for the South and 100 regions for the full sample). When computing the ratio, we add the covariances from the North and South regions assuming that they are independent following: $\text{cov}(R)/R^2=\text{cov}(\Delta\Sigma_N)/\Delta\Sigma_N^2+\text{cov}(\Delta\Sigma_S)/\Delta\Sigma_S^2$. The results from both regions are consistent with the difference being $5\pm8\%$. Lensing from SDSS yields the highest signal-to-noise measurements for LOWZ - hence if such differences are not detectable with SDSS, they are also not detectable with our other lensing data sets. Based on Figure \ref{fig:homoghist} and Figure \ref{fig:lowz_NS}, we conclude that differences between North and South are not a concern for the present study. Nonetheless, for completeness, we also perform a post-unblinding test on the impact of North versus South in Section \ref{postblindhomogtest}.

\begin{figure}
  \centering
  \includegraphics[width=\linewidth]{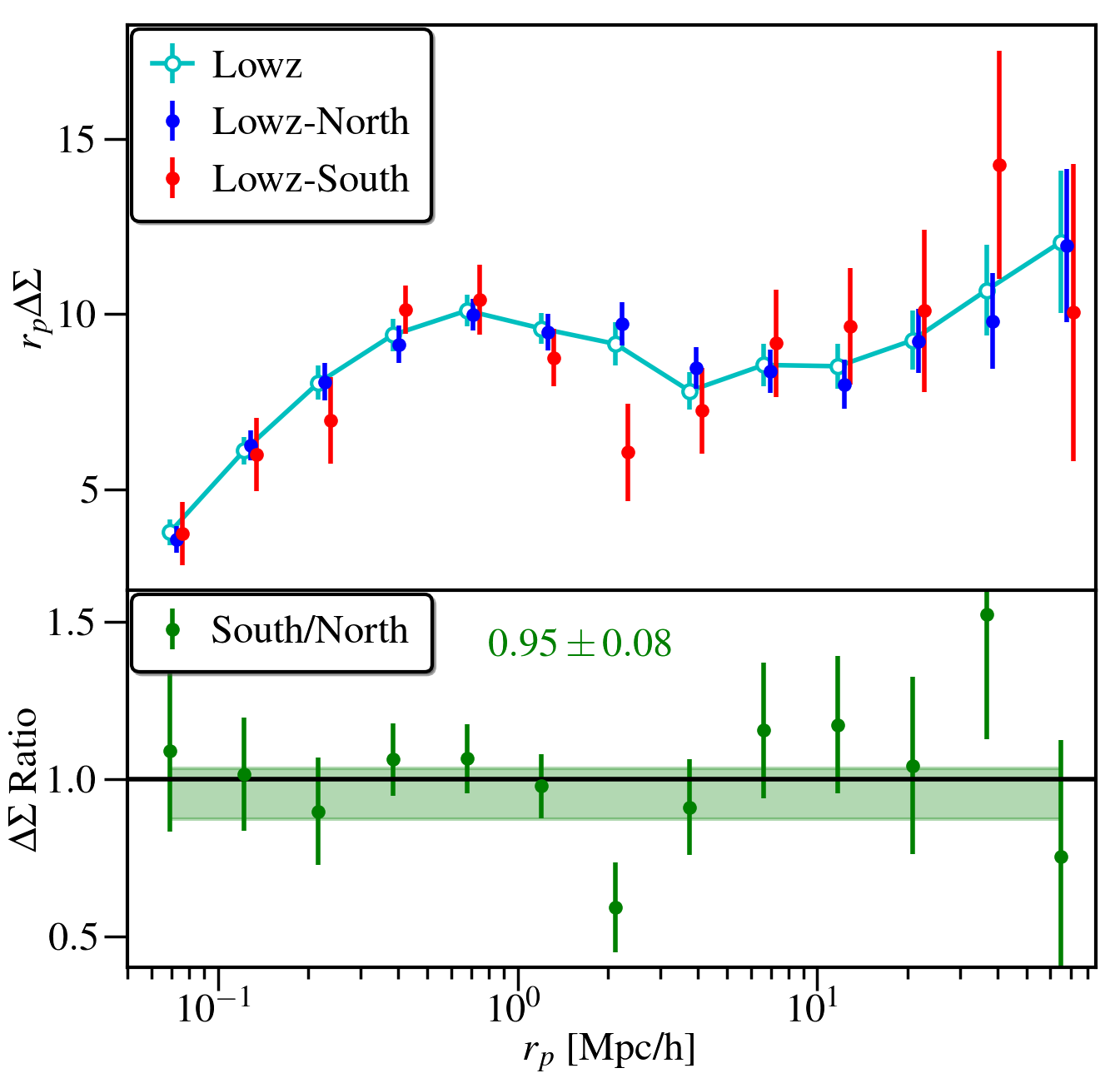}
  \caption{Impact of North versus South photometry differences on the $\Delta\Sigma$ signal for LOWZ. Upper panel: $\Delta\Sigma$ measured for the full LOWZ sample as well as for	the North and South regions separately. Signals are computed using the SDSS lensing catalogue. Lower Panel: ratio of $\Delta\Sigma$ from each hemisphere. The bands and the number quoted in the text are the mean values of the ratio obtained by fitting a constant to values between $0.5<r_{\rm p}<15$ Mpc. The amplitude of $\Delta\Sigma$ is consistent between both hemispheres.}
  \label{fig:lowz_NS}
\end{figure}

\subsection{Stellar density and seeing}

\citet[]{Ross:2012} investigated how the number density ($n_{\rm gal}$) of DR9 BOSS galaxies varied with stellar density, seeing, Galactic extinction, and sky background (in the imaging that was used for targeting). No effects were found for the LOWZ sample\footnote{with the exception of the LOWZE2 and LOWZE3 samples which are excluded from the present study.}. Larger effects were detected for CMASS. Differences between CMASS and LOWZ are explained by the fact that LOWZ galaxies are on average considerably brighter than CMASS galaxies \citep[][]{Tojeiro:2014aa}. 

For CMASS, stellar density ($n_{\rm star}$) was found to have the largest impact on $n_{\rm gal}$. The relationship between $n_{\rm gal}$ and $n_{\rm star}$ was found to depend on galaxy surface brightness where \ifib~ was used as a proxy for surface brightness. As the stellar density increases, on average, galaxies with lower magnitudes in a 2$\arcsec$ fiber are lost from the sample. In \citet{Reid:2016}, the functional form for $w_{\rm star}$ is given by:

\begin{equation}\label{eq:wstar}
w_{\rm star}(n_{\rm star},i_{\rm fib2})=(A_{\rm ifib2}+B_{\rm ifib2} n_{\rm s})^{-1}
\end{equation}

\noindent where $A_{\rm ifib2}$ and $B_{\rm ifib2}$ depend on $i_{\rm fib2}$. \citet[]{Ross:2012} found that applying $w_{\rm star}$ accounts for observed variations in $n_{\rm gal}$ with other quantities. 

\citet[]{Ross:2012} further show that the stellar density weights have a large impact on the clustering of CMASS which suggests that spatial variations in $n_{\rm star}$ might also be an important systematic effect for the present study. Clustering studies in BOSS adopt systematic weights that are designed to re-weight galaxies to a fixed overall number density. However, this scheme may not be appropriate for lensing. Indeed, the dark matter halo mass function displays a non linear relationship between number density and halo mass. Hence, weights designed to maintain a constant number density will not guarantee distributions of equal halo mass.

Figure \ref{fig:starfwhm} displays distributions of $n_{\rm star}$ for regions that overlap with lensing surveys (we study variations in $n_{\rm star}$ directly rather than $w_{\rm star}$). There are clear differences in $n_{\rm star}$ distributions across lensing surveys. 

\begin{figure*}
\centering
\includegraphics[width=13cm]{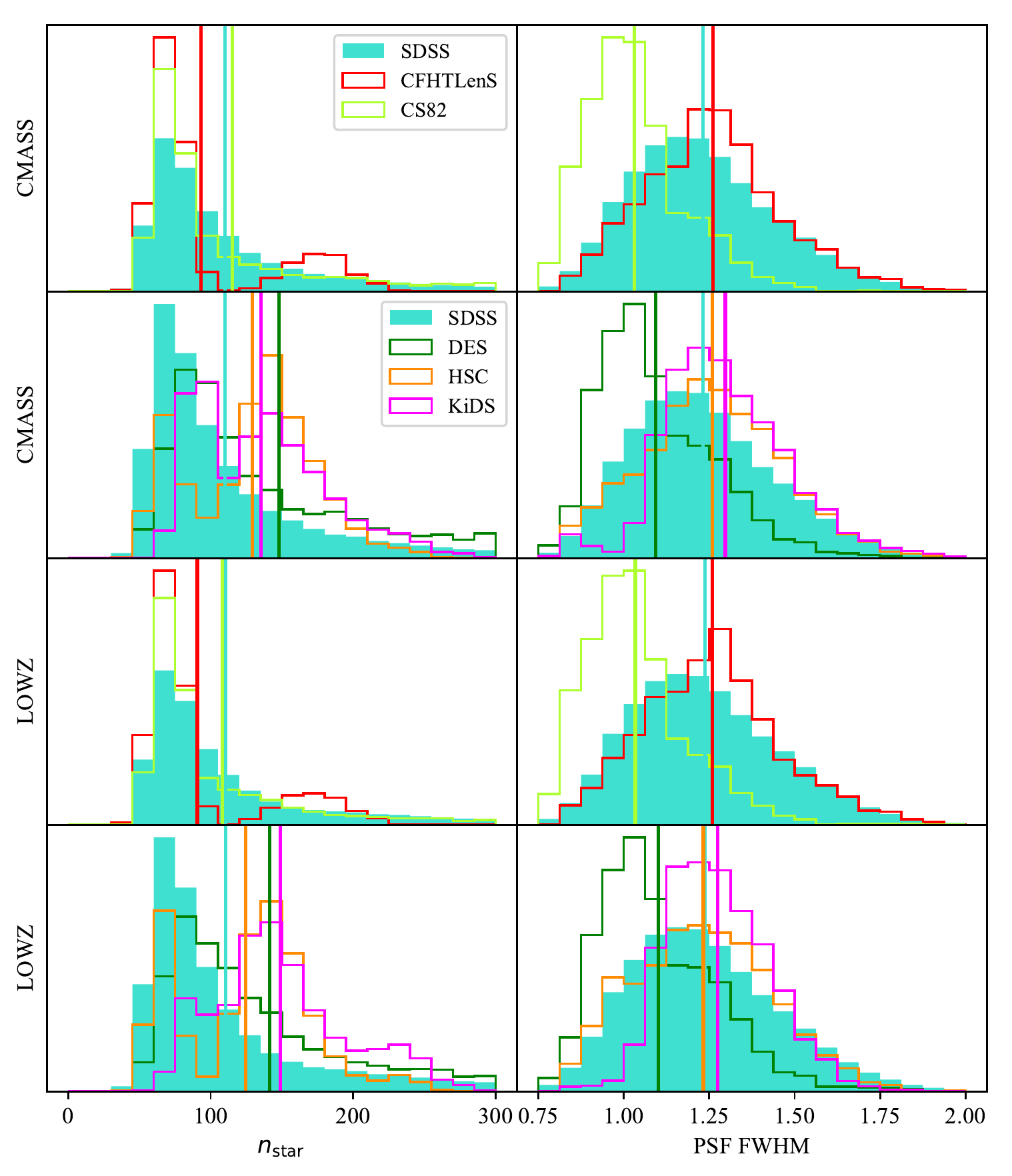}
\caption{The distributions of stellar density (left) and $i$-band seeing FWHM (right) from the BOSS targeting catalogues are not homogeneous in the regions of overlap with lensing surveys. Vertical lines indicate the mean value for each survey.}
\label{fig:starfwhm}
\end{figure*}

As originally detailed in \citet[]{Ross:2011aa}, the observed density of the CMASS sample also correlates with local seeing because of the star galaxy separation cuts. There is no detected corresponding effect for LOWZ\footnote{except for the LOWZE3 sample which we removed from our analysis.}. For CMASS, the effect is such that in poor seeing conditions, the number density decreases because compact galaxies are classified as stars and are removed from the sample. BOSS clustering studies employ a systematic weight, $w_{\rm see}$, to account for the seeing dependence of $n_g$ for CMASS. 

We tried a variety of tests to reweigh surveys to different effective distributions in $n_{\rm star}$ and seeing. However, we found such tests to be limited by a) the small sizes of the lensing surveys at hand, and b) strong spatial variations in the PSF associated with BOSS targeting. Figure \ref{fig:psfinhsc} in the Appendix shows an example of the spatial variation of the PSF in one of the HSC fields. Instead, we opted for a more straightforward post-unblinding test. This is described in Section \ref{postblindhomogtest} and Figures \ref{fig:ampsys_psf} and \ref{fig:ampsys_nstar}.

 \citet[][]{Ross:2017aa} do not find clear evidence for correlations between $n_{\rm gal}$ and sky background, airmass, or extinction. In the future, it would be instructive to investigate whether or not these quantities have an impact on the lensing, but we do not explore these aspects in this paper.
 
\section{Results}\label{Results}

The first goal of this paper is to search for trends in the data that could be due to systematic effects following the methodology outlined in Section \ref{trendmethod}. If found, such correlations could provide important clues as to the origins and level of systematic effects (including those effects that are ``known knowns"). Because we are specifically seeking to pin-point trends caused by systematic errors, in these tests, we use the reported statistical errors (Figures \ref{fig:ampsys_psf} through \ref{fig:ampzsourcemethod} display statistical errors). Section \ref{withsyseror} discusses trends including both statistical and estimated systematic errors. Our second goal is to use the measured spread between the amplitudes of $\Delta\Sigma$ as an empirical and end-to-end estimate of systematic errors following the methodology outlined in Section \ref{sigmasysmethod}. This estimate is then compared with the systematic errors as reported by each survey (Section \ref{sigsystests}). This test will help to determine if unknown systematic effects are present in the data. Unless mentioned otherwise, all of the results in this section were blinded according to the scheme presented in Section \ref{blinding}. It was agreed before unblinding that any homogeneity trends in the lens samples greater than 3$\sigma$ would be discussed. Our lensing signals and code used to make the main figures are available at \url{https://github.com/alexieleauthaud/lensingwithoutborders}.

\subsection{Comparison of $\Delta \Sigma$ and computation of amplitudes}

Figure \ref{fig:dsres} displays $\Delta\Sigma$ for the four lens samples. Results from SDSS are only shown for LOWZ because it was agreed before unblinding that SDSS might not be able to measure an accurate lensing signal for CMASS. Table \ref{lensingsn} gives the overall signal-to-noise of the various measurements in each of the radial ranges. SDSS and HSC have the highest signal-to-noise for LOWZ and HSC has the highest signal-to-noise for CMASS. 

We fit the overall amplitudes of each of the lensing signals using $\Delta\Sigma_{\rm HOD}$ as reference and for each of the three radial ranges (see Section \ref{afit}). The results are shown in  Figure \ref{fig:amps}.

As an additional test, we also compute the amplitudes using only pairs of surveys (when using pairs, the weighting scheme changes, see Section \ref{afit}). We use SDSS for LOWZ and HSC for CMASS as the reference survey. Results are unchanged when using a pair-wise weighting scheme, demonstrating that our amplitude fits are robust to the specifics of individual survey covariances.

\begin{figure*}
\centering
\includegraphics[width=14cm]{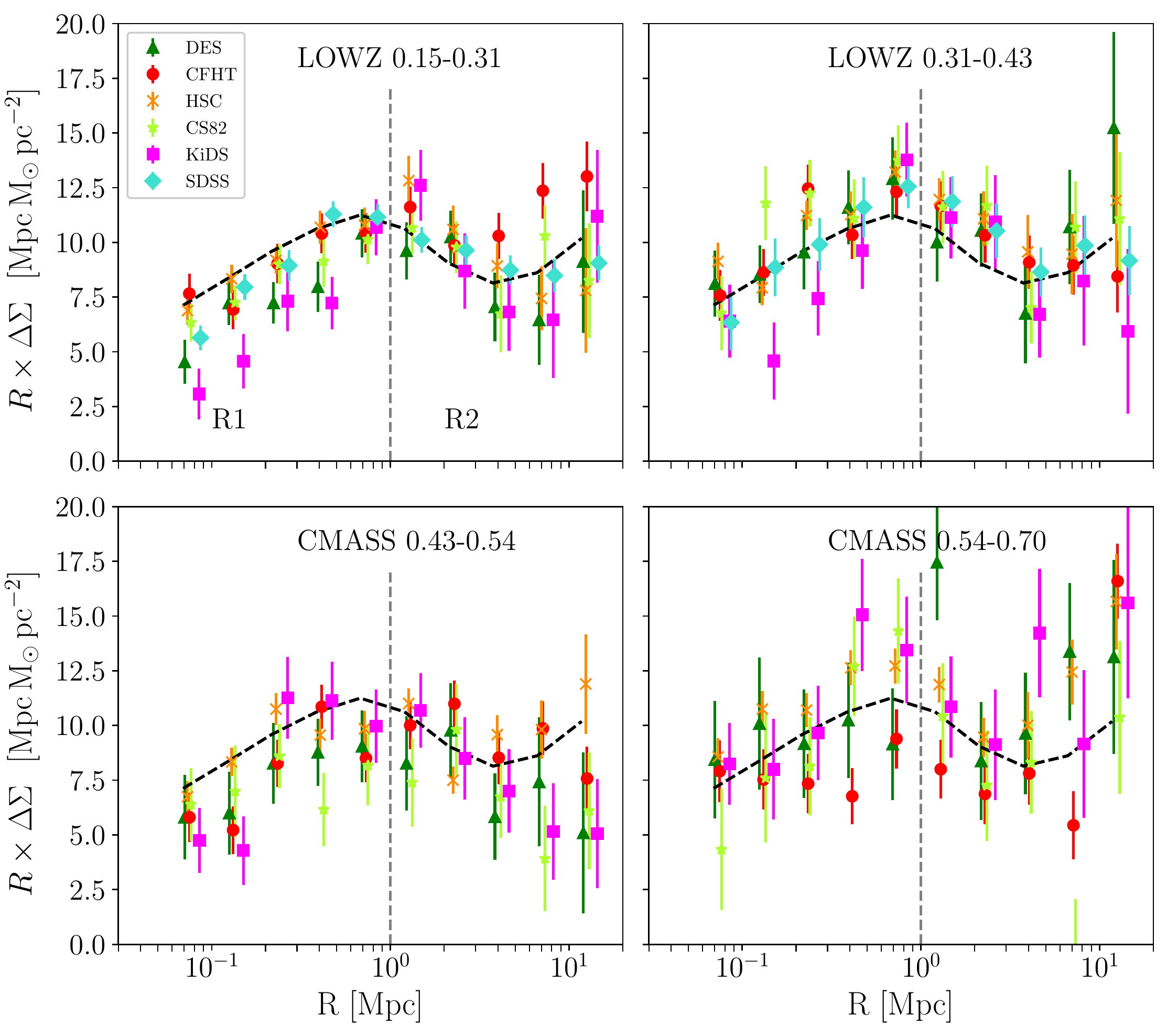}
\caption{Galaxy-galaxy lensing signal around four LRG lens samples from six different lensing surveys with statistical uncertainties. Dashed vertical grey lines delineate our two scale cuts (R$_1$=[0.05,1] Mpc and R$_2$=[1,15] Mpc). Dashed grey lines show $\Delta\Sigma_{\rm HOD}$ (the predicted shape of $\Delta\Sigma$ based on an HOD to CMASS clustering as described in Section \ref{afit}). This paper does not test for amplitude shifts between lensing and clustering and the amplitude of $\Delta\Sigma_{\rm HOD}$ has been arbitrarily normalised to match the lensing signals. Data points have been shifted slightly along the $x$-axis for visual clarity.}
\label{fig:dsres}
\end{figure*}

\begin{figure*}
\centering
\includegraphics[width=14cm]{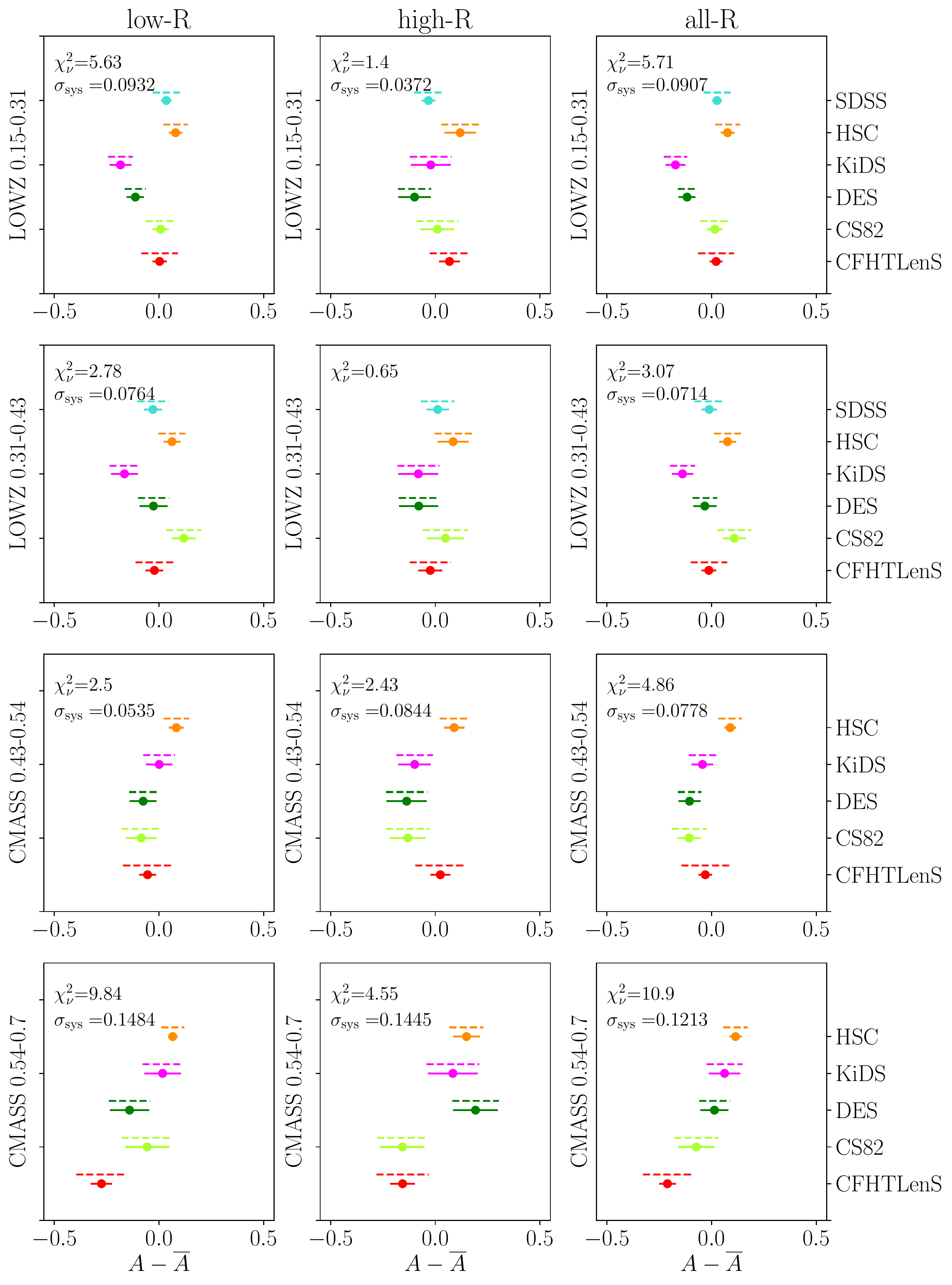}
\caption{Amplitude fits to lensing data in three different radial ranges. Left columns show the amplitude fits for $r<1$ Mpc. Middle columns show the results for $r>1$ Mpc. Columns on the right correspond to the full radial range. The top rows show data for the two LOWZ samples and lower rows show the data for the two CMASS samples. We show $A-\overline{A}$ where A is the fitted amplitude and $\overline{A}$ is the mean amplitude for all surveys. Dashed lines show the sum in quadrature of the statistical and the reported systematic error. In each panel, $\chi^2_{\nu}$ is the reduced chi square of the data points and $\sigma_{\rm sys}$ is the value of the estimated $\sigma_{\rm sys}$ that yields $\chi^2_{\nu}=1$. When $\chi^2_{\nu}<1$, upper limits are indicated for $\sigma_{\rm sys}$ in Figure \ref{fig:sigmasys}.}
\label{fig:amps}
\end{figure*}

\begin{table*}
  \caption{Signal-to-noise ratio of our $\Delta\Sigma$ measurements.}
\begin{tabular}{@{}lc|ccc|ccc|ccc|ccc}
\hline
      &    \multicolumn{3}{c|}{LOWZ 0.15-0.31}   & \multicolumn{3}{c|}{LOWZ 0.31-0.43} & \multicolumn{3}{c|}{CMASS 0.43-0.54} & \multicolumn{3}{c|}{CMASS 0.54-0.7} \\
\hline
Survey & R1 & R2 & All R &  R1 &  R2 &  All R& R1 & R2 & All R & R1 & R2 & All R\\
\hline
CFHTLenS  & 9.98 & 9.78 & 9.88 & 9.34 & 7.64 & 8.49 & 7.17 & 7.60 & 7.38 & 5.96 & 5.95 & 5.95 \\
CS82  & 8.56 & 6.38 & 7.47 & 6.70 & 5.30 & 6.00 & 4.26 & 3.15 & 3.71 & 3.83 & 2.74 & 3.29 \\
DES  & 7.12 & 5.15 & 6.13 & 6.18 & 4.25 & 5.22 & 4.39 & 2.97 & 3.68 & 3.68 & 4.07 & 3.87 \\
HSC  & 13.53 & 7.55 & 10.54  & 12.39 & 7.39 & 9.89 & 12.20 & 9.60 & 10.90 & 13.61 & 9.36 & 11.48\\
KiDS  & 5.25 & 4.54 & 4.90 & 4.80 & 3.64 & 4.21 & 4.72 & 3.64 & 4.18 & 4.55 & 3.63 & 4.10\\
SDSS  & 14.80 & 13.27 & 14.03 & 8.28 & 7.66 & 7.97 & - & - & - & - & - & -\\
\hline
\hline
\end{tabular}
\label{lensingsn}
\end{table*}

\subsection{Effective lens redshift}\label{effectivelensz}

Each of the lensing catalogues has a different mean source redshift, with SDSS being the most shallow, and HSC the deepest. As a result, each survey imparts a different lens weight on each of the different samples. The effective lens redshift (see Section \ref{zeff}) for each of the bins is computed for each survey in Section \ref{computeds}. However, due to the relatively narrow lens redshift bins, we find only minor differences in the effective lens redshifts among surveys. Differences in $z_{\rm eff}$ are always less than $\Delta z=0.02$. Using the Stripe 82 Massive Galaxy Catalogue \citep[][]{Bundy:2015}, we estimate how much the mean stellar mass of galaxies varies across this $\Delta z$. We find that $\Delta M^*$ variations are less than 0.015 dex over this the maximum $\Delta z$. As discussed in Section \ref{galactichemisphere} and Appendix \ref{onep_flux_calibration} such differences are not a concern for the present study.

\subsection{Tests related to homogeneity of lens samples}\label{postblindhomogtest}

We investigate whether or not the amplitude of the lensing signals vary according to spatially varying properties of the lens samples. We show post-unblinding results for variations in the amplitudes as a function of $n_{\rm star}$, the PSF full-width at half-maximum (FWHM) of the imaging used in BOSS targeting, as well as tests related to position on the sky (North versus South). For these tests, we include measurements both with, and without, the BOSS weights, $w_{\rm tot}$. Although BOSS did not find any trends in number density variations for LOWZ with $n_{\rm star}$ or PSF FWHM, we include test for LOWZ here for completeness. We fit the trends with a linear relation with slope $\beta$ and comment on whether or not the slope, $\beta$, is consistent with zero. It was agreed before unblinding that trends greater than 3$\sigma$ would be considered significant. In all figures, trends greater or equal to 3$\sigma$ are highlighted in red.

\subsubsection{SDSS PSF}

Figure \ref{fig:ampsys_psf} shows the amplitudes of each survey as a function of the PSF FWHM of the imaging used in BOSS targeting. No significant trends are found, suggesting that number density variations of BOSS with the SDSS PSF do not correlate with the halo mass properties of CMASS galaxies.

\subsubsection{Galactic hemisphere}

Figure \ref{fig:ampns} shows the amplitudes from each survey as a function of the fraction of area in the Northern galactic cap.  No significant trends are found for L1 and L2 which is consistent with the pre-blinding tests we carried out with SDSS (see Appendix \ref{missingredshiftsappendix}).

\new{A trend is found for C2 (4.0$\sigma$ for low-R, 1.6$\sigma$ for high-R, and 3.5$\sigma$ for all-R). However, there is a mild correlation between $n_{\rm star}$ and $A_{\rm North}/(A_{\rm South}+A_{\rm North})$ where $A_{\rm North}$ is the area in the North and $A_{\rm South}$ is the area in the South. A trend is also found in C2 with $n_{\rm star}$ and the $n_{\rm star}$ trend is more significant than the one in Figure \ref{fig:ampns}. For this reason, we will consider  $n_{\rm star}$ to be the driving trend in this redshift range.}

\subsubsection{Lensing amplitudes versus stellar density}\label{amplitudeversusnstar}

Figure \ref{fig:starfwhm} shows that the $n_{\rm star}$ distribution varies considerably between, and within, each of the lensing surveys (due to the location of each of survey on the sky in relation to the Galactic Plane).  The density of stars varies between $50<n_{\rm star}<300$ overall, and the mean value for each survey varies between $90<n_{\rm star}<150$. \new{The mean value of $n_{\rm star}$ is different for LOWZ and CMASS because they cover slightly different areas}.

Figure \ref{fig:ampsys_nstar} shows the lensing amplitudes from each survey as a function of $n_{\rm star}$. No trends are found for L2 and C1 but trends are found in L1 and C2. For L1, the lensing signal is found to decrease with $n_{\rm star}$ with a slope of $\beta \sim 0.0025$. The trend is 3.1$\sigma$ for low-R, 1.2$\sigma$ for high-R, and 3.4$\sigma$ for all-R. For C2, the amplitude of the lensing signal is found to increase with $n_{\rm star}$ with a slope of $\beta \sim 0.007$. The trend is 5$\sigma$ for low-R, 3.9$\sigma$ for high-R, and 6$\sigma$ for all-R. Applying the BOSS $w_{\rm tot}$ weights does not impact the lensing amplitudes and thus does not correct for this effect.

After unblinding, we decided to test the sensitivity of the trend to data points at the extremities (at low $n_{\rm star}$ and high $n_{\rm star}$). For L1, when removing the low $n_{\rm star}$ data point (CFHTLenS) the trend is detected at 4$\sigma$. When removing the high $n_{\rm star}$ data point (KiDS) the trend is not detected. The L1 trend is therefore sensitive to the data point at high $n_{\rm star}$. For C2, when removing the low $n_{\rm star}$ data point (CFHTLenS) the trend is not detected. When removing the high $n_{\rm star}$ data point (DES) we obtain positive slopes at 6$\sigma$ for low-R, 3.4$\sigma$ for high-R, and 6.5$\sigma$ for all-R. The C2 results are therefore sensitive to the data point at low $n_{\rm star}$. 

For L1, given that both lenses and sources are bright, it seems difficult to imagine how $n_{\rm star}$ could impact the lensing amplitudes. This is, however, the redshift range where the lensing signals have the highest signal-to-noise (see Table \ref{lensingsn}). One possible explanation for the trend in L1 is that the statistical signal-to-noise of the lensing signals are slightly overestimated in this redshift regime. Unlike other trends found in this paper, this L1 trend disappears if we use the internal variance in the data rather than the reported statistical lensing errors. The trends drop below 3$\sigma$ when the errors are inflated by 20\%. The statistical errors on the lensing signals could be underestimated or the cross-covariance between surveys (which we have neglected) could be playing a role. This aspect will be discussed further in Section \ref{jointcovariance}. Underestimated statistical errors would result in an overestimate of the global systematic error. Since our primary goal is to rule out large systematic errors, ignoring the possibility of underestimated statistical lensing errors (or joint covariance) will lead to conservative conclusions. Finally, another possibility is that the significance of the trend is overestimated (see explanation in Figure \ref{fig:staterrors})

The C2 trend is of larger interest because the overall trend is stronger and this is also the regime where both the lenses and the source are fainter and trends with $n_{\rm star}$ are more plausible. \report{There are four possible explanations for the C2 trend. The first explanation is that it could be an inhomogeneity in the lens sample that correlates with halo mass. This is plausible because we know from \citet[][]{Ross:2017aa} that the number density of CMASS galaxies varies with $n_{\rm star}$ with number density variations that are strongest in the C2 bin. The second explanation is that instead of being a variation in the halo masses of the lenses, this could be a systematic in the source sample that correlates with $n_{\rm star}$ (or another quantity that correlates with $n_{\rm star}$ such as galactic extinction) and leads to biased estimates of $\Delta\Sigma$. For example, one possibility could be errors in star-galaxy separation in source catalogues}. The third possibility is the significance of the trend could be overestimated and that there is in fact no trend with $n_{\rm star}$ (see Figure \ref{fig:staterrors} for an explanation on why this might occur). The fourth possibility is that the statistical lensing errors are underestimated. However, unlike the trend found in the L1 bin, the C2 trend does not vanish when the internal variance between the data points are used rather than the reported errors. For this reason, underestimated statistical lensing errors seem unlikely to be the full story for C2.

In the first case, the variations of $\Delta\Sigma$ with $n_{\rm star}$ would not be counted as a source systematic because the variations would reflect true variations in halo mass. In the second case, the spread in signals would be interpreted as evidence for a lensing systematic that correlates with $n_{\rm star}$. In the third case, the spread would be interpreted as a source systematic, but one that does not correlate with $n_{\rm star}$. 

After unblinding we designed a test to attempt to differentiate between the first and the second explanation. Figure \ref{fig:starfwhm} shows that HSC has a bimodal distribution in $n_{\rm star}$ (because of where the HSC fields are located with respect to the Milky Way). Taking advantage of this fact, together with the fact that HSC has the highest signal-to-noise at this redshift range, we divided the HSC field into two separate regions, one with low mean $n_{\rm star}$ and one with high mean $n_{\rm star}$. The resulting lensing amplitudes are shown in Figure \ref{fig:amps_hsc_internal}. When we use only HSC, interestingly, we do not find evidence for any trend with $n_{\rm star}$. This would disfavour the first explanation in favour of the other two possible explanations. However, follow-up work will be required to convincingly disentangle between the three scenarios described above \report{because we cannot rule out with this work alone the possibility that a combination of lens inhomogeneity and source systematics simply happen to cancel out in the HSC data in Figure \ref{fig:amps_hsc_internal}}. \eric{An interesting test to carry out would be to test the correlation separately for bright and faint stars}. Further discussion on possible explanations for this trend is also presented in Appendix \ref{nsappendix}.

\begin{figure*}
\centering
\includegraphics[width=14cm]{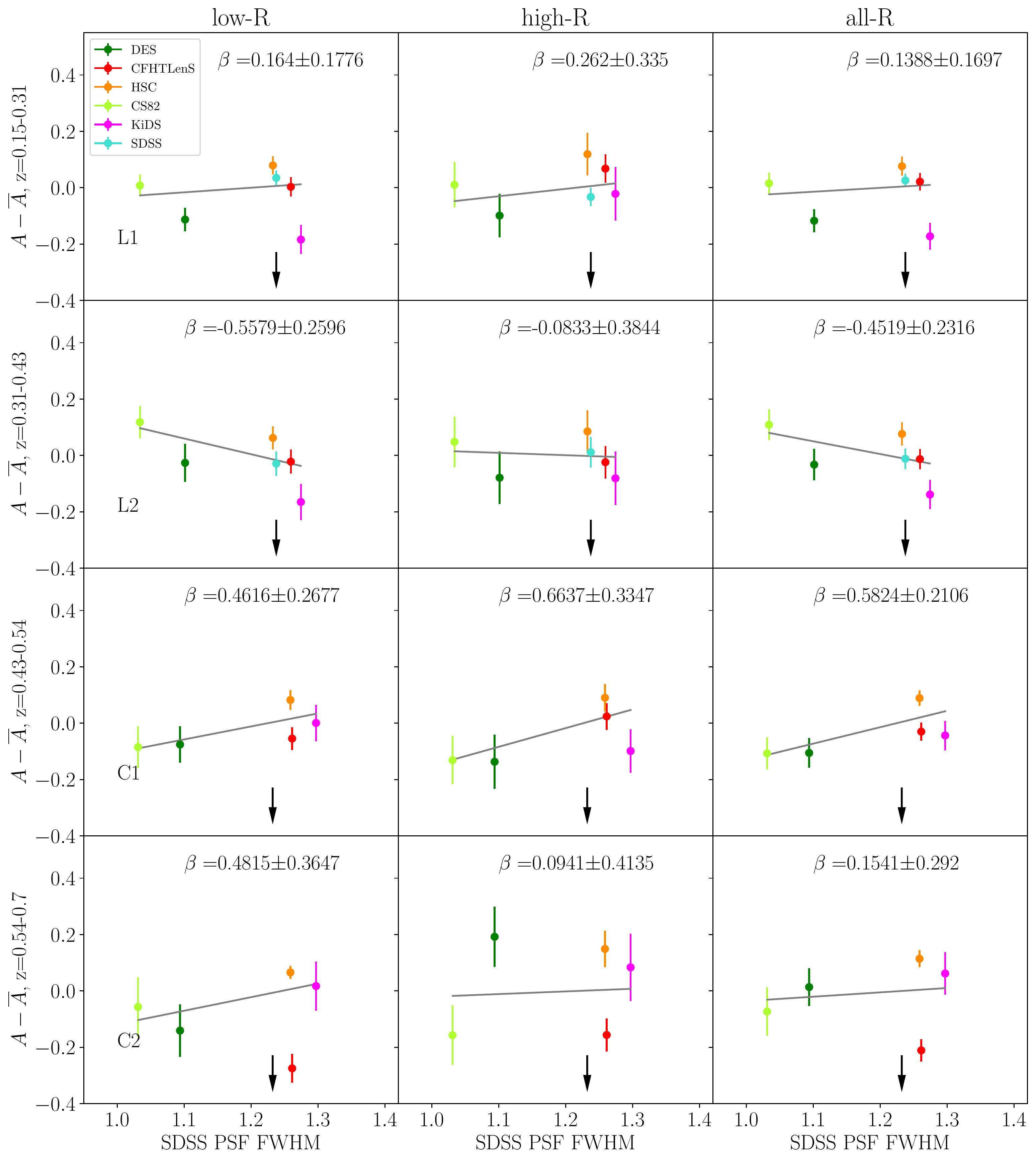}
\caption{Amplitude of $\Delta\Sigma$ versus PSF of SDSS as given in the BOSS targeting catalogue.  The two upper rows correspond to LOWZ and the two bottom rows correspond to CMASS.  The vertical arrow indicates the average BOSS value for the SDSS PSF. We use the statistical errors to search for trends (see \ref{trendmethod} and Figure \ref{fig:staterrors}) and Figures \ref{fig:ampsys_psf} through \ref{fig:ampzsourcemethod} display statistical errors. The results are fit with a linear relation with slope $\beta$. No trends are found between the amplitude of the lensing signal and the SDSS PSF.}
\label{fig:ampsys_psf}
\end{figure*}

\begin{figure*}
\centering
\includegraphics[width=14cm]{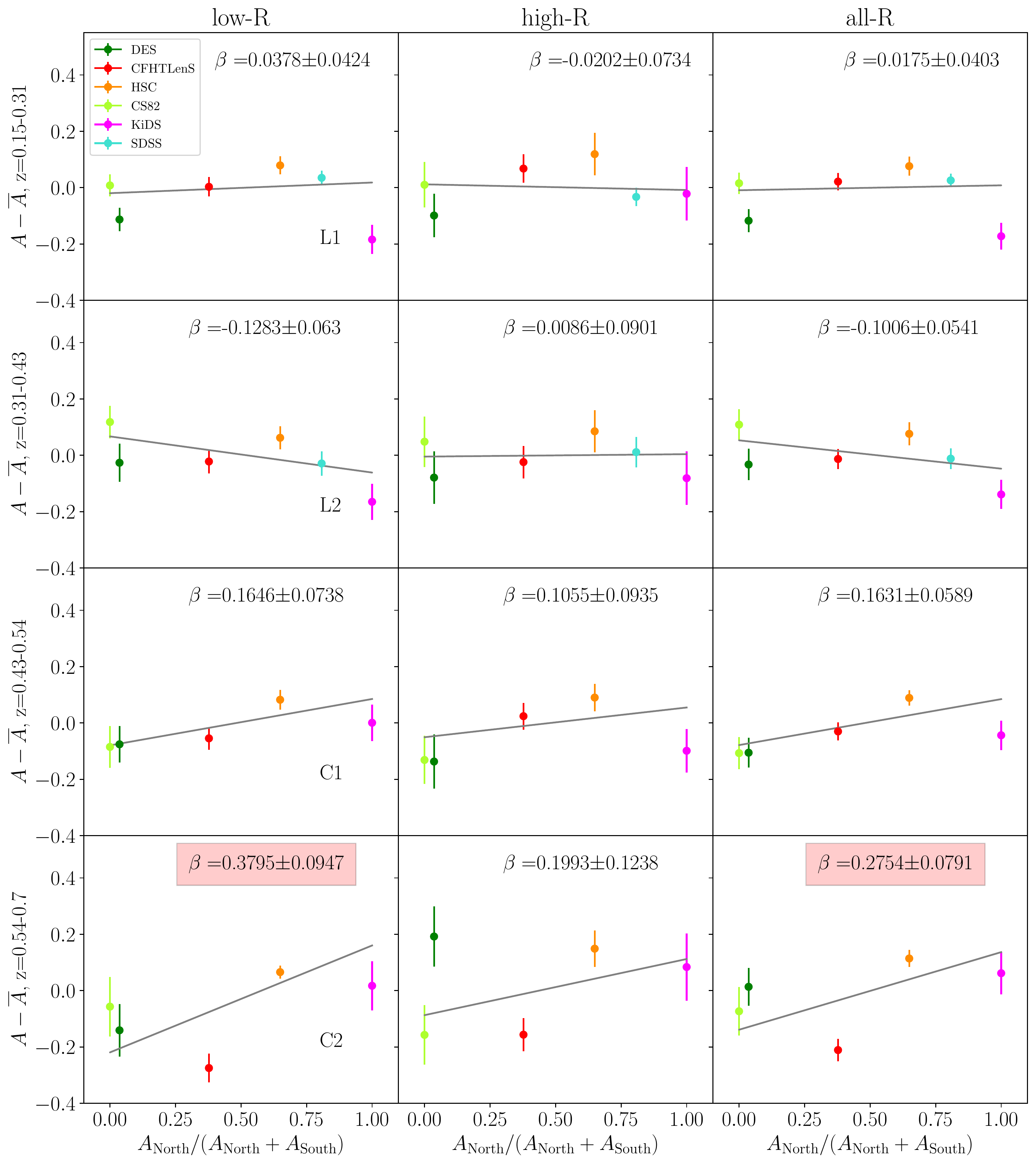}
\caption{Amplitude of $\Delta\Sigma$ versus fraction of area in the Northern galactic cap. A trend is found in C2. However, the trend is less significant than the trend with $n_{\rm star}$ which is the driving trend in this redshift range.}
\label{fig:ampns}
\end{figure*}

\begin{figure*}
\centering
\includegraphics[width=14cm]{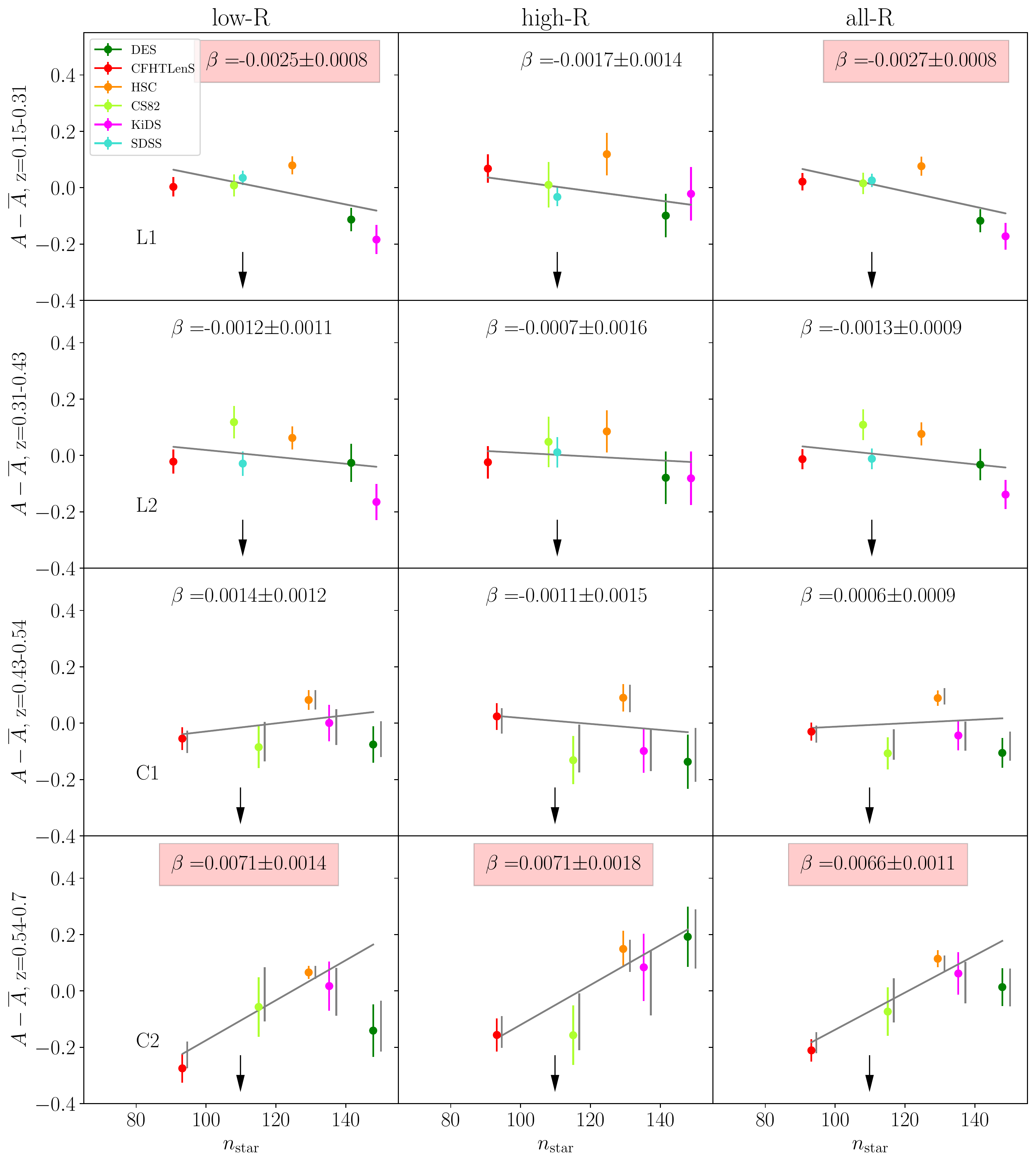}
\caption{Amplitude of $\Delta\Sigma$ versus stellar density, $n_{\rm star}$. The vertical arrow indicates the average value for $n_{\rm star}$ for BOSS galaxies. Solid lines correspond to the amplitudes for signals computed with $w_{\rm tot}$. In the bottom panels, grey data points indicate the lensing amplitudes with no weighting applied (no $w_{\rm tot}$ weight). Grey data points are slightly offset along the $x$-axis for visual clarity. Applying the boss $w_{\rm tot}$ weights does not impact the amplitude of $\Delta\Sigma$. The trend is fit with a linear function with slope $\beta$. A red box indicates a value of beta detected to be non zero at greater than 3$\sigma$. No trends are found for L2 and C1, but a trend is found for L1 and C2. We interpret the L1 trend as possibly arising from underestimated lensing errors. For C2, the trend is 5$\sigma$ for low-R, 3.9$\sigma$ for high-R, and 6$\sigma$ for all-R but the detection of this trend is sensitive to the low $n_{\rm star}$ data point. }
\label{fig:ampsys_nstar}
\end{figure*}

\begin{figure*}
\centering
\includegraphics[width=14cm]{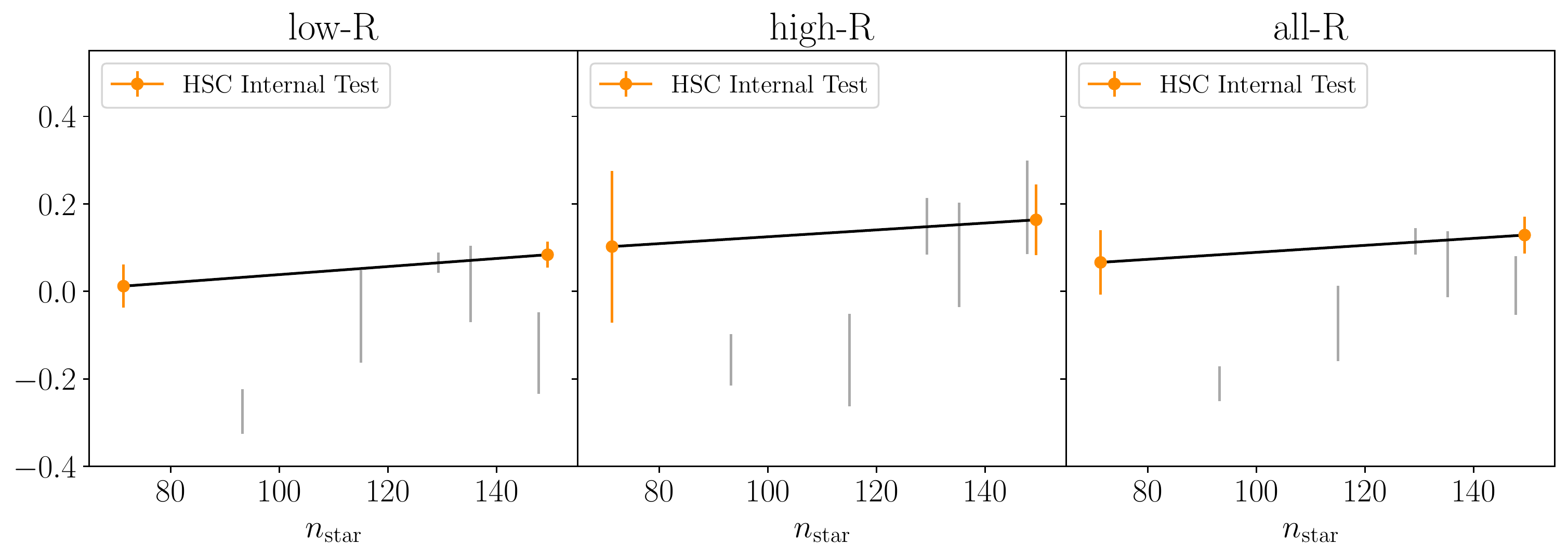}
\caption{A test to determine if the correlation between the amplitudes of the lensing signals with $n_{\rm star}$ is due to lens inhomogeneity (and hence a genuine variation in halo mass) or due to a lensing source systematic. Grey data points show the amplitudes of the given surveys (same format as in Figure \ref{fig:ampsys_nstar}). Left is low-R, middle is high-R, and right is all-R. The two orange data points show the lensing signal as measured internally using just HSC and dividing the survey into two separate areas with low and high $n_{\rm star}$. The internal trend with HSC is flat. This suggests that the trend observed in Figure \ref{fig:ampsys_nstar} is not an intrinsic variation in the halo masses of CMASS across the sky.}
\label{fig:amps_hsc_internal}
\end{figure*}

\subsection{Scaling of amplitudes with mean source redshift}\label{meansourcez}

Figure \ref{fig:ampzsource} displays the lensing amplitudes as a function of the  mean source redshift for each survey. No trends are found for $z_{\rm s}$ for L1 and for L2 but trends are found for C1 and C2.  For C1, a trend with slope of $\beta \sim 0.8$ is detected at 2.9$\sigma$, 2.9$\sigma$, and 4.3$\sigma$ respectively for low-R, high-R, and all-R. For C2, a trend with similar slope values is detected at 4.2$\sigma$, 1.5$\sigma$, and 3.5$\sigma$ respectively for low-R, high-R, and all-R. We remind readers that we have used the statistical errors on $\Delta\Sigma$ to constrain this trend (see Section \ref{trendmethod}).

After unblinding, we decided to also test to determine how sensitive the trend is to data points at the extremities (at low $z_{\rm s}$ and high $z_{\rm s}$). For both C1 and C2, the significance of the detection of a positive slope remains similar even if we remove the low $z_{\rm s}$ data point (CS82). However, the trend is no longer detected in either C1 or C2 after the removal of the high $z_{\rm s}$ data point (HSC). We conclude that the correlation between the amplitude of the lensing signals is sensitive to the high $z_{\rm s}$ data point (HSC).

The absence of redshift trends in L1 and L2 is not inconsistent with the results for C1 and C2. The impact of effects such as blending, photo-$z$ calibration, and shear calibration will become more apparent with fainter source galaxies and when the sources are more closely located behind the lens galaxies. The source galaxies that dominate the measurements for L1 and L2 will be brighter galaxies at lower redshift than for C1 and C2. The absence of a significant trend in L1 and L2 points to effects that predominantly affect the fainter and higher redshift samples. 

The C1 bin is high redshift for lensing with SDSS and so initially we had decided to not compute lensing with SDSS in this bin. However, after unblinding, we searched for an additional way to test the $z_{\rm s}$ trend in C1. We decided to compute the lensing amplitude with SDSS in the C1 bin. We have shown the SDSS data point in cyan in Figure \ref{fig:ampzsource} but we do not use this SDSS data point in any of our fits. 

HSC has the largest dynamic range in source redshifts and so it is interesting to consider testing for source redshift dependent trends by dividing the source sample. In fact this exact test was already carried out in \citet[][]{Speagle:2019aa} who performed tests with HSC that divided the source sample into a high redshift ($z_{\rm s}>1$) and a low redshift ($z_{\rm s}<1$) subset (section 7.5 in \citealt{Speagle:2019aa}). They used the same CMASS and LOW galaxies as here, but in redshift bins of $0.2<z<0.4$ and $0.4<z<0.6$ and $0.6<z<0.8$. No significant trends (more than 3$\sigma$) were found. In the lens redshift bin $0.4<z<0.6$, which corresponds roughly to C1 and C2 here, there was a 2$\sigma$ detection of the opposite trend than reported here (\citealt[][]{Speagle:2019aa} found the signals from the low source redshift bin to be systematically higher than those from higher redshift). We conclude that this test is hard to carry out in a statistically significant manner yet with any single survey, underscoring the utility of harnessing the power of multiple surveys.

\kidsnew{We use a Monte Carlo test to asses the likelihood of these trends arising both from our use of statistical errors in the fits and from the look elsewhere effect \citep[e.g.,][]{Lyons2008}\footnote{\report{the phenomenon where an apparently statistically significant observation may arise by chance because of the large size of the parameter space that is searched}.} . We create a series of Monte Carlo tests in which there is no amplitude variation with redshift ($\beta=0$). We perturb the amplitudes according to the sum in quadrature of each surveys statistical and systematic error. We fit the data points using the statistical errors only. We asses how often one trend will occur in the four low-R tests, and how often two trends will occur in the four all-R  bins. We find a 1\% probability of having one trend occur for the low-R tests. For the high-R tests, the expected probability for two trends is much less than 1\%.}

\citet[][]{Prat2018} also perform a similar shear ratio test for DES Y1 (see Figure 12 in \citealt{Prat2018}). However, they did not use lens/source combination’s whose redshift would have corresponded to a shear ratio test for C1 or C2 and so would not have seen the effects described here.

\citet[][]{Giblin2021} performed the most recent shear ratio test for KiDS using CMASS galaxies (see their figure 11).  Although the methods are not directly comparable because we use $\Delta \Sigma$ and \citet[][]{Giblin2021} use $\gamma_{\rm t}$, we can compare to the redshift evolution of the KiDS galaxy-galaxy lensing signal for the C2 bin. \citet[][]{Giblin2021} find a fully consistent galaxy-galaxy lensing signal for sources at a range of different photometric redshifts behind the BOSS lenses.   They conclude, however, that this test is fairly insensitive to redshift and shear calibration errors once uncertainty in the amplitude of the intrinsic alignment of galaxies is included in the analysis (see Section \ref{zsource-discussion}).

Finally, after unblinding, we also decided to assess whether or not the $z_{\rm s}$ trend might arise from the different methodologies used to compute $\Delta\Sigma$. Figure \ref{fig:ampzsourcemethod} shows the correlation between the lensing amplitudes and the $z_{\rm s}$ trend for those bins in which a trend was detected at more than 3$\sigma$. Data points are colour-coded according to different methodologies as detailed in Table \ref{wlsurveytable}. Figure \ref{fig:ampzsourcemethod} does not reveal any obvious relationship between the lensing amplitudes and the methodology employed. This suggests instead that the trend is more likely to be caused by some intrinsic bias in the source sample that correlates with survey depth.

\begin{figure*}
\centering
\includegraphics[width=14cm]{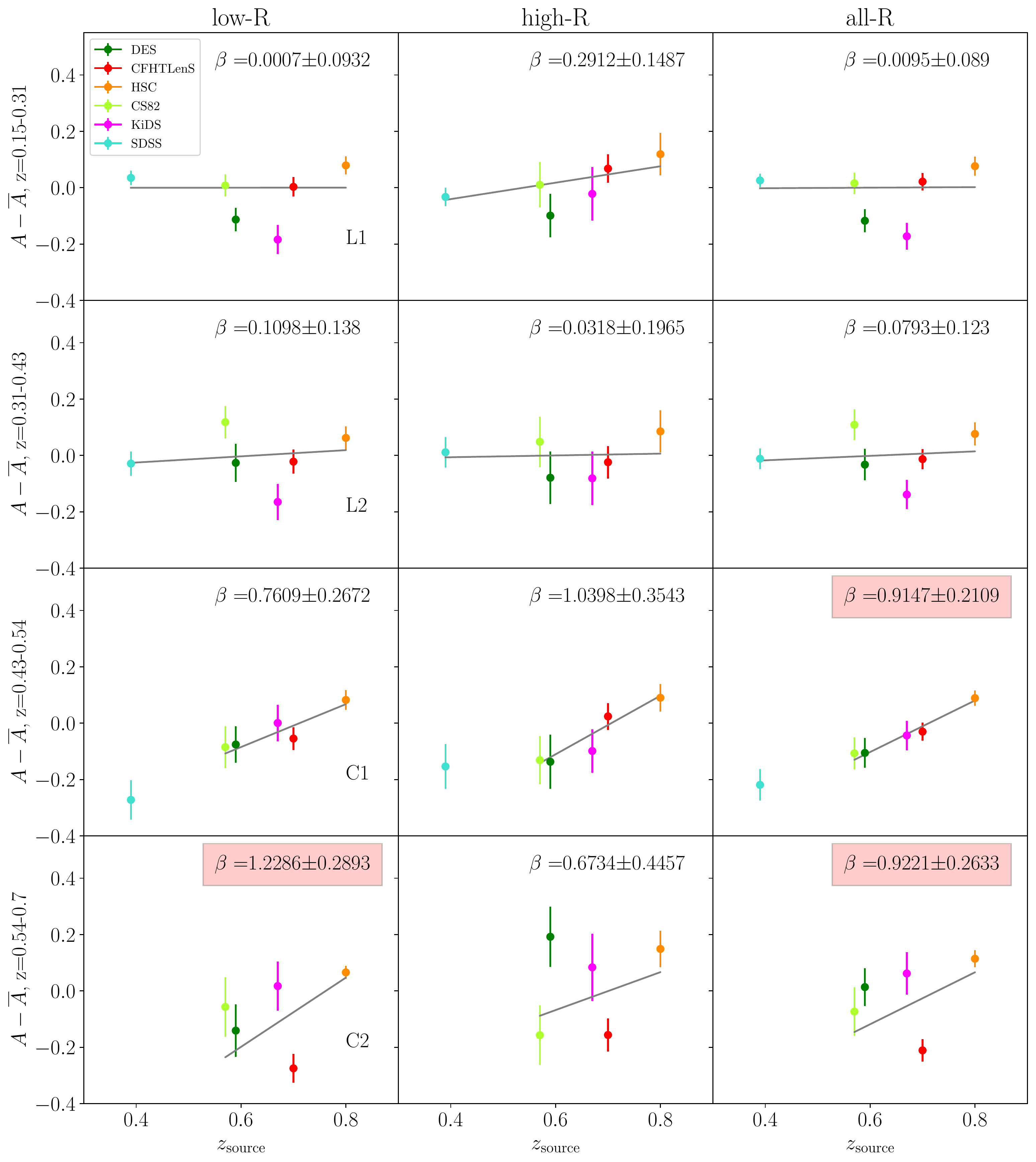}
\caption{Amplitude of $\Delta\Sigma$ versus mean redshift of source galaxies (i.e., depth of lensing surveys). We find a correlation between the amplitude of the lensing signal and the mean source redshift of the lensing surveys for C1 and C2. The amplitude shift between the two extremities (CS82 and HSC) in the C1 bin is of order $\sim$ 25\% but is consistent with the systematic errors reported by each of the lensing surveys if we assume that the $z_{\rm s}$ trend dominates the error budget and that some surveys have experienced a 1.5$\sigma$ upward or downward systematic shift (see Table \ref{syserrtable} and discussion in Section \ref{zsource-discussion}). After unblinding, we decided to add a SDSS data point to the C1 bin. We show the SDSS result here but do not include it in the fits. The C2 bin is the noisiest because shallow lensing surveys (e.g., DES, CS82) have a limited number of source galaxies above the C2 lens redshift range. A tentative trend with $n_{\rm star}$ also adds scatter to the C2 bin.}
\label{fig:ampzsource}
\end{figure*}

\begin{figure*}
\centering
\includegraphics[width=14cm]{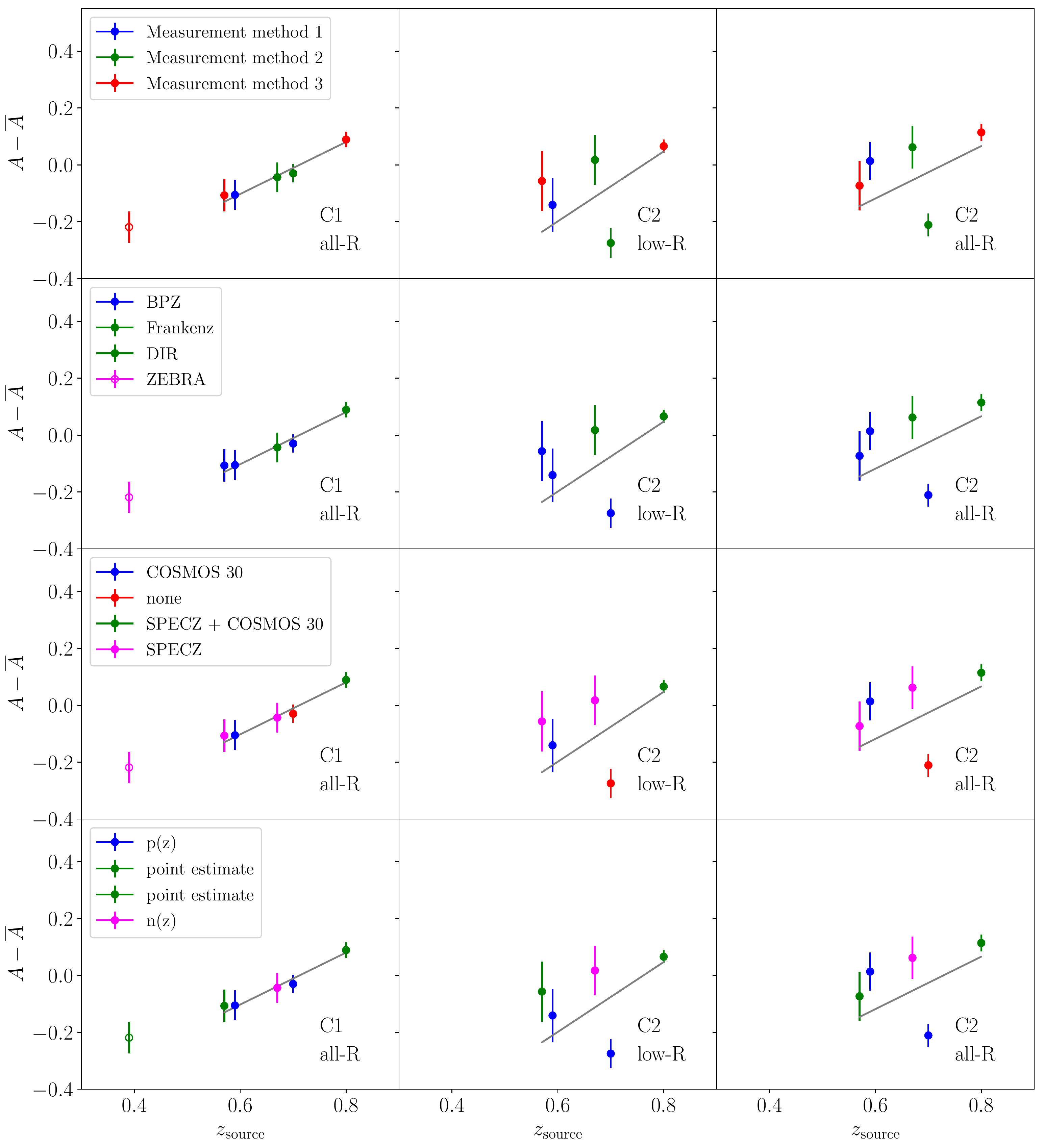}
\caption{Amplitude of $\Delta\Sigma$ versus mean redshift of source galaxies for bins in which a trend greater than 3$\sigma$ was detected in Figure \ref{fig:ampzsource}. Data points are color coded using different choices for the computation of $\Delta\Sigma$ as summarised in Table \ref{wlsurveytable}. In the first row, data points are color coded according to overall similarity between methods (see Section \ref{computeds}). In the second row, data points are color coded according to the photo-$z$ computation method. In the third row, data points are color coded according to the calibration samples used to ensure unbiased photo-$z$'s. In the fourth row, data points are color coded according to the redshift adopted to compute $\Sigma_c$. We do not find evidence that any of these methodology or photo-$z$ choices determine the trend. The SDSS data point was added after unblinding for C1, is shown using an open symbol, and was not used in the fit.}
\label{fig:ampzsourcemethod}
\end{figure*}

\subsection{Trends including systematic errors}\label{withsyseror}

We have also investigated these trends using the sum in quadrature of the statistical and the reported systematic errors. In this case, all trends except the L1 all-R trend in Figure \ref{fig:ampsys_nstar} drop below 3$\sigma$. The L1 all-R trend drops from 3.4$\sigma$ to 3.1$\sigma$ (just above our predefined limit for claiming a detection). This is consistent with our interpretation that: a) the L1 trend in Figure \ref{fig:ampsys_nstar} may result from underestimated statistical errors and from neglecting the covariance between the measurements (see Section \ref{amplitudeversusnstar}), and b) that the trends highlighted in the previous sections are globally consistent with the reported systematic error budgets (see also next Section). The trends could originate from the ``known knowns" types of systematic errors accounted for by existing error budgets.

\subsection{Empirical estimates of the systematic error}\label{sigsystests}

We now present empirical systematic error estimates for the ensemble of lensing data following the methodology outined in Section \ref{sigmasysmethod}. We will assume a single systematic error for all data and per test (one test corresponds to one panel in Figure \ref{fig:amps}).

Table \ref{syserrtable} and Figure \ref{fig:sigmasys} compare the values of the systematic errors reported by each of the surveys to our empirical estimates. \kidsnew{As described in Section \ref{sigmasysmethod}, Monte Carlo tests were used to show that the $\sigma_{\rm sys}$ value that we estimate is close to the mean systematic error among surveys. For all redshift and radial ranges, we find good agreement between our empirically estimated values and the reported systematic errors. Differences between the estimated and reported systematic errors are always less than 3$\sigma$ (our pre-blinding determined criterion).}

 \kidsnew{For L1, L2, and C1 (lenses at $z<0.54$) we find excellent agreement between our empirical estimates and those reported by lensing surveys. No evidence is found for large unknown systematic errors. Using the values for $\sigma_{\rm sys}$ derived over the full radial range (all-R), we find estimated values of $\sigma_{\rm sys}=0.09 \pm _{0.02}^{0.06}$, $\sigma_{\rm sys}=0.07 \pm _{0.02}^{0.05}$, and $\sigma_{\rm sys}=0.08\pm _{0.02}^{0.07}$ for L1, L2, and C1 respectively. These are in good agreement with the mean reported systematic errors which are 0.05, 0.05, and 0.056 for L1, L2, and C1 respectively.}

\kidsnew{For the C2 bin (lenses between $0.54<z_{\rm L}<0.7$ and sources at $z_{\rm s} \gtrapprox 0.7$) the estimated value is  $\sigma_{\rm sys}=0.12 \pm _{0.03}^{0.09}$ which is higher than values found for L1, L2, and C1. However, the difference with the mean reported systematic error ($\overline{\sigma}_{\rm sys}=0.058$) only has a significance of 2$\sigma$ which does not meet out predefined 3$\sigma$ threshold for claiming a detection. } The interpretation of this redshift range is also rendered difficult given the possibility of a trend with $n_{\rm star}$ that could explain a large fraction of this spread. In Section \ref{amplitudeversusnstar} we discussed three possible explanations for the origin of the $n_{\rm star}$ trend. \report{This trend could be due to intrinsic halo mass variations of the CMASS sample across the sky. If this is the case, it would be appropriate to subtract out this effect when reporting $\sigma_{\rm sys}$. If on the other hand, the variations with $n_{\rm star}$ are not related to intrinsic halo mass variations, the spread amongst data points should count as a source systematic}. Since Figure \ref{fig:amps_hsc_internal} seems to favour the second explanation, we quote raw values (without subtracting out the $n_{\rm star}$ trend). However, it is clear that more detailed follow-up work will be required to fully resolve the question of the $n_{\rm star}$ trend. Further discussion is presented in Appendix \ref{nsappendix}.

For the C1 bin (lenses between $0.43<z_{\rm L}<0.54$ and sources at $z_{\rm s} \gtrsim 0.54$) we find that the amplitude of $\Delta\Sigma$ correlates with $z_{\rm s}$ with variations between surveys reaching a maximum difference of 23\% (between HSC and CS82). These variations are consistent with the reported systematic errors in this redshift range but only if we also assume that the $z_{\rm s}$ trend dominates the systematic error budget. The reported systematic errors for C1 are between 2\% and 11\% and the estimated values are between 5\% and 8\% (see $\sigma_{\rm sys}$ for C1 in Table \ref{syserrtable}). Assuming an ensemble value of $\sigma_{\rm sys}$=0.073, Figure \ref{fig:ampzsource} can be understood by assuming that surveys have experienced a 1.5$\sigma$ upwards or downwards shift depending on the value of $z_{\rm s}$. This would imply that the correlation with $z_{\rm s}$ is the dominant term in the systematic error budget in this redshift range. In other terms, $\sigma_{\rm sys}$ and $z_{\rm s}$ are strongly correlated. Possible explanations for the origin of the $z_{\rm s}$ are discussed in Section \ref{discussion}. 

For C1 and C2, CFHTLenS has the largest reported systematic errors (11\%). This fact, combined with the observation that the $n_{\rm star}$ trend is sensitive to the removal of this data point, led us to decide post-blinding to also quote values for C1 and C2 without the CFHTLenS data point. The removal of CFHTLenS has little impact on the results for for C1 but does reduce the estimated values of $\sigma_{\rm sys}$ for C2 and $\sigma_{\rm sys}$ drops from $\sigma_{\rm sys}=0.12 \pm _{0.03}^{0.09}$ to $\sigma_{\rm sys}=0.06\pm _{0.05}^{0.11}$. The 68 and 95\% upper confidence levels are, however, relatively unchanged. Overall C2 remains less constraining than the other redshift bins because the lensing data are noisier in this redshift range.

In the context of the ``lensing is low" phenomenon in which models of the galaxy-halo connection applied to the clustering of CMASS and LOWZ over-predict $\Delta\Sigma$ by 20-30\% \citep[][]{Leauthaud:2017aa, Lange2019,Singh:2020, Lange2020}, it is interesting to consider to what degree large unknown systematic errors are ruled out by these results. Overall, systematic errors greater than 15\%, 13\%, and 14\% are ruled out at 68\% confidence level for L1, L2, and C1 respectively. Systematic errors greater than 24\%, 20\%, and 26\% are ruled out at 95\% confidence level for L1, L2, and C1 respectively. These constraints could be made tighter by combining between lens bins -- but we have not attempted this here. In summary, \report{a 25\% systematic error is ruled out at 95\% confidence level in three fairly independent redshift bins}. A more detailed discussion on this topic is presented in Section \ref{discussion}. 

It is also interesting in Table \ref{syserrtable} and Figure \ref{fig:sigmasys} to consider whether or not there is a radial dependence to $\sigma_{\rm sys}$. When considering all lens samples and comparing the ``low-R" results and the ``high-R" results, we do not see evidence for a radial dependence in $\sigma_{\rm sys}$. The systematic that leads to the $z_{\rm s}$ trend identified in Figure \ref{fig:ampzsource} (which was found to be radially independent) is likely to be a more dominant systematic than boost factor corrections. 

The systematic errors estimates presented in this paper are conservative in the sense that we have included data from earlier lensing surveys. Lensing methods have been evolving rapidly and the most recent lensing surveys probably have errors that are smaller than the ensemble values reported here. Again, one of our goals in the paper is to rule out large unknown sources of systematic error. For this, our conservative ensemble estimate is sufficient.

These results are the first empirically derived tests on the systematic errors on $\Delta\Sigma$ and present an important sanity check on the validity of reported $\Delta\Sigma$ lensing signals.  However, the uncertainties on the reported $\sigma_{\rm sys}$ are still large compared to the values of $\sigma_{\rm sys}$ reported by the lensing surveys. In Section \ref{discussion} we discuss how $\sigma_{\rm sys}$ could be better constrained in future implementations of these tests. 

\begin{figure*}
\centering
\includegraphics[width=13.5cm]{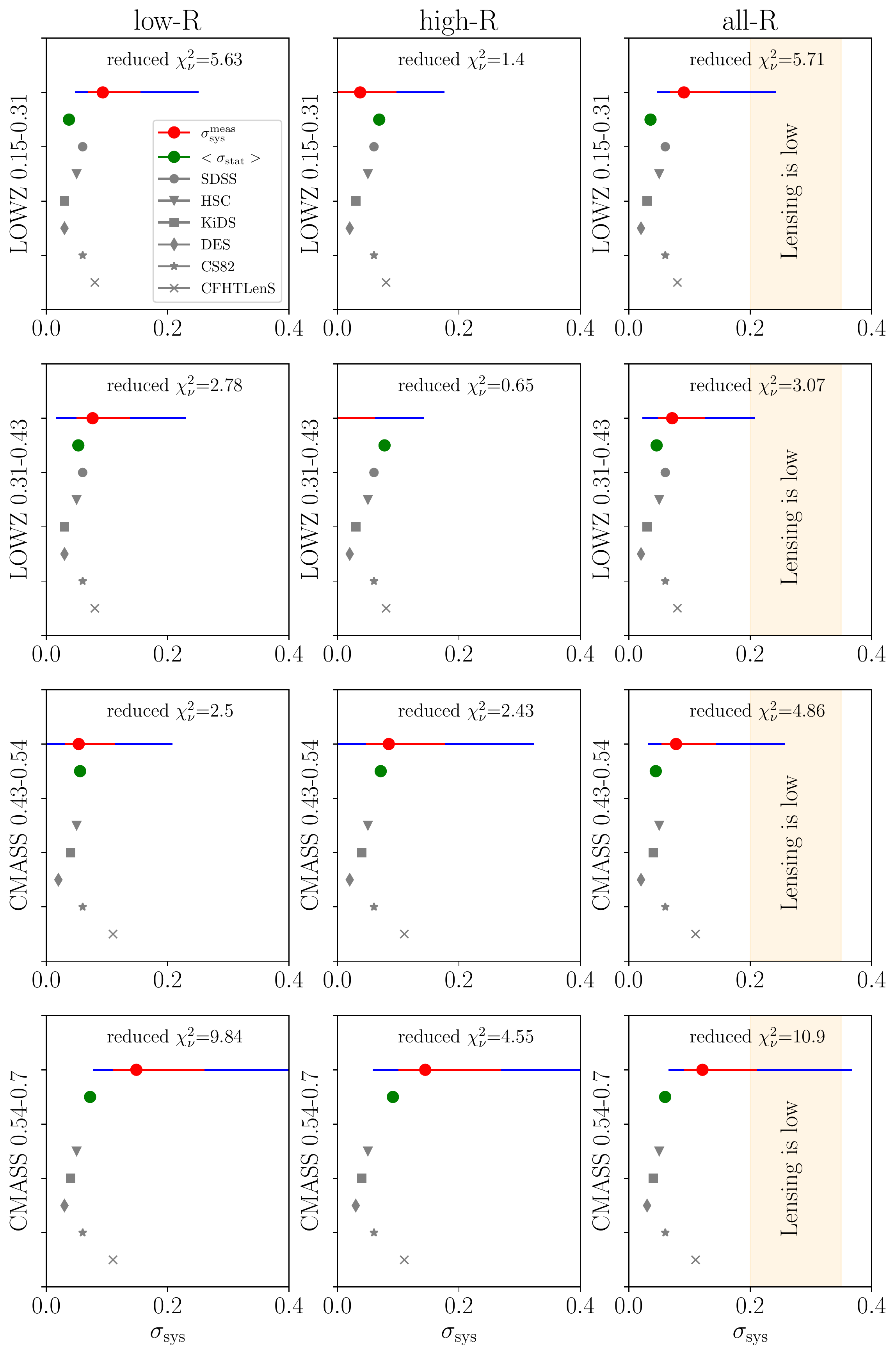}
\caption{Comparison between reported systematic errors and our empirically estimated values. Red lines correspond to 68\% confidence errors on $\sigma_{\rm sys}$ and blue lines to 95\% confidence. For L1, L2, and C1 we find good agreement between our empirical estimates and those reported by lensing surveys. Generally speaking, systematic errors greater than 15\% are ruled out at 68\% confidence. At $z_{\rm L}<0.54$, systematic errors greater than 30\% are ruled out in most bins at 95\% confidence. For C2 (lenses between $0.54<z_{\rm L}<0.7$ and sources at $z_{\rm s} \gtrapprox 0.7$) $\sigma_{\rm sys}$ is larger than reported values. A large fraction of this scatter could be explained by a correlation between the lensing amplitudes and $n_{\rm star}$. The C2 bin is also the noisiest because there are fewer sources galaxies with which to perform the measurements in this lens range and the upper limits on $\sigma_{\rm sys}$ are therefore not very constraining. The ``lensing is low effect" corresponds to a 20\% to 35\% shift in amplitudes and is highlighted by the orange shaded region. 
We do not find strong evidence for boost factors impacting $\sigma_{\rm sys}$ -- the signature of this would be larger $\sigma_{\rm sys}$ values at low-R. Rather,  $\sigma_{\rm sys}$ is found to be fairly independent of radial scales (albeit with large uncertainties on $\sigma_{\rm sys}$).  The green data point shows the mean statistical error across all surveys. Galaxy-galaxy lensing measurements are entering an era in which systematic errors will dominate error budgets.}
\label{fig:sigmasys}
\end{figure*}

\begin{figure*}
\centering
\includegraphics[width=13.5cm]{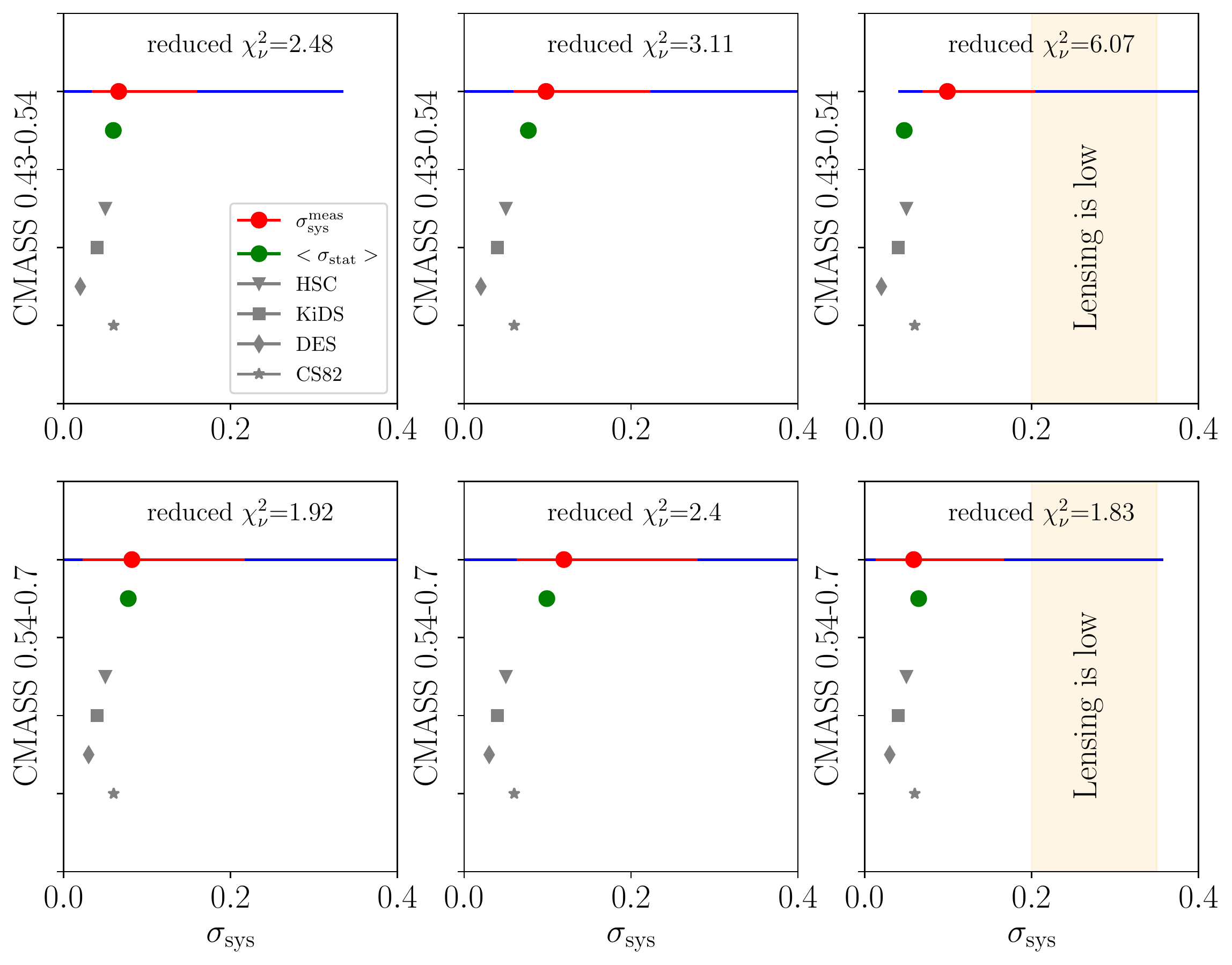}
\caption{Same as Figure \ref{fig:sigmasys} for C1 and C2 but without the CFHTLenS data points (which have the largest reported systematic errors in this redshift range). There is some evidence that removing the CFHTLenS data points reduces the variance in the C2 bin with $\chi_{\nu}^2$ dropping from 9.84 to 1.92 for low-R, from 4.55 to 2.4 for high-R, and from 10.9 to 1.83 for all-R. However, the 68 and 95\% upper limits are similar to those reported in Figure \ref{fig:sigmasys} and are less constraining than for L1 and L2. The ``lensing is low effect" corresponds to a 20\% to 35\% shift in amplitudes and is highlighted by the orange shaded region. When the CFHTLenS data points are removed, the upper limits on $\sigma_{\rm sys}$ for C1 and C2 are not constraining enough to rule out 20\% amplitude shifts at more than 1$\sigma$ because only 4 surveys are used to constrain $\sigma_{\rm sys}$ instead of 5.}
\label{fig:sigmasys_nocfhtlens}
\end{figure*}

\begin{table*}
  \caption{Comparison of reported and estimated systematic errors. The systematic that we consider is a multiplicative systematic on the amplitude of $\Delta\Sigma$ (see Equation \ref{syserrequation}) . We assume that this systematic error is Gaussian with width $\sigma_{\rm sys}$ and can take on a different value for each survey. Reported errors on the estimate of $\sigma_{\rm sys}$ are 68\% confidence. We also quote the 68 and 95\% upper confidence limits on $\sigma_{\rm sys}$. Our global estimate of $\sigma_{\rm sys}$ correspond roughly to the mean systematic error among surveys. For C1 and C2, CFHTLenS reports the largest systematic error. For this reason, combined with the fact that the $n_{\rm star}$ trend is also found to depend on this data point, we also report the values of $\sigma_{\rm sys}$ without CFHTLenS. \mmnew{This does not have a large impact on C1 but it does reduce the difference seen in C2 and brings the estimated values into closer agreement with the mean reported value. } For both C1 and C2, the 68 and 95\% confidence limits are not as constraining as for L1 and L2 because of the higher lens redshift range and because 5 surveys are used to constrain $\sigma_{\rm sys}$ rather than 6.}
\begin{tabular}{@{}lc|ccc|ccc}
\hline
      &    \multicolumn{3}{c|}{LOWZ 0.15-0.31}   & \multicolumn{3}{c|}{LOWZ 0.31-0.43}  \\
\hline
Survey & R1 & R2 & All R &  R1 &  R2 &  All R\\
\hline
CFHTLenS  & 0.08 & 0.08 & 0.08 & 0.08 & 0.08 & 0.08  \\
CS82  & 0.06 & 0.06 & 0.06 & 0.06 & 0.06 & 0.06  \\
DES  & 0.03 & 0.02 & 0.02 & 0.03 & 0.02  &  0.02  \\
HSC  & 0.05 & 0.05 & 0.05 & 0.05 & 0.05 & 0.05 \\
KiDS  &  0.03 &  0.03 &  0.03 & 0.03 & 0.03 & 0.03  \\
SDSS  & 0.06 & 0.06 & 0.06 & 0.06 & 0.06 & 0.06 \\
\hline
\textbf{Mean reported $\sigma_{\rm sys}$}  & 0.052  & 0.050  & 0.050 & 0.052 & 0.050 & 0.050 \\
\hline
   \multicolumn{7}{c|}{Ensemble Systematic Error}   \\
\hline
\textbf{Estimated} &
$0.09 \pm _{0.02}^{0.06}$&
$0.04 \pm _{0.04}^{0.06}$&
$0.09 \pm _{0.02}^{0.06}$&
$0.08 \pm _{0.03}^{0.06}$&
$<0.06$&
$0.07 \pm _{0.02}^{0.05}$&\\
1$\sigma$ upper confidence limit&
0.16&
0.1&
0.15&
0.14&
0.06&
0.13&\\
2$\sigma$ upper confidence limit&
0.25&
0.18&
0.24&
0.23&
0.14&
0.2&\\
\hline
\hline
      &    \multicolumn{3}{c|}{CMASS 0.43-0.54}   & \multicolumn{3}{c|}{CMASS 0.54-0.7}  \\
\hline
Survey & R1 & R2 & All R &  R1 &  R2 &  All R\\
\hline
CFHTLenS & 0.11 & 0.11 & 0.11 & 0.11 & 0.11 & 0.11 \\
CS82  & 0.06 & 0.06 & 0.06 & 0.06 & 0.06 & 0.06 \\
DES  & 0.02 & 0.02  &  0.02& 0.03  & 0.03  &  0.03\\
HSC  & 0.05 & 0.05 & 0.05 & 0.05 & 0.05 & 0.05 \\
KiDS  &  0.04 & 0.04 & 0.04& 0.04 & 0.04 & 0.04 &  \\
\hline
\textbf{Mean reported $\sigma_{\rm sys}$}  & 0.056  & 0.056  & 0.056 & 0.058 & 0.058 & 0.058 \\
\hline
   \multicolumn{7}{c|}{Ensemble Systematic Error}   \\
\hline
\textbf{Estimated}   &
$0.05\pm _{0.02}^{0.06}$&
$0.08\pm _{0.04}^{0.09}$&
$0.08\pm _{0.02}^{0.07}$&
$0.15\pm _{0.04}^{0.11}$&
$0.14\pm _{0.04}^{0.12}$&
$0.12 \pm _{0.03}^{0.09}$\\
1$\sigma$ upper confidence limit&
0.11&
0.18&
0.14&
0.26&
0.27&
0.21&\\
2$\sigma$ upper confidence limit&
0.2&
0.32&
0.26&
0.46&
0.48&
0.37&\\
\hline
   \multicolumn{7}{c|}{C1 and C2 without CFHTLenS}   \\
\hline
\textbf{Mean reported $\sigma_{\rm sys}$}  & 0.043  & 0.043  & 0.043 & 0.045 & 0.045 & 0.045 \\
\hline
\textbf{Estimated}&
$0.07\pm _{0.03}^{0.09}$&
$0.1\pm _{0.04}^{0.12}$&
$0.1\pm _{0.03}^{0.1}$&
$0.08\pm _{0.06}^{0.14}$&
$0.12\pm _{0.06}^{0.16}$&
$0.06\pm _{0.05}^{0.11}$\\
1$\sigma$ upper confidence limit&
0.16&
0.22&
0.2&
0.2&
0.28&
0.17&\\
2$\sigma$ upper confidence limit&
0.33&
0.46&
0.41&
0.46&
0.598&
0.36&\\
\hline
\end{tabular}
\label{syserrtable}
\end{table*}

\subsection{Comparison with statistical error}\label{statcomparison}

Finally, we present a comparison between the current statistical errors on $\Delta\Sigma$ and the systematic errors. Green data points in Figure \ref{fig:sigmasys} indicate the mean statistical error on $\Delta\Sigma$ (averaged over all lensing surveys). When considering the full radial range, the mean statistical error is 3.6\%, 4.6\%, 4.4\%, and 6\% respectively for L1, L2, C1, and C2. Individual surveys have different constraining power with signal-to-noise ratios given in Table \ref{lensingsn}. Generally speaking, as can be seen by comparing the values of $\overline{\sigma}_{\rm sys}$ with the systematic errors in Figure \ref{fig:sigmasys}, galaxy-galaxy lensing measurements are entering an era in which systematic errors will dominate error budgets and identifying the origin of the $z_{\rm s}$ trend found in Figure \ref{fig:ampzsource} is a top priority.

\section{Discussion}\label{discussion}

We discuss variations in $\Delta\Sigma$ with $z_{\rm s}$, outline a number of considerations for future implementations of these tests, discuss the connection with cosmic shear, and implications for the ``lensing is low" effect.

\subsection{Amplitude of $\Delta\Sigma$ versus source redshift}\label{zsource-discussion}

Figure \ref{fig:ampzsource} finds the amplitude of $\Delta\Sigma$ to correlate with the mean source redshift of the lensing survey for the C1 and C2 bins when considering statistical errors (see Section \ref{trendmethod}). The trend is such that the amplitude shift between the two extremities (CS82 and HSC) is of order $\sim$ 23\% and is independent of radial scales. While we see a trend that is detected with high significance, the spread caused by this trend ($\sim$ 23\% between CS82 and HSC) is consistent with the systematic errors reported by each of the lensing surveys (see test in Section \ref{withsyseror}) if we assume that the $z_{\rm s}$ trend is the leading term in the systematic error budget and some surveys have experienced a 1.5$\sigma$ upwards/downward shift. This effect could be attributed to photometric redshift calibration or other effects that correlate with the depth of lensing surveys. We now outline several plausible explanations for the origin of this trend. We have rank ordered this list from top to bottom beginning with what we consider to be the most plausible explanation and ending with the least likely.

\begin{enumerate}

    \item The impact of unrecognised (and uncorrected) blends is a possible important source of systematic error. The impact of unrecognised blends could correlate with the depth of lensing surveys. The trends discussed here could therefore be a manifestation of errors in the multiplicative shear calibration bias or redshift distributions due to blending. The lensing data used here include varying levels of sophistication to account for galaxy blends, and the interplay between galaxy blends and redshift distributions. In DES, \citet[][]{MacCrann2020} find that blending has the largest impact on high redshift source samples and is one of the dominant systematic errors for DES Y3. It is therefore quite plausible that the trend we find here is caused by unaccounted for unrecognised blends. It is not clear how unrecognised blends would impact different surveys. On the one hand, surveys with higher mean source redshifts (e.g. HSC) could be more impacted by blends. On the other hand, HSC also has the best seeing and should, in principle, have a better ability to disentangle galaxy blends. In the future, it will be interesting to consider a joint analysis of both $z_{\rm s}$ and survey specific PSF to see which has the dominant impact.
    
    \item Both HSC and DES rely on the COSMOS 30-band catalogue \citep{Laigle:2016aa} in calibrating their photometric redshifts. It is possible that there are biases in the COSMOS 30-band catalogue \citep{Hildebrandt2020, Joudaki2020,Myles2020}. Similarly, CFHTLenS redshifts have been found to be biased \citep{Choi:2015}. HSC, CS82, and KiDS also use similar spectroscopic galaxy catalogues (e.g., DEEP2, VVDS, Primus, etc.) to calibrate redshifts and/or to perform $f_{\rm bias}$ corrections. Unidentified selection effects in these spectroscopic catalogues that cannot be corrected for by re-weighting schemes could also be at play \citep[][]{Masters:2015aa,Gruen2017,Wright2020cfht}. However, Figure \ref{fig:ampzsourcemethod} did not reveal any pattern suggesting that photo-$z$ calibration choices determine the trend.
        
    \item With imperfect photometric redshifts, gg-lensing can be impacted by the so-called ``gI" intrinsic alignments (this is the correlation between the intrinsic shape of a source galaxy (I) and the position of the lens galaxy (g), see \citealt{joachimi2015} and references therein).  Intrinsic alignments could explain the $z_{\rm s}$ trend because deeper surveys (where source galaxies are on average further behind lens galaxies and therefore are less likely to be physically associated with the lens) will have reduced contamination from gI intrinsic alignment.  However, \citet[][]{blazek2012} used BOSS galaxies to quantify the level of gI contamination for gg-lensing and concluded that with stringent photo-z cuts, the expected gI contamination was at the 1-2\% percent level which would be too low to explain the trend observed here.

    \item The magnification of galaxies can impact the amplitude of $\Delta\Sigma$. \report{In particular, magnification due to the lens sample can have an important effect \citep[][]{Unruh2020, Wietersheim2021}. This effect can be calculated analytically given the slope of the galaxy luminosity function at the cut that determines the lens selection (for example, see equation 13 in \citealt{Unruh2020})}. For CMASS, the $i_{\rm cmod}$ cut limits the CMASS sample in this redshift range. Appendix \ref{lensmagnification} shows that CMASS has a slope, $\alpha$, that is steeper than 1. Because $\alpha > 1$, lens magnification would induce a positive bias in the average tangential shear $\gamma_t$ that doesn't strongly depend on $z_{\rm s}$. Because the critical surface density for a given lens sample is anti-correlated with $z_{\rm s}$, the lens magnification impact on $\Delta\Sigma=\gamma_{\rm t}\Sigma_{\rm crit}$ should also decrease with $z_{\rm s}$. Since the net effect of lens magnification is positive, this means the measured $\Delta\Sigma$ (without correction for lens magnification) should decrease with $z_{\rm s}$. This is the opposite of what we find.
    
    \item Amplitude offsets could be caused by dilution effects and should be investigated further. However, when estimates of the dilution factor are made using spectroscopic calibration samples, they are generally found to be relatively small (a few percent).
\end{enumerate}

The assumption of an incorrect cosmology does bias measurements of $\Delta\Sigma$, but this bias is only a very weak function of $z_{\rm s}$ and we have checked that this cannot explain the trends seen in Figure \ref{fig:ampzsource}. Boost factor corrections would cause a scale-dependent trend, unlike the one found.

Overall, unrecognised blends and redshift calibration seem to be the most plausible explanations. These effects represent two of the biggest challenges for weak lensing analyses. The trend identified in Figure \ref{fig:ampzsource} warrants further investigation as a leading term in the systematic error budget for galaxy-galaxy lensing measurements for $z_{\rm s}>0.43$.

\subsection{On the possibility of underestimated statistical errors}\label{errs-discussion}

Some of the trends found in this paper could be the result of underestimated statistical errors. This is a likely explanation for the correlation between the lensing amplitude and $n_{\rm star}$ for the L1 bin. Indeed, we have neglected the covariance between lensing surveys (this is discussed further in Section \ref{jointcovariance}). An underestimation of the lensing statistical errors would result in an overestimation of $\sigma_{\rm sys}$. Since the main goal of this paper is to place upper limits on $\sigma_{sys}$, underestimated lensing errors would lead to conservative conclusions.

\subsection{Considerations for future implementations of these tests}

The types of empirically motivated tests presented here will become more precise as lensing surveys expand and the overlap between surveys grows. Furthermore, DESI will both provide a larger number of galaxies to use for lenses and DESI lens samples will also extend in redshift to $z>0.7$. Samples from 4MOST \citep[4-metre Multi-Object Spectroscopic Telescope,][]{deJONG2019} will also provide additional lens samples with spectroscopic redshifts. Here we discuss a number of considerations for future implementations of this work. 

\subsubsection{Inhomogeneity with angular clustering}

In future work, a deeper understanding of inhomogeneity effects could also be gained by investigating the angular clustering, $w(\theta)$ in the regions of overlap with lensing surveys, as well as studying the properties of BOSS or DESI galaxies (e.g. stellar masses, sizes) as a function of various observing conditions (SDSS PSF, stellar density, etc.) using a deeper catalogue than SDSS (e.g. the HSC catalogue).

\subsubsection{Joint covariance}\label{jointcovariance}

In the present analysis, we have assumed that the errors on $\Delta\Sigma$ for measurements by different surveys are uncorrelated. In reality, there is likely to be a small positive correlation between the signals measured by different surveys as discussed in more detail below. The assumption of uncorrelated errors is thus conservative in terms of avoiding false positive detections of discrepancies. Future implementations of this program will need to account for the cross correlations between surveys, especially as the overlap regions between surveys increases. There will be two sources of correlation of statistical errors. One is correlated large-scale structure in the regions of overlap. The second will be correlated shape noise for sources that are in common between surveys. Differences in depth and binning will reduce the number of common sources in overlap regions and will have to be modelled correctly. For the correlated large-scale structure component, it will be possible to use mock catalogues that are lensed by the same large-scale structure which have the respective $dn/dz$ for each survey. For the correlated shape noise component, a data-driven approach will be possible. The correlated shape noise component can be estimated by taking galaxies in common between two given surveys, rotating them by the same angle, and measuring the correlated shape noise. 

\subsubsection{Tightening constraint on $\sigma_{\rm sys}$}

The constraints on $\sigma_{\rm sys}$ from this paper are still relatively weak because of the limited overlap between lensing surveys and BOSS. In future tests, it will be desirable to tighten the constraints on $\sigma_{\rm sys}$. One way to tighten constraints on $\sigma_{\rm sys}$ will be to combine across redshift bins to make a more stringent test. However, this will require modeling the cross-covariance between lens bins (due to the overlap in the source populations). Also, if present, inhomogeneity effects must first be characterised and removed (or ruled out) before attempting to combine across redshift bins.

\subsubsection{Disentangling biases}

The approach presented here provides guidance on the source of any differences that are found (e.g. boost factors vs. photo-$z$ bias) but unavoidable degeneracies will also be present (e.g. calibration bias versus photo-$z$s and methodology). As such, tests on simulations will also be important to disentangle factors that contribute to $\sigma_{\rm sys}$. Such tests are already underway within the context of the DESI collaboration (Lange et al in prep).

Future implementation of these tests should also consider how to optimally make use of the overlap regions between surveys. For example, using areas of common overlap avoids the issue of inhomogeneous lens samples. However, using the full area will always yield tighter constraints on $\sigma_{\rm sys}$. On the other hand, DESI will be much more homogeneous \citep[][]{Kitanidis2020} and inhomogeneity may not be an issue for DESI. \report{It will also be interesting to consider joint fits for different effects using radially dependant functional form with the expected scaling for various effects (e.g. one could assume a specific radially dependent functional form for the impact of boost factors)}.

\subsection{Connection with cosmic shear}

The ensemble systematic error estimates presented here are relevant for cosmic shear measurements in the sense that the same source galaxies used to measure $\Delta\Sigma$ are also used in cosmic shear measurements. Systematic errors in the source sample that correlate with survey depth will affect both $\Delta\Sigma$ as well as cosmic shear. However, it isn't trivial to directly translate the numbers found here into systematic errors for a given cosmic shear tomographic bin. For example, $\Delta\Sigma$ will have a different sensitivity to photo-$z$ errors than cosmic shear. Follow-up work is warranted to study the connection between the numbers reported here for $\Delta\Sigma$ and cosmic shear. 

It is also important to note that trends found in this paper could also originate from different methodologies used to compute $\Delta\Sigma$. While teams have performed numerous tests of their methods, it is still possible that trends could originate from different ways in which photo-$z$'s are used for example (although Figure \ref{fig:ampzsourcemethod} did not reveal anything obvious). To rule out this possibility, the methodology employed by SDSS, HSC, and CS82 (as implemented in the \texttt{dsigma} pipeline) will undergo extensive testing in the DESI Lensing Mock Challenge (Lange et al in prep.) and the next implementation of Lensing Without Borders will employ a single measurement pipeline. Ruling out methodology differences will also be important in the effort to build the connection between the numbers presented here and cosmic shear.

\subsection{Implication for the ``Lensing is Low" effect}

It has been observed that models of the galaxy-halo connection applied to the BOSS CMASS and LOWZ samples over-predict $\Delta\Sigma$ by 20-30\% \citep[][]{Leauthaud:2017aa, Lange2019,Singh:2020, Lange2020}. What are the implications of the finding of this paper with regards to ``lensing is low"? 

\citet[][]{Leauthaud:2017aa} studied the CMASS sample over the full redshift range $0.45<z<0.7$. For the range $0.45<z<0.54$ we find estimated values for the ensemble systematic error to be of order 5-8\% in good agreement with those reported by the lensing surveys. For the redshift range $0.54<z<0.7$ we find values that are larger: 12-15\% but also with large uncertainties on the reported values. Our results in the range $0.54<z<0.7$ also depend on the interpretation of the amplitude-$n_{\rm star}$ trend (\report{because if the $n_{\rm star}$ trend is due to lens inhomogeneity then the variance of the amplitudes in this redshift range should not be attributed to source systematics}). This redshift range will therefore require further investigation and constraints on $\sigma_{\rm sys}$ are not tight enough to draw any conclusions at $z_{\rm L}>0.54$.

\citet[][]{Lange2019} find the  ``lensing is low" effect in the LOWZ redshift range using CFHTLenS. \citet[][]{Lange2020} also find the same effect using LOWZ in stellar mass bins using lensing from SDSS. In this redshift range, we find $\sigma_{sys}$ values that are in good agreement with those reported by lensing surveys. 

Using a fully empirical method, we do not find evidence for large (20-30\% level) unknown systematic errors. Using the values for $\sigma_{\rm sys}$ derived over the full radial range, systematic errors greater than 15\%, 13\%, and 14\% are ruled out at 68\% confidence level for L1, L2, and C1 respectively. Systematic errors greater than 24\%, 20\%, and 26\% are ruled out at 95\% confidence level for L1, L2, and C1 respectively. Given that the observed ``lensing is low" effect is at the 20\% to 35\% level (see e.g., figure 5 in \citealt[][]{Lange2019}), this suggests that it is difficult to explain the ``lensing is low" effect via lensing systematic errors alone (e.g., see Figure \ref{fig:sigmasys}).  A follow-up paper will explore the ``lensing is low" effect with BOSS clustering data and updated lensing data (Amon et al. in prep).


\section{Summary and Conclusions}\label{conclusions}

Lensing Without Borders is a a cross-survey collaboration created to assess the consistency of lensing signals computed with different data-sets and to perform empirically motivated tests of lensing systematic errors. Our main tests are based on the premise that the gg-lensing signal ($\Delta\Sigma$) around BOSS galaxies is a physical quantity. The amplitude of $\Delta\Sigma$ should be independent of the lensing data used to perform the measurement. The excess spread (above the expected statistical uncertainties) in the amplitude of  $\Delta\Sigma$ amongst lensing surveys can be used to estimate an ensemble systematic error, $\sigma_{\rm sys}$.  We estimate $\sigma_{\rm sys}$ via a blind comparison of the amplitude of $\Delta\Sigma$ using four lens sample from BOSS and using the sources catalogues and methodologies from six distinct lensing surveys (SDSS, CS82, CFHTLenS, DES, HSC, KiDS). Our main results are summarised below.

\begin{itemize}
    \item \kidsnew{For all redshift and radial ranges, we find good agreement between our empirically estimated values and the reported systematic errors (see Table \ref{syserrtable} and Figure \ref{fig:sigmasys}). Differences between the estimated and reported systematic errors are always less than 2.3$\sigma$}.    
    
     \item \kidsnew{Estimated values for $\sigma_{\rm sys}$ are largest in the C2 bin with values that are $\sim$2$\sigma$ larger than reported values. But differences remain below 3$\sigma$ (our pre-blinding determined criterion for claiming a detection). There are also other effects (see below) that complicate interpretation in this redshift range.}
    
    \item For lenses with $0.43<z_{\rm L}<0.54 $ (source galaxies with $z_{\rm s} \gtrapprox 0.5$) we detect a correlation between the $\Delta\Sigma$ amplitudes and the depth of lensing surveys (detected at 3-4$\sigma$ using statistical errors). This correlation explains most of the scatter between the lensing measurements. Section \ref{zsource-discussion} presents several explanations for this trend. Two likely candidates are unrecognised blends and photometric redshift calibration. Investigating the origin of this trend is key as it will be a leading term in the systematic error budget for gg-lensing measurements at $z_{\rm s} \gtrapprox 0.5$. 
    
    \item For lenses between $0.54<z_{\rm L}<0.7$ (source galaxies with $z_{\rm s} \gtrapprox 0.7$) we find  a correlation between the amplitude of $\Delta\Sigma$ and the foreground stellar density, $n_{\rm star}$ \report{(4$\sigma$ to 6$\sigma$ using statistical errors)}. We raise three possible explanations for this trend (see Appendix \ref{nsappendix}). We perform a test using HSC data (Figure \ref{fig:amps_hsc_internal}) that leads us to favour an explanation in which the trend originates from a background source systematic (blending, shear calibration, etc. ), or that the trend is a statistical fluke, however, the picture could also be more complicated \citep[e.g., see ][]{Singh:2020}. Further work will be required to draw conclusive statements about the origin of the dispersion among lensing measurements in this redshift range.
    \item The combined effects of the $n_{\rm star}$ and the $z_{\rm s}$ (survey depth) trend explain a majority of the observed scatter in the lensing amplitudes at $z>0.43$. Investigating the origin of both the $n_{\rm star}$ and the $z_{\rm s}$ trends will be key for gg-lensing measurements in the redshift range $z_{\rm L}>0.43$.
    \item \report{All trends except the L1 all-R trend in Figure \ref{fig:ampsys_nstar} drop below 3$\sigma$ when the estimated systematic errors are summed in quadrature with the statistical errors. This is consistent with our assessment that trends with $z_{\rm s}$ and $n_{\rm star}$ are within the reported systematic error budgets (see Appendix \ref{statcomparison} and Figure \ref{fig:staterrors}). These trends could originate from ``known known" (and thus accounted for) sources of systematic error. }
    \item Our systematic errors estimates do not appear to be strongly radially dependent. This suggests that boost factors are not a dominant cause of spread between the measurements.   
    \item We compare our empirical estimates of the systematic error to current statistical constraints on the amplitude of gg-lensing (Figure \ref{fig:sigmasys}). We find that current measurements have $\sigma_{\rm stat}\sim \sigma_{\rm sys}$. This is a statement about the ensemble of data - individual surveys may have lower systematics than this. Understanding the origin of systematic errors, and reducing uncertainty in our corrections for these errors, will be the next key challenge facing gg-lensing measurements.
    \item Using a fully empirical method, we do not find evidence for large (20-30\% level) unknown systematic errors. Using the values for $\sigma_{\rm sys}$ derived over the full radial range, systematic errors greater than 15\%, 13\%, and 14\% are ruled out at 68\% confidence level for L1, L2, and C1 respectively. Systematic errors greater than 24\%, 20\%, and 26\% are ruled out at 95\% confidence level for L1, L2, and C1 respectively. These constraints could be made tighter by combining between lens bins -- but we have not attempted this here. The overlap between lensing surveys and BOSS limits the constraining power of our tests, but the methods developed here will become more powerful as DESI comes online, and as the coverage of lensing surveys continue to grow. 
\end{itemize}

We have provided the first direct and empirically motivated test of the consistency of the gg-lensing amplitude across lensing surveys and developed a framework for such comparisons. We do not find evidence for large unknown systematic errors in these lensing data. However, systematic errors that are common between different lensing surveys cannot be tested with our methodology. 

Our results  are relevant for cosmic shear measurements because the same source galaxies used to measure $\Delta\Sigma$ are also used in cosmic shear measurements. However, cosmic shear and galaxy-galaxy lensing are different measurements (for example they are sensitive to photo-$z$'s in different ways) and so the numbers reported here cannot be directly applied to cosmic shear.

For lenses in the range $z_{\rm L}<0.54$ we find $\sigma_{\rm sys}$ of order 4-9\% in good agreement with reported values and systematic errors of 25\% are ruled out at 2$\sigma$ in three different lens bins. In this same redshift range, the  ``lensing is low" effect is at 20\% -35\% (e.g., figure 5 in \citealt[][]{Lange2020}). We conclude that for lenses below $z_{\rm L}<0.54$, it is difficult to explain the ``lensing is low"  mis-match with clustering \citep{Leauthaud:2017aa, Lange2019, Singh:2020, Lange2020} via lensing systematic errors alone (Figure \ref{fig:sigmasys}). Constraints on $\sigma_{\rm sys}$ are not tight enough to draw any conclusions at $z_{\rm L}>0.54$. The ``lensing is low" effect in relation to clustering will be explored in greater detail in a companion paper (Amon et al in prep).

\section*{Acknowledgements}

\textit{Author contributions}:  The authorship list reflects the two lead authors (AL, AA), those who measured the lensing signals and contributed key ideas to the paper (SS, DG, UL, SH, NR, TV, YL, CH, CB) followed by an alphabetical group.

This paper has gone through internal review by the DES, HSC, and KiDS collaborations.

We acknowledge use of the lux supercomputer at UC Santa Cruz, funded by NSF MRI grant AST 1828315. This material is based on work supported by the U.D Department of Energy, Office of Science, Office of High Energy Physics under Award Number DE-SC0019301. AL acknowledges support from the David and Lucille Packard foundation, and from the Alfred .P Sloan foundation.

CH acknowledges support from the European Research Council under grant number 647112, and support from the Max Planck Society and the Alexander von Humboldt Foundation in the framework of the Max Planck-Humboldt Research Award endowed by the Federal Ministry of Education and Research. KK acknowledges support from the Royal Society and Imperial College. HH is supported by a Heisenberg grant of the Deutsche Forschungsgemeinschaft (Hi 1495/5-1) as well as an ERC Consolidator Grant (No. 770935). AHW is supported by an European Research Council Consolidator Grant (No. 770935).

Funding for the DES Projects has been provided by the U.S. Department of Energy, the U.S. National Science Foundation, the Ministry of Science and Education of Spain, 
the Science and Technology Facilities Council of the United Kingdom, the Higher Education Funding Council for England, the National Center for Supercomputing 
Applications at the University of Illinois at Urbana-Champaign, the Kavli Institute of Cosmological Physics at the University of Chicago, 
the Center for Cosmology and Astro-Particle Physics at the Ohio State University,
the Mitchell Institute for Fundamental Physics and Astronomy at Texas A\&M University, Financiadora de Estudos e Projetos, 
Funda{\c c}{\~a}o Carlos Chagas Filho de Amparo {\`a} Pesquisa do Estado do Rio de Janeiro, Conselho Nacional de Desenvolvimento Cient{\'i}fico e Tecnol{\'o}gico and 
the Minist{\'e}rio da Ci{\^e}ncia, Tecnologia e Inova{\c c}{\~a}o, the Deutsche Forschungsgemeinschaft and the Collaborating Institutions in the Dark Energy Survey. 

The Collaborating Institutions are Argonne National Laboratory, the University of California at Santa Cruz, the University of Cambridge, Centro de Investigaciones Energ{\'e}ticas, 
Medioambientales y Tecnol{\'o}gicas-Madrid, the University of Chicago, University College London, the DES-Brazil Consortium, the University of Edinburgh, 
the Eidgen{\"o}ssische Technische Hochschule (ETH) Z{\"u}rich, 
Fermi National Accelerator Laboratory, the University of Illinois at Urbana-Champaign, the Institut de Ci{\`e}ncies de l'Espai (IEEC/CSIC), 
the Institut de F{\'i}sica d'Altes Energies, Lawrence Berkeley National Laboratory, the Ludwig-Maximilians Universit{\"a}t M{\"u}nchen and the associated Excellence Cluster Universe, 
the University of Michigan, NFS's NOIRLab, the University of Nottingham, The Ohio State University, the University of Pennsylvania, the University of Portsmouth, 
SLAC National Accelerator Laboratory, Stanford University, the University of Sussex, Texas A\&M University, and the OzDES Membership Consortium.

Based in part on observations at Cerro Tololo Inter-American Observatory at NSF's NOIRLab (NOIRLab Prop. ID 2012B-0001; PI: J. Frieman), which is managed by the Association of Universities for Research in Astronomy (AURA) under a cooperative agreement with the National Science Foundation.

The DES data management system is supported by the National Science Foundation under Grant Numbers AST-1138766 and AST-1536171.
The DES participants from Spanish institutions are partially supported by MICINN under grants ESP2017-89838, PGC2018-094773, PGC2018-102021, SEV-2016-0588, SEV-2016-0597, and MDM-2015-0509, some of which include ERDF funds from the European Union. IFAE is partially funded by the CERCA program of the Generalitat de Catalunya.
Research leading to these results has received funding from the European Research
Council under the European Union's Seventh Framework Program (FP7/2007-2013) including ERC grant agreements 240672, 291329, and 306478.
We  acknowledge support from the Brazilian Instituto Nacional de Ci\^encia
e Tecnologia (INCT) do e-Universo (CNPq grant 465376/2014-2).

This manuscript has been authored by Fermi Research Alliance, LLC under Contract No. DE-AC02-07CH11359 with the U.S. Department of Energy, Office of Science, Office of High Energy Physics.

The Hyper Suprime-Cam (HSC) collaboration includes the astronomical communities of Japan and Taiwan, and Princeton University.  The HSC instrumentation and software were developed by the National Astronomical Observatory of Japan (NAOJ), the Kavli Institute for the Physics and Mathematics of the Universe (Kavli IPMU), the University of Tokyo, the High Energy Accelerator Research Organisation (KEK), the Academia Sinica Institute for Astronomy and Astrophysics in Taiwan (ASIAA), and Princeton University.  Funding was contributed by the FIRST program from the Japanese Cabinet Office, the Ministry of Education, Culture, Sports, Science and Technology (MEXT), the Japan Society for the Promotion of Science (JSPS), Japan Science and Technology Agency  (JST), the Toray Science  Foundation, NAOJ, Kavli IPMU, KEK, ASIAA, and Princeton University.
 
This paper makes use of software developed for the Large Synoptic Survey Telescope. We thank the LSST Project for making their code available as free software at  \url{http://dm.lsst.org}
 
This paper is based [in part] on data collected at the Subaru Telescope and retrieved from the HSC data archive system, which is operated by Subaru Telescope and Astronomy Data Center (ADC) at NAOJ. Data analysis was in part carried out with the cooperation of Center for Computational Astrophysics (CfCA), NAOJ.
 
The Pan-STARRS1 Surveys (PS1) and the PS1 public science archive have been made possible through contributions by the Institute for Astronomy, the University of Hawaii, the Pan-STARRS Project Office, the Max Planck Society and its participating institutes, the Max Planck Institute for Astronomy, Heidelberg, and the Max Planck Institute for Extraterrestrial Physics, Garching, The Johns Hopkins University, Durham University, the University of Edinburgh, the Queen’s University Belfast, the Harvard-Smithsonian Center for Astrophysics, the Las Cumbres Observatory Global Telescope Network Incorporated, the National Central University of Taiwan, the Space Telescope Science Institute, the National Aeronautics and Space Administration under grant No. NNX08AR22G issued through the Planetary Science Division of the NASA Science Mission Directorate, the National Science Foundation grant No. AST-1238877, the University of Maryland, Eotvos Lorand University (ELTE), the Los Alamos National Laboratory, and the Gordon and Betty Moore Foundation.

Based on observations made with ESO Telescopes at the La Silla Paranal Observatory under programme IDs 177.A-3016, 177.A-3017, 177.A-3018 and 179.A-2004, and on data products produced by the KiDS consortium. The KiDS production team acknowledges support from: Deutsche Forschungsgemeinschaft, ERC, NOVA and NWO-M grants; Target; the University of Padova, and the University Federico II (Naples).

This work was supported by the Department of Energy, Laboratory Directed Research and Development program at SLAC National Accelerator Laboratory, under contract DE-AC02-76SF00515 and as part of the Panofsky Fellowship awarded to DG. MM is partially funded by FAPERJ, CNPq and CONICET. BM acknowledges support from the Brazilian funding agency FAPERJ.

\section*{Data Availability}

Our lensing signals and code used to make the main figures are available at \url{https://github.com/alexieleauthaud/lensingwithoutborders}.


\bibliographystyle{mnras}

\bibliography{all_refs.bib,SDSS.bib,all_refs_AA.bib,DES.bib}

\begin{thebibliography}{}
\makeatletter
\relax
\def\mn@urlcharsother{\let\do\@makeother \do\$\do\&\do\#\do\^\do\_\do\%\do\~}
\def\mn@doi{\begingroup\mn@urlcharsother \@ifnextchar [ {\mn@doi@}
  {\mn@doi@[]}}
\def\mn@doi@[#1]#2{\def\@tempa{#1}\ifx\@tempa\@empty \href
  {http://dx.doi.org/#2} {doi:#2}\else \href {http://dx.doi.org/#2} {#1}\fi
  \endgroup}
\def\mn@eprint#1#2{\mn@eprint@#1:#2::\@nil}
\def\mn@eprint@arXiv#1{\href {http://arxiv.org/abs/#1} {{\tt arXiv:#1}}}
\def\mn@eprint@dblp#1{\href {http://dblp.uni-trier.de/rec/bibtex/#1.xml}
  {dblp:#1}}
\def\mn@eprint@#1:#2:#3:#4\@nil{\def\@tempa {#1}\def\@tempb {#2}\def\@tempc
  {#3}\ifx \@tempc \@empty \let \@tempc \@tempb \let \@tempb \@tempa \fi \ifx
  \@tempb \@empty \def\@tempb {arXiv}\fi \@ifundefined
  {mn@eprint@\@tempb}{\@tempb:\@tempc}{\expandafter \expandafter \csname
  mn@eprint@\@tempb\endcsname \expandafter{\@tempc}}}

\bibitem[\protect\citeauthoryear{{Abazajian} et~al.,}{{Abazajian}
  et~al.}{2004}]{Abazajian:2004}
{Abazajian} K.,  et~al., 2004, \mn@doi [\aj] {10.1086/421365}, \href
  {http://adsabs.harvard.edu/abs/2004AJ....128..502A} {128, 502}

\bibitem[\protect\citeauthoryear{{Abbott} et~al.,}{{Abbott}
  et~al.}{2018}]{desy1kp}
{Abbott} T.~M.~C.,  et~al., 2018, \mn@doi [\prd] {10.1103/PhysRevD.98.043526},
  \href {https://ui.adsabs.harvard.edu/abs/2018PhRvD..98d3526A} {98, 043526}

\bibitem[\protect\citeauthoryear{{Ahn} et~al.,}{{Ahn} et~al.}{2014}]{Ahn:2014}
{Ahn} C.~P.,  et~al., 2014, \mn@doi [\apjs] {10.1088/0067-0049/211/2/17}, \href
  {http://adsabs.harvard.edu/abs/2014ApJS..211...17A} {211, 17}

\bibitem[\protect\citeauthoryear{{Aihara} et~al.,}{{Aihara}
  et~al.}{2011}]{Aihara:2011}
{Aihara} H.,  et~al., 2011, \mn@doi [\apjs] {10.1088/0067-0049/193/2/29}, \href
  {http://adsabs.harvard.edu/abs/2011ApJS..193...29A} {193, 29}

\bibitem[\protect\citeauthoryear{{Aihara} et~al.,}{{Aihara}
  et~al.}{2018a}]{Aihara:2018}
{Aihara} H.,  et~al., 2018a, \mn@doi [\pasj] {10.1093/pasj/psx066}, \href
  {http://adsabs.harvard.edu/abs/2018PASJ...70S...4A} {70, S4}

\bibitem[\protect\citeauthoryear{{Aihara} et~al.,}{{Aihara}
  et~al.}{2018b}]{Aihara:2018aa}
{Aihara} H.,  et~al., 2018b, \mn@doi [Publications of the Astronomical Society
  of Japan] {10.1093/pasj/psx081}, \href
  {https://ui.adsabs.harvard.edu/abs/2018PASJ...70S...8A} {70, S8}

\bibitem[\protect\citeauthoryear{{Alam} et~al.,}{{Alam}
  et~al.}{2015}]{Alam:2015}
{Alam} S.,  et~al., 2015, \mn@doi [\apjs] {10.1088/0067-0049/219/1/12}, \href
  {http://adsabs.harvard.edu/abs/2015ApJS..219...12A} {219, 12}

\bibitem[\protect\citeauthoryear{{Alam} et~al.,}{{Alam}
  et~al.}{2017}]{Alam:2017aa}
{Alam} S.,  et~al., 2017, \mn@doi [\mnras] {10.1093/mnras/stx721}, \href
  {http://adsabs.harvard.edu/abs/2017MNRAS.470.2617A} {470, 2617}

\bibitem[\protect\citeauthoryear{{Amon} et~al.,}{{Amon} et~al.}{2018a}]{Amonir}
{Amon} A.,  et~al., 2018a, \mn@doi [\mnras] {10.1093/mnras/sty859}, \href
  {http://adsabs.harvard.edu/abs/2018MNRAS.477.4285A} {477, 4285}

\bibitem[\protect\citeauthoryear{{Amon} et~al.,}{{Amon}
  et~al.}{2018b}]{Amon2018}
{Amon} A.,  et~al., 2018b, \mn@doi [\mnras] {10.1093/mnras/sty1624}, \href
  {https://ui.adsabs.harvard.edu/abs/2018MNRAS.479.3422A} {479, 3422}

\bibitem[\protect\citeauthoryear{{Annis} et~al.,}{{Annis}
  et~al.}{2014}]{Annis2014}
{Annis} J.,  et~al., 2014, \mn@doi [\apj] {10.1088/0004-637X/794/2/120}, \href
  {https://ui.adsabs.harvard.edu/abs/2014ApJ...794..120A} {794, 120}

\bibitem[\protect\citeauthoryear{{Applegate} et~al.,}{{Applegate}
  et~al.}{2014}]{Applegate:2014aa}
{Applegate} D.~E.,  et~al., 2014, \mn@doi [\mnras] {10.1093/mnras/stt2129},
  \href {http://adsabs.harvard.edu/abs/2014MNRAS.439...48A} {439, 48}

\bibitem[\protect\citeauthoryear{{Asgari} et~al.,}{{Asgari}
  et~al.}{2019}]{Asgari:2019}
{Asgari} M.,  et~al., 2019, \mn@doi [\aap] {10.1051/0004-6361/201834379}, \href
  {https://ui.adsabs.harvard.edu/abs/2019A&A...624A.134A} {624, A134}

\bibitem[\protect\citeauthoryear{{Asgari} et~al.,}{{Asgari}
  et~al.}{2021}]{Asgari2020}
{Asgari} M.,  et~al., 2021, \mn@doi [\aap] {10.1051/0004-6361/202039070}, \href
  {https://ui.adsabs.harvard.edu/abs/2021A&A...645A.104A} {645, A104}

\bibitem[\protect\citeauthoryear{{Ben{\'{\i}}tez}}{{Ben{\'{\i}}tez}}{2000}]{Benitez:2000}
{Ben{\'{\i}}tez} N.,  2000, \mn@doi [\apj] {10.1086/308947}, \href
  {http://adsabs.harvard.edu/cgi-bin/nph-bib_query?bibcode=2000ApJ...536..571B&db_key=AST}
  {536, 571}

\bibitem[\protect\citeauthoryear{{Blake} et~al.,}{{Blake}
  et~al.}{2016}]{Blake2016}
{Blake} C.,  et~al., 2016, \mn@doi [\mnras] {10.1093/mnras/stv2875}, \href
  {https://ui.adsabs.harvard.edu/abs/2016MNRAS.456.2806B} {456, 2806}

\bibitem[\protect\citeauthoryear{{Blake} et~al.,}{{Blake}
  et~al.}{2020}]{Blake2020}
{Blake} C.,  et~al., 2020, \mn@doi [\aap] {10.1051/0004-6361/202038505}, \href
  {https://ui.adsabs.harvard.edu/abs/2020A&A...642A.158B} {642, A158}

\bibitem[\protect\citeauthoryear{{Blazek}, {Mandelbaum}, {Seljak}  \&
  {Nakajima}}{{Blazek} et~al.}{2012}]{blazek2012}
{Blazek} J.,  {Mandelbaum} R.,  {Seljak} U.,   {Nakajima} R.,  2012, \mn@doi
  [\jcap] {10.1088/1475-7516/2012/05/041}, \href
  {https://ui.adsabs.harvard.edu/abs/2012JCAP...05..041B} {2012, 041}

\bibitem[\protect\citeauthoryear{{Bolton} et~al.,}{{Bolton}
  et~al.}{2012}]{Bolton:2012}
{Bolton} A.~S.,  et~al., 2012, \mn@doi [\aj] {10.1088/0004-6256/144/5/144},
  \href {http://adsabs.harvard.edu/abs/2012AJ....144..144B} {144, 144}

\bibitem[\protect\citeauthoryear{{Bosch} et~al.,}{{Bosch}
  et~al.}{2018}]{Bosch:2018aa}
{Bosch} J.,  et~al., 2018, \mn@doi [Publications of the Astronomical Society of
  Japan] {10.1093/pasj/psx080}, \href
  {https://ui.adsabs.harvard.edu/abs/2018PASJ...70S...5B} {70, S5}

\bibitem[\protect\citeauthoryear{{Boulade} et~al.,}{{Boulade}
  et~al.}{2003}]{Boulade:2003}
{Boulade} O.,  et~al., 2003, in {Iye} M.,  {Moorwood} A.~F.~M.,  eds,  Society
  of Photo-Optical Instrumentation Engineers (SPIE) Conference Series Vol.
  4841, Society of Photo-Optical Instrumentation Engineers (SPIE) Conference
  Series. pp 72--81, \mn@doi{10.1117/12.459890}

\bibitem[\protect\citeauthoryear{{Bridle} et~al.,}{{Bridle}
  et~al.}{2009}]{Bridle2009}
{Bridle} S.,  et~al., 2009, \mn@doi [Annals of Applied Statistics]
  {10.1214/08-AOAS222}, \href
  {http://adsabs.harvard.edu/abs/2009AnApS...3....6B} {3, 6}

\bibitem[\protect\citeauthoryear{{Buchs} et~al.,}{{Buchs}
  et~al.}{2019}]{Buchs:2019}
{Buchs} R.,  et~al., 2019, \mn@doi [\mnras] {10.1093/mnras/stz2162}, \href
  {https://ui.adsabs.harvard.edu/abs/2019MNRAS.489..820B} {489, 820}

\bibitem[\protect\citeauthoryear{{Bundy} et~al.,}{{Bundy}
  et~al.}{2015}]{Bundy:2015}
{Bundy} K.,  et~al., 2015, \mn@doi [\apjs] {10.1088/0067-0049/221/1/15}, \href
  {http://adsabs.harvard.edu/abs/2015ApJS..221...15B} {221, 15}

\bibitem[\protect\citeauthoryear{{Cacciato}, {van den Bosch}, {More}, {Mo}  \&
  {Yang}}{{Cacciato} et~al.}{2013}]{Cacciato:2013}
{Cacciato} M.,  {van den Bosch} F.~C.,  {More} S.,  {Mo} H.,   {Yang} X.,
  2013, \mn@doi [\mnras] {10.1093/mnras/sts525}, \href
  {http://adsabs.harvard.edu/abs/2013MNRAS.430..767C} {430, 767}

\bibitem[\protect\citeauthoryear{{Carlstrom} et~al.,}{{Carlstrom}
  et~al.}{2011}]{Carlstrom2011}
{Carlstrom} J.~E.,  et~al., 2011, \mn@doi [\pasp] {10.1086/659879}, \href
  {https://ui.adsabs.harvard.edu/abs/2011PASP..123..568C} {123, 568}

\bibitem[\protect\citeauthoryear{{Choi} et~al.,}{{Choi}
  et~al.}{2016}]{Choi:2015}
{Choi} A.,  et~al., 2016, \mn@doi [\mnras] {10.1093/mnras/stw2241}, \href
  {https://ui.adsabs.harvard.edu/abs/2016MNRAS.463.3737C} {463, 3737}

\bibitem[\protect\citeauthoryear{{Coe}, {Ben{\'{\i}}tez}, {S{\'a}nchez}, {Jee},
  {Bouwens}  \& {Ford}}{{Coe} et~al.}{2006}]{Coe:2006}
{Coe} D.,  {Ben{\'{\i}}tez} N.,  {S{\'a}nchez} S.~F.,  {Jee} M.,  {Bouwens} R.,
    {Ford} H.,  2006, \mn@doi [\aj] {10.1086/505530}, \href
  {http://adsabs.harvard.edu/cgi-bin/nph-bib_query?bibcode=2006AJ....132..926C&db_key=AST}
  {132, 926}

\bibitem[\protect\citeauthoryear{{DES Collaboration} et~al.,}{{DES
  Collaboration} et~al.}{2021}]{desy3}
{DES Collaboration} et~al., 2021, arXiv e-prints, \href
  {https://ui.adsabs.harvard.edu/abs/2021arXiv210513549D} {p. arXiv:2105.13549}

\bibitem[\protect\citeauthoryear{{DESI Collaboration} et~al.,}{{DESI
  Collaboration} et~al.}{2016}]{DESI2016}
{DESI Collaboration} et~al., 2016, arXiv e-prints, \href
  {https://ui.adsabs.harvard.edu/abs/2016arXiv161100036D} {p. arXiv:1611.00036}

\bibitem[\protect\citeauthoryear{{Davis} et~al.,}{{Davis}
  et~al.}{2017}]{Davis2018}
{Davis} C.,  et~al., 2017, arXiv e-prints, \href
  {https://ui.adsabs.harvard.edu/abs/2017arXiv171002517D} {p. arXiv:1710.02517}

\bibitem[\protect\citeauthoryear{{Dawson} et~al.,}{{Dawson}
  et~al.}{2013}]{Dawson:2013}
{Dawson} K.~S.,  et~al., 2013, \mn@doi [\aj] {10.1088/0004-6256/145/1/10},
  \href {http://adsabs.harvard.edu/abs/2013AJ....145...10D} {145, 10}

\bibitem[\protect\citeauthoryear{{Drlica-Wagner} et~al.,}{{Drlica-Wagner}
  et~al.}{2018}]{DrlicaWagner2018}
{Drlica-Wagner} A.,  et~al., 2018, \mn@doi [\apjs] {10.3847/1538-4365/aab4f5},
  \href {https://ui.adsabs.harvard.edu/abs/2018ApJS..235...33D} {235, 33}

\bibitem[\protect\citeauthoryear{{Dunkley} et~al.}{{Dunkley}
  et~al.}{2009}]{Dunkley:2009}
{Dunkley} J.,  et~al., 2009, \mn@doi [\apjs] {10.1088/0067-0049/180/2/306},
  \href {http://adsabs.harvard.edu/abs/2009ApJS..180..306D} {180, 306}

\bibitem[\protect\citeauthoryear{{Dvornik} et~al.,}{{Dvornik}
  et~al.}{2018}]{Dvornik2018}
{Dvornik} A.,  et~al., 2018, \mn@doi [\mnras] {10.1093/mnras/sty1502}, \href
  {https://ui.adsabs.harvard.edu/abs/2018MNRAS.479.1240D} {479, 1240}

\bibitem[\protect\citeauthoryear{{Edge}, {Sutherland}, {Kuijken}, {Driver},
  {McMahon}, {Eales}  \& {Emerson}}{{Edge} et~al.}{2013}]{Edge2013}
{Edge} A.,  {Sutherland} W.,  {Kuijken} K.,  {Driver} S.,  {McMahon} R.,
  {Eales} S.,   {Emerson} J.~P.,  2013, The Messenger, \href
  {https://ui.adsabs.harvard.edu/abs/2013Msngr.154...32E} {154, 32}

\bibitem[\protect\citeauthoryear{{Eisenstein} et~al.,}{{Eisenstein}
  et~al.}{2001}]{Eisenstein:2001}
{Eisenstein} D.~J.,  et~al., 2001, \mn@doi [\aj] {10.1086/323717}, \href
  {http://adsabs.harvard.edu/abs/2001AJ....122.2267E} {122, 2267}

\bibitem[\protect\citeauthoryear{{Eisenstein} et~al.,}{{Eisenstein}
  et~al.}{2011}]{Eisenstein:2011}
{Eisenstein} D.~J.,  et~al., 2011, \mn@doi [\aj] {10.1088/0004-6256/142/3/72},
  \href {http://adsabs.harvard.edu/abs/2011AJ....142...72E} {142, 72}

\bibitem[\protect\citeauthoryear{{Erben} et~al.,}{{Erben}
  et~al.}{2005}]{Erben2005}
{Erben} T.,  et~al., 2005, \mn@doi [Astronomische Nachrichten]
  {10.1002/asna.200510396}, \href
  {https://ui.adsabs.harvard.edu/abs/2005AN....326..432E} {326, 432}

\bibitem[\protect\citeauthoryear{{Erben} et~al.,}{{Erben}
  et~al.}{2009}]{Erben:2009}
{Erben} T.,  et~al., 2009, \mn@doi [\aap] {10.1051/0004-6361:200810426}, \href
  {http://adsabs.harvard.edu/abs/2009A%26A...493.1197E} {493, 1197}

\bibitem[\protect\citeauthoryear{{Erben} et~al.,}{{Erben}
  et~al.}{2013}]{Erben:2013}
{Erben} T.,  et~al., 2013, \mn@doi [\mnras] {10.1093/mnras/stt928}, \href
  {http://adsabs.harvard.edu/abs/2013MNRAS.433.2545E} {433, 2545}

\bibitem[\protect\citeauthoryear{{Feldmann} et~al.,}{{Feldmann}
  et~al.}{2006}]{Feldman2006}
{Feldmann} R.,  et~al., 2006, \mn@doi [\mnras]
  {10.1111/j.1365-2966.2006.10930.x}, \href
  {http://adsabs.harvard.edu/abs/2006MNRAS.372..565F} {372, 565}

\bibitem[\protect\citeauthoryear{{Fenech Conti}, {Herbonnet}, {Hoekstra},
  {Merten}, {Miller}  \& {Viola}}{{Fenech Conti}
  et~al.}{2017}]{fenech-conti/etal:2016}
{Fenech Conti} I.,  {Herbonnet} R.,  {Hoekstra} H.,  {Merten} J.,  {Miller} L.,
    {Viola} M.,  2017, \mn@doi [\mnras] {10.1093/mnras/stx200}, \href
  {http://adsabs.harvard.edu/abs/2017MNRAS.467.1627F} {467, 1627}

\bibitem[\protect\citeauthoryear{{Finkbeiner} et~al.,}{{Finkbeiner}
  et~al.}{2016}]{Finkbeiner:2016}
{Finkbeiner} D.~P.,  et~al., 2016, \mn@doi [\apj] {10.3847/0004-637X/822/2/66},
  \href {http://adsabs.harvard.edu/abs/2016ApJ...822...66F} {822, 66}

\bibitem[\protect\citeauthoryear{{Flaugher} et~al.,}{{Flaugher}
  et~al.}{2015}]{Flaugher2015}
{Flaugher} B.,  et~al., 2015, \mn@doi [\aj] {10.1088/0004-6256/150/5/150},
  \href {https://ui.adsabs.harvard.edu/abs/2015AJ....150..150F} {150, 150}

\bibitem[\protect\citeauthoryear{{Ford} et~al.,}{{Ford}
  et~al.}{2015}]{Ford:2015}
{Ford} J.,  et~al., 2015, \mn@doi [\mnras] {10.1093/mnras/stu2545}, \href
  {https://ui.adsabs.harvard.edu/abs/2015MNRAS.447.1304F} {447, 1304}

\bibitem[\protect\citeauthoryear{{Fukugita}, {Ichikawa}, {Gunn}, {Doi},
  {Shimasaku}  \& {Schneider}}{{Fukugita} et~al.}{1996}]{Fukugita:1996}
{Fukugita} M.,  {Ichikawa} T.,  {Gunn} J.~E.,  {Doi} M.,  {Shimasaku} K.,
  {Schneider} D.~P.,  1996, \mn@doi [\aj] {10.1086/117915}, \href
  {http://adsabs.harvard.edu/abs/1996AJ....111.1748F} {111, 1748}

\bibitem[\protect\citeauthoryear{{Gatti} et~al.,}{{Gatti}
  et~al.}{2018a}]{GattiVielzeuf2018}
{Gatti} M.,  et~al., 2018a, \mn@doi [\mnras] {10.1093/mnras/sty466}, \href
  {https://ui.adsabs.harvard.edu/abs/2018MNRAS.477.1664G} {477, 1664}

\bibitem[\protect\citeauthoryear{{Gatti} et~al.,}{{Gatti}
  et~al.}{2018b}]{Gatti2017}
{Gatti} M.,  et~al., 2018b, \mn@doi [\mnras] {10.1093/mnras/sty466}, \href
  {https://ui.adsabs.harvard.edu/abs/2018MNRAS.477.1664G} {477, 1664}

\bibitem[\protect\citeauthoryear{Giblin et~al.,}{Giblin
  et~al.}{2021}]{Giblin2021}
Giblin B.,  et~al., 2021, \mn@doi [Astronomy & Astrophysics]
  {10.1051/0004-6361/202038850}, 645, A105

\bibitem[\protect\citeauthoryear{{G{\'o}rski}, {Hivon}, {Banday}, {Wand elt},
  {Hansen}, {Reinecke}  \& {Bartelmann}}{{G{\'o}rski}
  et~al.}{2005}]{gorski2005}
{G{\'o}rski} K.~M.,  {Hivon} E.,  {Banday} A.~J.,  {Wand elt} B.~D.,  {Hansen}
  F.~K.,  {Reinecke} M.,   {Bartelmann} M.,  2005, \mn@doi [\apj]
  {10.1086/427976}, \href
  {https://ui.adsabs.harvard.edu/abs/2005ApJ...622..759G} {622, 759}

\bibitem[\protect\citeauthoryear{{Gruen} \& {Brimioulle}}{{Gruen} \&
  {Brimioulle}}{2017}]{Gruen2017}
{Gruen} D.,  {Brimioulle} F.,  2017, \mn@doi [\mnras] {10.1093/mnras/stx471},
  \href {https://ui.adsabs.harvard.edu/abs/2017MNRAS.468..769G} {468, 769}

\bibitem[\protect\citeauthoryear{{Gruen} et~al.,}{{Gruen}
  et~al.}{2014}]{Gruen2014}
{Gruen} D.,  et~al., 2014, \mn@doi [\mnras] {10.1093/mnras/stu949}, \href
  {https://ui.adsabs.harvard.edu/abs/2014MNRAS.442.1507G} {442, 1507}

\bibitem[\protect\citeauthoryear{{Gunn} et~al.,}{{Gunn}
  et~al.}{1998}]{Gunn:1998}
{Gunn} J.~E.,  et~al., 1998, \mn@doi [\aj] {10.1086/300645}, \href
  {http://adsabs.harvard.edu/abs/1998AJ....116.3040G} {116, 3040}

\bibitem[\protect\citeauthoryear{{Gunn} et~al.,}{{Gunn}
  et~al.}{2006}]{Gunn:2006}
{Gunn} J.~E.,  et~al., 2006, \mn@doi [\aj] {10.1086/500975}, \href
  {http://adsabs.harvard.edu/abs/2006AJ....131.2332G} {131, 2332}

\bibitem[\protect\citeauthoryear{{Heymans} et~al.,}{{Heymans}
  et~al.}{2006}]{Heymans2006}
{Heymans} C.,  et~al., 2006, \mn@doi [\mnras]
  {10.1111/j.1365-2966.2006.10198.x}, \href
  {http://adsabs.harvard.edu/abs/2006MNRAS.368.1323H} {368, 1323}

\bibitem[\protect\citeauthoryear{{Heymans} et~al.,}{{Heymans}
  et~al.}{2012}]{Heymans:2012}
{Heymans} C.,  et~al., 2012, \mn@doi [\mnras]
  {10.1111/j.1365-2966.2012.21952.x}, \href
  {http://adsabs.harvard.edu/abs/2012MNRAS.427..146H} {427, 146}

\bibitem[\protect\citeauthoryear{{Heymans} et~al.,}{{Heymans}
  et~al.}{2021}]{Heymans2021}
{Heymans} C.,  et~al., 2021, \mn@doi [\aap] {10.1051/0004-6361/202039063},
  \href {https://ui.adsabs.harvard.edu/abs/2021A&A...646A.140H} {646, A140}

\bibitem[\protect\citeauthoryear{{Hikage} et~al.,}{{Hikage}
  et~al.}{2019}]{Hikage:2019aa}
{Hikage} C.,  et~al., 2019, \mn@doi [\pasj] {10.1093/pasj/psz010}, \href
  {http://adsabs.harvard.edu/abs/2019PASJ..tmp...22H} {}

\bibitem[\protect\citeauthoryear{{Hildebrandt} et~al.,}{{Hildebrandt}
  et~al.}{2010}]{Hildebrandt2010}
{Hildebrandt} H.,  et~al., 2010, \mn@doi [\aap] {10.1051/0004-6361/201014885},
  \href {https://ui.adsabs.harvard.edu/abs/2010A&A...523A..31H} {523, A31}

\bibitem[\protect\citeauthoryear{{Hildebrandt} et~al.,}{{Hildebrandt}
  et~al.}{2012}]{Hildebrandt:2012}
{Hildebrandt} H.,  et~al., 2012, \mn@doi [\mnras]
  {10.1111/j.1365-2966.2012.20468.x}, \href
  {http://adsabs.harvard.edu/abs/2012MNRAS.421.2355H} {421, 2355}

\bibitem[\protect\citeauthoryear{{Hildebrandt} et~al.,}{{Hildebrandt}
  et~al.}{2016}]{Hildebrandt2016}
{Hildebrandt} H.,  et~al., 2016, \mn@doi [\mnras] {10.1093/mnras/stw2013},
  \href {https://ui.adsabs.harvard.edu/abs/2016MNRAS.463..635H} {463, 635}

\bibitem[\protect\citeauthoryear{Hildebrandt et~al.,}{Hildebrandt
  et~al.}{2020}]{Hildebrandt2020}
Hildebrandt H.,  et~al., 2020, \mn@doi [Astronomy & Astrophysics]
  {10.1051/0004-6361/201834878}, 633, A69

\bibitem[\protect\citeauthoryear{{Hirata} \& {Seljak}}{{Hirata} \&
  {Seljak}}{2003}]{Hirata:2003aa}
{Hirata} C.,  {Seljak} U.,  2003, \mn@doi [Monthly Notices of the Royal
  Astronomical Society] {10.1046/j.1365-8711.2003.06683.x}, \href
  {https://ui.adsabs.harvard.edu/abs/2003MNRAS.343..459H} {343, 459}

\bibitem[\protect\citeauthoryear{{Hirata} et~al.,}{{Hirata}
  et~al.}{2004}]{Hirata:2004aa}
{Hirata} C.~M.,  et~al., 2004, \mn@doi [\mnras]
  {10.1111/j.1365-2966.2004.08090.x}, \href
  {http://adsabs.harvard.edu/abs/2004MNRAS.353..529H} {353, 529}

\bibitem[\protect\citeauthoryear{{Hoyle} et~al.,}{{Hoyle}
  et~al.}{2018}]{Hoyle2018}
{Hoyle} B.,  et~al., 2018, \mn@doi [\mnras] {10.1093/mnras/sty957}, \href
  {https://ui.adsabs.harvard.edu/abs/2018MNRAS.478..592H} {478, 592}

\bibitem[\protect\citeauthoryear{{Huang} et~al.,}{{Huang}
  et~al.}{2018a}]{Huang:2018ac}
{Huang} S.,  et~al., 2018a, \mn@doi [Publications of the Astronomical Society
  of Japan] {10.1093/pasj/psx126}, \href
  {https://ui.adsabs.harvard.edu/abs/2018PASJ...70S...6H} {70, S6}

\bibitem[\protect\citeauthoryear{{Huang}, {Leauthaud}, {Greene}, {Bundy},
  {Lin}, {Tanaka}, {Miyazaki}  \& {Komiyama}}{{Huang}
  et~al.}{2018b}]{Huang:2018}
{Huang} S.,  {Leauthaud} A.,  {Greene} J.~E.,  {Bundy} K.,  {Lin} Y.-T.,
  {Tanaka} M.,  {Miyazaki} S.,   {Komiyama} Y.,  2018b, \mn@doi [\mnras]
  {10.1093/mnras/stx3200}, \href
  {http://adsabs.harvard.edu/abs/2018MNRAS.475.3348H} {475, 3348}

\bibitem[\protect\citeauthoryear{{Huang}, {Eifler}, {Mandelbaum}  \&
  {Dodelson}}{{Huang} et~al.}{2019}]{Huang:2019aa}
{Huang} H.-J.,  {Eifler} T.,  {Mandelbaum} R.,   {Dodelson} S.,  2019, \mn@doi
  [\mnras] {10.1093/mnras/stz1714}, \href
  {https://ui.adsabs.harvard.edu/abs/2019MNRAS.tmp.1672H} {p.~1672}

\bibitem[\protect\citeauthoryear{{Huff} \& {Mandelbaum}}{{Huff} \&
  {Mandelbaum}}{2017}]{Huff2017}
{Huff} E.,  {Mandelbaum} R.,  2017, arXiv e-prints, \href
  {https://ui.adsabs.harvard.edu/abs/2017arXiv170202600H} {}

\bibitem[\protect\citeauthoryear{{Jee}, {Tyson}, {Schneider}, {Wittman},
  {Schmidt}  \& {Hilbert}}{{Jee} et~al.}{2013}]{Jee:2013}
{Jee} M.~J.,  {Tyson} J.~A.,  {Schneider} M.~D.,  {Wittman} D.,  {Schmidt} S.,
   {Hilbert} S.,  2013, \mn@doi [\apj] {10.1088/0004-637X/765/1/74}, \href
  {http://adsabs.harvard.edu/abs/2013ApJ...765...74J} {765, 74}

\bibitem[\protect\citeauthoryear{{Joachimi} et~al.,}{{Joachimi}
  et~al.}{2015}]{joachimi2015}
{Joachimi} B.,  et~al., 2015, \mn@doi [\ssr] {10.1007/s11214-015-0177-4}, \href
  {https://ui.adsabs.harvard.edu/abs/2015SSRv..193....1J} {193, 1}

\bibitem[\protect\citeauthoryear{{Joudaki} et~al.,}{{Joudaki}
  et~al.}{2020}]{Joudaki2020}
{Joudaki} S.,  et~al., 2020, \mn@doi [\aap] {10.1051/0004-6361/201936154},
  \href {https://ui.adsabs.harvard.edu/abs/2020A&A...638L...1J} {638, L1}

\bibitem[\protect\citeauthoryear{{Kannawadi} et~al.,}{{Kannawadi}
  et~al.}{2019}]{Kannawadi2019}
{Kannawadi} A.,  et~al., 2019, \mn@doi [\aap] {10.1051/0004-6361/201834819},
  \href {https://ui.adsabs.harvard.edu/abs/2019A&A...624A..92K} {624, A92}

\bibitem[\protect\citeauthoryear{{Kilbinger} et~al.,}{{Kilbinger}
  et~al.}{2017}]{Kilbinger:2017}
{Kilbinger} M.,  et~al., 2017, \mn@doi [\mnras] {10.1093/mnras/stx2082}, \href
  {https://ui.adsabs.harvard.edu/abs/2017MNRAS.472.2126K} {472, 2126}

\bibitem[\protect\citeauthoryear{{Kitanidis} et~al.,}{{Kitanidis}
  et~al.}{2020}]{Kitanidis2020}
{Kitanidis} E.,  et~al., 2020, \mn@doi [\mnras] {10.1093/mnras/staa1621}, \href
  {https://ui.adsabs.harvard.edu/abs/2020MNRAS.496.2262K} {496, 2262}

\bibitem[\protect\citeauthoryear{{Kneib} et~al.,}{{Kneib}
  et~al.}{2003}]{Kneib:2003aa}
{Kneib} J.-P.,  et~al., 2003, \mn@doi [\apj] {10.1086/378633}, \href
  {http://adsabs.harvard.edu/abs/2003ApJ...598..804K} {598, 804}

\bibitem[\protect\citeauthoryear{{Komiyama} et~al.,}{{Komiyama}
  et~al.}{2018}]{Komiyama:2018}
{Komiyama} Y.,  et~al., 2018, \mn@doi [\pasj] {10.1093/pasj/psx069}, \href
  {https://ui.adsabs.harvard.edu/abs/2018PASJ...70S...2K} {70, S2}

\bibitem[\protect\citeauthoryear{{Krause} et~al.,}{{Krause}
  et~al.}{2017}]{Krause:2017}
{Krause} E.,  et~al., 2017, arXiv e-prints, \href
  {https://ui.adsabs.harvard.edu/abs/2017arXiv170609359K} {p. arXiv:1706.09359}

\bibitem[\protect\citeauthoryear{{Kuijken} et~al.,}{{Kuijken}
  et~al.}{2015}]{Kuijken:2015}
{Kuijken} K.,  et~al., 2015, \mn@doi [\mnras] {10.1093/mnras/stv2140}, \href
  {http://adsabs.harvard.edu/abs/2015MNRAS.454.3500K} {454, 3500}

\bibitem[\protect\citeauthoryear{{LSST Science Collaboration} et~al.,}{{LSST
  Science Collaboration} et~al.}{2009}]{LSST-Science-Collaboration:2009}
{LSST Science Collaboration} et~al., 2009, preprint, \href
  {http://adsabs.harvard.edu/abs/2009arXiv0912.0201L} {} (\mn@eprint {arXiv}
  {0912.0201})

\bibitem[\protect\citeauthoryear{{Laigle} et~al.,}{{Laigle}
  et~al.}{2016}]{Laigle:2016aa}
{Laigle} C.,  et~al., 2016, \mn@doi [\apjs] {10.3847/0067-0049/224/2/24}, \href
  {http://adsabs.harvard.edu/abs/2016ApJS..224...24L} {224, 24}

\bibitem[\protect\citeauthoryear{{Lange}, {Yang}, {Guo}, {Luo}  \& {van den
  Bosch}}{{Lange} et~al.}{2019}]{Lange2019}
{Lange} J.~U.,  {Yang} X.,  {Guo} H.,  {Luo} W.,   {van den Bosch} F.~C.,
  2019, \mn@doi [\mnras] {10.1093/mnras/stz2124}, \href
  {https://ui.adsabs.harvard.edu/abs/2019MNRAS.488.5771L} {488, 5771}

\bibitem[\protect\citeauthoryear{{Lange}, {Leauthaud}, {Singh}, {Guo}, {Zhou},
  {Smith}  \& {Cyr-Racine}}{{Lange} et~al.}{2021}]{Lange2020}
{Lange} J.~U.,  {Leauthaud} A.,  {Singh} S.,  {Guo} H.,  {Zhou} R.,  {Smith}
  T.~L.,   {Cyr-Racine} F.-Y.,  2021, \mn@doi [\mnras] {10.1093/mnras/stab189},
  \href {https://ui.adsabs.harvard.edu/abs/2021MNRAS.502.2074L} {502, 2074}

\bibitem[\protect\citeauthoryear{{Laureijs} et~al.,}{{Laureijs}
  et~al.}{2011}]{Laureijs:2011}
{Laureijs} R.,  et~al., 2011, preprint, \href
  {http://adsabs.harvard.edu/abs/2011arXiv1110.3193L} {} (\mn@eprint {arXiv}
  {1110.3193})

\bibitem[\protect\citeauthoryear{{Leauthaud} et~al.,}{{Leauthaud}
  et~al.}{2012}]{Leauthaud:2012a}
{Leauthaud} A.,  et~al., 2012, \mn@doi [\apj] {10.1088/0004-637X/744/2/159},
  \href {http://adsabs.harvard.edu/abs/2012ApJ...744..159L} {744, 159}

\bibitem[\protect\citeauthoryear{{Leauthaud} et~al.,}{{Leauthaud}
  et~al.}{2017}]{Leauthaud:2017aa}
{Leauthaud} A.,  et~al., 2017, \mn@doi [\mnras] {10.1093/mnras/stx258}, \href
  {http://adsabs.harvard.edu/abs/2017MNRAS.467.3024L} {467, 3024}

\bibitem[\protect\citeauthoryear{{Lee} et~al.,}{{Lee} et~al.}{2019}]{Lee2019}
{Lee} S.,  et~al., 2019, \mn@doi [\mnras] {10.1093/mnras/stz2288}, \href
  {https://ui.adsabs.harvard.edu/abs/2019MNRAS.489.2887L} {489, 2887}

\bibitem[\protect\citeauthoryear{{Lima}, {Cunha}, {Oyaizu}, {Frieman}, {Lin}
  \& {Sheldon}}{{Lima} et~al.}{2008}]{Lima:2008}
{Lima} M.,  {Cunha} C.~E.,  {Oyaizu} H.,  {Frieman} J.,  {Lin} H.,   {Sheldon}
  E.~S.,  2008, \mn@doi [\mnras] {10.1111/j.1365-2966.2008.13510.x}, \href
  {http://adsabs.harvard.edu/abs/2008MNRAS.390..118L} {390, 118}

\bibitem[\protect\citeauthoryear{Luis~Bernal \& Peacock}{Luis~Bernal \&
  Peacock}{2018}]{Luis_Bernal_2018}
Luis~Bernal J.,  Peacock J.~A.,  2018, \mn@doi [Journal of Cosmology and
  Astroparticle Physics] {10.1088/1475-7516/2018/07/002}, 2018, 002–002

\bibitem[\protect\citeauthoryear{Lyons}{Lyons}{2008}]{Lyons2008}
Lyons L.,  2008, \mn@doi [The Annals of Applied Statistics]
  {10.1214/08-aoas163}, 2

\bibitem[\protect\citeauthoryear{{MacCrann} et~al.,}{{MacCrann}
  et~al.}{2021}]{MacCrann2020}
{MacCrann} N.,  et~al., 2021, \mn@doi [\mnras] {10.1093/mnras/stab2870}, \href
  {https://ui.adsabs.harvard.edu/abs/2021MNRAS.tmp.2655M} {}

\bibitem[\protect\citeauthoryear{{Mandelbaum}, {Tasitsiomi}, {Seljak},
  {Kravtsov}  \& {Wechsler}}{{Mandelbaum} et~al.}{2005}]{Mandelbaum:2005a}
{Mandelbaum} R.,  {Tasitsiomi} A.,  {Seljak} U.,  {Kravtsov} A.~V.,
  {Wechsler} R.~H.,  2005, \mn@doi [\mnras] {10.1111/j.1365-2966.2005.09417.x},
  \href
  {http://adsabs.harvard.edu/cgi-bin/nph-bib_query?bibcode=2005MNRAS.362.1451M&db_key=AST}
  {362, 1451}

\bibitem[\protect\citeauthoryear{{Mandelbaum}, {Seljak}, {Cool}, {Blanton},
  {Hirata}  \& {Brinkmann}}{{Mandelbaum} et~al.}{2006}]{Mandelbaum:2006}
{Mandelbaum} R.,  {Seljak} U.,  {Cool} R.~J.,  {Blanton} M.,  {Hirata} C.~M.,
  {Brinkmann} J.,  2006, \mn@doi [\mnras] {10.1111/j.1365-2966.2006.10906.x},
  \href
  {http://adsabs.harvard.edu/cgi-bin/nph-bib_query?bibcode=2006MNRAS.372..758M&db_key=AST}
  {372, 758}

\bibitem[\protect\citeauthoryear{{Mandelbaum}, {Hirata}, {Leauthaud}, {Massey}
  \& {Rhodes}}{{Mandelbaum} et~al.}{2012}]{Mandelbaum:2012}
{Mandelbaum} R.,  {Hirata} C.~M.,  {Leauthaud} A.,  {Massey} R.~J.,   {Rhodes}
  J.,  2012, \mn@doi [\mnras] {10.1111/j.1365-2966.2011.20138.x}, \href
  {http://adsabs.harvard.edu/abs/2012MNRAS.420.1518M} {420, 1518}

\bibitem[\protect\citeauthoryear{{Mandelbaum}, {Slosar}, {Baldauf}, {Seljak},
  {Hirata}, {Nakajima}, {Reyes}  \& {Smith}}{{Mandelbaum}
  et~al.}{2013}]{Mandelbaum:2013}
{Mandelbaum} R.,  {Slosar} A.,  {Baldauf} T.,  {Seljak} U.,  {Hirata} C.~M.,
  {Nakajima} R.,  {Reyes} R.,   {Smith} R.~E.,  2013, \mn@doi [\mnras]
  {10.1093/mnras/stt572}, \href
  {http://adsabs.harvard.edu/abs/2013MNRAS.432.1544M} {432, 1544}

\bibitem[\protect\citeauthoryear{{Mandelbaum} et~al.,}{{Mandelbaum}
  et~al.}{2014}]{Mandelbaum2014}
{Mandelbaum} R.,  et~al., 2014, \mn@doi [\apjs] {10.1088/0067-0049/212/1/5},
  \href {http://adsabs.harvard.edu/abs/2014ApJS..212....5M} {212, 5}

\bibitem[\protect\citeauthoryear{{Mandelbaum} et~al.,}{{Mandelbaum}
  et~al.}{2018a}]{Mandelbaum:2018ab}
{Mandelbaum} R.,  et~al., 2018a, \mn@doi [Publications of the Astronomical
  Society of Japan] {10.1093/pasj/psx130}, \href
  {https://ui.adsabs.harvard.edu/abs/2018PASJ...70S..25M} {70, S25}

\bibitem[\protect\citeauthoryear{{Mandelbaum} et~al.,}{{Mandelbaum}
  et~al.}{2018b}]{Mandelbaum2018}
{Mandelbaum} R.,  et~al., 2018b, \mn@doi [\mnras] {10.1093/mnras/sty2420},
  \href {https://ui.adsabs.harvard.edu/abs/2018MNRAS.481.3170M} {481, 3170}

\bibitem[\protect\citeauthoryear{{Mandelbaum} et~al.,}{{Mandelbaum}
  et~al.}{2018c}]{Mandelbaum:2018aa}
{Mandelbaum} R.,  et~al., 2018c, \mn@doi [Monthly Notices of the Royal
  Astronomical Society] {10.1093/mnras/sty2420}, \href
  {https://ui.adsabs.harvard.edu/abs/2018MNRAS.481.3170M} {481, 3170}

\bibitem[\protect\citeauthoryear{{Massey} et~al.,}{{Massey}
  et~al.}{2007}]{Massey2007}
{Massey} R.,  et~al., 2007, \mn@doi [\mnras]
  {10.1111/j.1365-2966.2006.11315.x}, \href
  {http://adsabs.harvard.edu/abs/2007MNRAS.376...13M} {376, 13}

\bibitem[\protect\citeauthoryear{{Masters} et~al.,}{{Masters}
  et~al.}{2015}]{Masters:2015aa}
{Masters} D.,  et~al., 2015, \mn@doi [\apj] {10.1088/0004-637X/813/1/53}, \href
  {http://adsabs.harvard.edu/abs/2015ApJ...813...53M} {813, 53}

\bibitem[\protect\citeauthoryear{{McClintock} et~al.,}{{McClintock}
  et~al.}{2019}]{McClintock2019}
{McClintock} T.,  et~al., 2019, \mn@doi [\mnras] {10.1093/mnras/sty2711}, \href
  {https://ui.adsabs.harvard.edu/abs/2019MNRAS.482.1352M} {482, 1352}

\bibitem[\protect\citeauthoryear{{Melchior} et~al.,}{{Melchior}
  et~al.}{2015}]{Melchior:2015aa}
{Melchior} P.,  et~al., 2015, \mn@doi [\mnras] {10.1093/mnras/stv398}, \href
  {http://adsabs.harvard.edu/abs/2015MNRAS.449.2219M} {449, 2219}

\bibitem[\protect\citeauthoryear{{Miller} et~al.,}{{Miller}
  et~al.}{2013}]{Miller:2013}
{Miller} L.,  et~al., 2013, \mn@doi [\mnras] {10.1093/mnras/sts454}, \href
  {http://adsabs.harvard.edu/abs/2013MNRAS.429.2858M} {429, 2858}

\bibitem[\protect\citeauthoryear{{Miralda-Escude}}{{Miralda-Escude}}{1991}]{Miralda-Escude:1991}
{Miralda-Escude} J.,  1991, \mn@doi [\apj] {10.1086/169789}, \href
  {http://adsabs.harvard.edu/abs/1991ApJ...370....1M} {370, 1}

\bibitem[\protect\citeauthoryear{{Miyazaki} et~al.,}{{Miyazaki}
  et~al.}{2018}]{Miyazaki:2018}
{Miyazaki} S.,  et~al., 2018, \mn@doi [\pasj] {10.1093/pasj/psx063}, \href
  {https://ui.adsabs.harvard.edu/abs/2018PASJ...70S...1M} {70, S1}

\bibitem[\protect\citeauthoryear{{Myles} et~al.,}{{Myles}
  et~al.}{2021}]{Myles2020}
{Myles} J.,  et~al., 2021, \mn@doi [\mnras] {10.1093/mnras/stab1515}, \href
  {https://ui.adsabs.harvard.edu/abs/2021MNRAS.505.4249M} {505, 4249}

\bibitem[\protect\citeauthoryear{{Nakajima}, {Mandelbaum}, {Seljak}, {Cohn},
  {Reyes}  \& {Cool}}{{Nakajima} et~al.}{2012}]{Nakajima:2012aa}
{Nakajima} R.,  {Mandelbaum} R.,  {Seljak} U.,  {Cohn} J.~D.,  {Reyes} R.,
  {Cool} R.,  2012, \mn@doi [\mnras] {10.1111/j.1365-2966.2011.20249.x}, \href
  {http://adsabs.harvard.edu/abs/2012MNRAS.420.3240N} {420, 3240}

\bibitem[\protect\citeauthoryear{{Newman} et~al.,}{{Newman}
  et~al.}{2013}]{Newman:2013a}
{Newman} J.~A.,  et~al., 2013, \mn@doi [\apjs] {10.1088/0067-0049/208/1/5},
  \href {http://adsabs.harvard.edu/abs/2013ApJS..208....5N} {208, 5}

\bibitem[\protect\citeauthoryear{{Planck Collaboration} et~al.,}{{Planck
  Collaboration} et~al.}{2020}]{planck2020}
{Planck Collaboration} et~al., 2020, \mn@doi [\aap]
  {10.1051/0004-6361/201833880}, \href
  {https://ui.adsabs.harvard.edu/abs/2020A&A...641A...1P} {641, A1}

\bibitem[\protect\citeauthoryear{{Prat} et~al.,}{{Prat}
  et~al.}{2018}]{Prat2018}
{Prat} J.,  et~al., 2018, \mn@doi [\prd] {10.1103/PhysRevD.98.042005}, \href
  {https://ui.adsabs.harvard.edu/abs/2018PhRvD..98d2005P} {98, 042005}

\bibitem[\protect\citeauthoryear{{Reid} et~al.,}{{Reid}
  et~al.}{2016}]{Reid:2016}
{Reid} B.,  et~al., 2016, \mn@doi [\mnras] {10.1093/mnras/stv2382}, \href
  {http://adsabs.harvard.edu/abs/2016MNRAS.455.1553R} {455, 1553}

\bibitem[\protect\citeauthoryear{{Reyes}, {Mandelbaum}, {Gunn}, {Nakajima},
  {Seljak}  \& {Hirata}}{{Reyes} et~al.}{2012}]{Reyes:2012aa}
{Reyes} R.,  {Mandelbaum} R.,  {Gunn} J.~E.,  {Nakajima} R.,  {Seljak} U.,
  {Hirata} C.~M.,  2012, \mn@doi [\mnras] {10.1111/j.1365-2966.2012.21472.x},
  \href {http://adsabs.harvard.edu/abs/2012MNRAS.425.2610R} {425, 2610}

\bibitem[\protect\citeauthoryear{{Ross} et~al.,}{{Ross}
  et~al.}{2011}]{Ross:2011aa}
{Ross} A.~J.,  et~al., 2011, \mn@doi [\mnras]
  {10.1111/j.1365-2966.2011.19351.x}, \href
  {http://adsabs.harvard.edu/abs/2011MNRAS.417.1350R} {417, 1350}

\bibitem[\protect\citeauthoryear{{Ross} et~al.,}{{Ross}
  et~al.}{2012}]{Ross:2012}
{Ross} A.~J.,  et~al., 2012, \mn@doi [\mnras]
  {10.1111/j.1365-2966.2012.21235.x}, \href
  {http://adsabs.harvard.edu/abs/2012MNRAS.424..564R} {424, 564}

\bibitem[\protect\citeauthoryear{{Ross} et~al.,}{{Ross}
  et~al.}{2017}]{Ross:2017aa}
{Ross} A.~J.,  et~al., 2017, \mn@doi [\mnras] {10.1093/mnras/stw2372}, \href
  {http://adsabs.harvard.edu/abs/2017MNRAS.464.1168R} {464, 1168}

\bibitem[\protect\citeauthoryear{{Schirmer}, {Erben}, {Schneider}, {Wolf}  \&
  {Meisenheimer}}{{Schirmer} et~al.}{2004}]{Schirmer2004}
{Schirmer} M.,  {Erben} T.,  {Schneider} P.,  {Wolf} C.,   {Meisenheimer} K.,
  2004, \mn@doi [\aap] {10.1051/0004-6361:20041072}, \href
  {https://ui.adsabs.harvard.edu/abs/2004A&A...420...75S} {420, 75}

\bibitem[\protect\citeauthoryear{{Schlafly} \& {Finkbeiner}}{{Schlafly} \&
  {Finkbeiner}}{2011}]{Schlafly:2011aa}
{Schlafly} E.~F.,  {Finkbeiner} D.~P.,  2011, \mn@doi [\apj]
  {10.1088/0004-637X/737/2/103}, \href
  {http://adsabs.harvard.edu/abs/2011ApJ...737..103S} {737, 103}

\bibitem[\protect\citeauthoryear{{Schlafly}, {Finkbeiner}, {Schlegel},
  {Juri{\'c}}, {Ivezi{\'c}}, {Gibson}, {Knapp}  \& {Weaver}}{{Schlafly}
  et~al.}{2010}]{Schlafly:2010aa}
{Schlafly} E.~F.,  {Finkbeiner} D.~P.,  {Schlegel} D.~J.,  {Juri{\'c}} M.,
  {Ivezi{\'c}} {\v Z}.,  {Gibson} R.~R.,  {Knapp} G.~R.,   {Weaver} B.~A.,
  2010, \mn@doi [\apj] {10.1088/0004-637X/725/1/1175}, \href
  {http://adsabs.harvard.edu/abs/2010ApJ...725.1175S} {725, 1175}

\bibitem[\protect\citeauthoryear{{Schlegel}, {Finkbeiner}  \&
  {Davis}}{{Schlegel} et~al.}{1998}]{Schlegel:1998}
{Schlegel} D.~J.,  {Finkbeiner} D.~P.,   {Davis} M.,  1998, \mn@doi [\apj]
  {10.1086/305772}, \href {http://adsabs.harvard.edu/abs/1998ApJ...500..525S}
  {500, 525}

\bibitem[\protect\citeauthoryear{{Sheldon} \& {Huff}}{{Sheldon} \&
  {Huff}}{2017}]{Sheldon2017}
{Sheldon} E.~S.,  {Huff} E.~M.,  2017, \mn@doi [\apj]
  {10.3847/1538-4357/aa704b}, \href
  {https://ui.adsabs.harvard.edu/abs/2017ApJ...841...24S} {841, 24}

\bibitem[\protect\citeauthoryear{{Sheldon} et~al.}{{Sheldon}
  et~al.}{2004}]{Sheldon:2004}
{Sheldon} E.~S.,  et~al., 2004, \mn@doi [\aj] {10.1086/383293}, \href
  {http://adsabs.harvard.edu/cgi-bin/nph-bib_query?bibcode=2004AJ....127.2544S&db_key=AST}
  {127, 2544}

\bibitem[\protect\citeauthoryear{{Simet} \& {Mandelbaum}}{{Simet} \&
  {Mandelbaum}}{2015}]{Simet:2015}
{Simet} M.,  {Mandelbaum} R.,  2015, \mn@doi [\mnras] {10.1093/mnras/stv313},
  \href {http://adsabs.harvard.edu/abs/2015MNRAS.449.1259S} {449, 1259}

\bibitem[\protect\citeauthoryear{{Simet}, {McClintock}, {Mandelbaum}, {Rozo},
  {Rykoff}, {Sheldon}  \& {Wechsler}}{{Simet} et~al.}{2016}]{Simet:2016}
{Simet} M.,  {McClintock} T.,  {Mandelbaum} R.,  {Rozo} E.,  {Rykoff} E.,
  {Sheldon} E.,   {Wechsler} R.~H.,  2016, preprint, \href
  {http://adsabs.harvard.edu/abs/2016arXiv160306953S} {} (\mn@eprint {arXiv}
  {1603.06953})

\bibitem[\protect\citeauthoryear{{Singh}, {Mandelbaum}, {Seljak}, {Slosar}  \&
  {Vazquez Gonzalez}}{{Singh} et~al.}{2017}]{Singh:2017}
{Singh} S.,  {Mandelbaum} R.,  {Seljak} U.,  {Slosar} A.,   {Vazquez Gonzalez}
  J.,  2017, \mn@doi [\mnras] {10.1093/mnras/stx1828}, \href
  {http://adsabs.harvard.edu/abs/2017MNRAS.471.3827S} {471, 3827}

\bibitem[\protect\citeauthoryear{{Singh}, {Mandelbaum}, {Seljak},
  {Rodr{\'\i}guez-Torres}  \& {Slosar}}{{Singh} et~al.}{2018}]{Singh:2018aa}
{Singh} S.,  {Mandelbaum} R.,  {Seljak} U.,  {Rodr{\'\i}guez-Torres} S.,
  {Slosar} A.,  2018, arXiv e-prints, \href
  {https://ui.adsabs.harvard.edu/abs/2018arXiv181106499S} {p. arXiv:1811.06499}

\bibitem[\protect\citeauthoryear{{Singh}, {Mandelbaum}, {Seljak},
  {Rodr{\'\i}guez-Torres}  \& {Slosar}}{{Singh} et~al.}{2020}]{Singh:2020}
{Singh} S.,  {Mandelbaum} R.,  {Seljak} U.,  {Rodr{\'\i}guez-Torres} S.,
  {Slosar} A.,  2020, \mn@doi [\mnras] {10.1093/mnras/stz2922}, \href
  {https://ui.adsabs.harvard.edu/abs/2020MNRAS.491...51S} {491, 51}

\bibitem[\protect\citeauthoryear{{Smee} et~al.,}{{Smee}
  et~al.}{2013}]{Smee:2013}
{Smee} S.~A.,  et~al., 2013, \mn@doi [\aj] {10.1088/0004-6256/146/2/32}, \href
  {http://adsabs.harvard.edu/abs/2013AJ....146...32S} {146, 32}

\bibitem[\protect\citeauthoryear{Speagle et~al.,}{Speagle
  et~al.}{2019}]{Speagle:2019aa}
Speagle J.~S.,  et~al., 2019, \mn@doi [Monthly Notices of the Royal
  Astronomical Society] {10.1093/mnras/stz2968}, 490, 5658–5677

\bibitem[\protect\citeauthoryear{Spergel et~al.,}{Spergel
  et~al.}{2015}]{WFIRST}
Spergel D.,  et~al., 2015, Wide-Field InfrarRed Survey Telescope-Astrophysics
  Focused Telescope Assets WFIRST-AFTA 2015 Report (\mn@eprint {arXiv}
  {1503.03757})

\bibitem[\protect\citeauthoryear{{Stoughton} et~al.,}{{Stoughton}
  et~al.}{2002}]{Stoughton:2002}
{Stoughton} C.,  et~al., 2002, \mn@doi [\aj] {10.1086/324741}, \href
  {http://adsabs.harvard.edu/abs/2002AJ....123..485S} {123, 485}

\bibitem[\protect\citeauthoryear{{Tanaka} et~al.,}{{Tanaka}
  et~al.}{2018}]{Tanaka:2018aa}
{Tanaka} M.,  et~al., 2018, \mn@doi [Publications of the Astronomical Society
  of Japan] {10.1093/pasj/psx077}, \href
  {https://ui.adsabs.harvard.edu/abs/2018PASJ...70S...9T} {70, S9}

\bibitem[\protect\citeauthoryear{{The Dark Energy Survey Collaboration}
  et~al.,}{{The Dark Energy Survey Collaboration}
  et~al.}{2015}]{The-Dark-Energy-Survey-Collaboration:2015}
{The Dark Energy Survey Collaboration} et~al., 2015, preprint, \href
  {http://adsabs.harvard.edu/abs/2015arXiv150705552T} {} (\mn@eprint {arXiv}
  {1507.05552})

\bibitem[\protect\citeauthoryear{{Tojeiro} et~al.,}{{Tojeiro}
  et~al.}{2014}]{Tojeiro:2014aa}
{Tojeiro} R.,  et~al., 2014, \mn@doi [\mnras] {10.1093/mnras/stu371}, \href
  {http://adsabs.harvard.edu/abs/2014MNRAS.440.2222T} {440, 2222}

\bibitem[\protect\citeauthoryear{{Troxel} et~al.,}{{Troxel}
  et~al.}{2018a}]{Troxel2018}
{Troxel} M.~A.,  et~al., 2018a, \mn@doi [\prd] {10.1103/PhysRevD.98.043528},
  \href {https://ui.adsabs.harvard.edu/abs/2018PhRvD..98d3528T} {98, 043528}

\bibitem[\protect\citeauthoryear{{Troxel} et~al.,}{{Troxel}
  et~al.}{2018b}]{Troxel:2018}
{Troxel} M.~A.,  et~al., 2018b, \mn@doi [\mnras] {10.1093/mnras/sty1889}, \href
  {https://ui.adsabs.harvard.edu/abs/2018MNRAS.479.4998T} {479, 4998}

\bibitem[\protect\citeauthoryear{{Unruh}, {Schneider}, {Hilbert}, {Simon},
  {Martin}  \& {Puertas}}{{Unruh} et~al.}{2020}]{Unruh2020}
{Unruh} S.,  {Schneider} P.,  {Hilbert} S.,  {Simon} P.,  {Martin} S.,
  {Puertas} J.~C.,  2020, \mn@doi [\aap] {10.1051/0004-6361/201936915}, \href
  {https://ui.adsabs.harvard.edu/abs/2020A&A...638A..96U} {638, A96}

\bibitem[\protect\citeauthoryear{{Varga} et~al.,}{{Varga}
  et~al.}{2019}]{Varga2019}
{Varga} T.~N.,  et~al., 2019, \mn@doi [\mnras] {10.1093/mnras/stz2185}, \href
  {https://ui.adsabs.harvard.edu/abs/2019MNRAS.489.2511V} {489, 2511}

\bibitem[\protect\citeauthoryear{{Wilson}, {Kaiser}, {Luppino}  \&
  {Cowie}}{{Wilson} et~al.}{2001}]{Wilson:2001}
{Wilson} G.,  {Kaiser} N.,  {Luppino} G.~A.,   {Cowie} L.~L.,  2001, \mn@doi
  [\apj] {10.1086/321441}, \href
  {http://adsabs.harvard.edu/cgi-bin/nph-bib_query?bibcode=2001ApJ...555..572W&db_key=AST}
  {555, 572}

\bibitem[\protect\citeauthoryear{{Wright} et~al.,}{{Wright}
  et~al.}{2019}]{Wright2018}
{Wright} A.~H.,  et~al., 2019, \mn@doi [\aap] {10.1051/0004-6361/201834879},
  \href {https://ui.adsabs.harvard.edu/abs/2019A&A...632A..34W} {632, A34}

\bibitem[\protect\citeauthoryear{{Wright}, {Hildebrandt}, {van den Busch}  \&
  {Heymans}}{{Wright} et~al.}{2020}]{Wright2020cfht}
{Wright} A.~H.,  {Hildebrandt} H.,  {van den Busch} J.~L.,   {Heymans} C.,
  2020, \mn@doi [\aap] {10.1051/0004-6361/201936782}, \href
  {https://ui.adsabs.harvard.edu/abs/2020A&A...637A.100W} {637, A100}

\bibitem[\protect\citeauthoryear{{Xia} et~al.,}{{Xia}
  et~al.}{2020}]{Filaments2020}
{Xia} Q.,  et~al., 2020, \mn@doi [\aap] {10.1051/0004-6361/201936678}, \href
  {https://ui.adsabs.harvard.edu/abs/2020A&A...633A..89X} {633, A89}

\bibitem[\protect\citeauthoryear{{York} et~al.,}{{York}
  et~al.}{2000}]{York:2000}
{York} D.~G.,  et~al., 2000, \mn@doi [\aj] {10.1086/301513}, \href
  {http://adsabs.harvard.edu/abs/2000AJ....120.1579Y} {120, 1579}

\bibitem[\protect\citeauthoryear{{Zuntz}, {Kacprzak}, {Voigt}, {Hirsch}, {Rowe}
   \& {Bridle}}{{Zuntz} et~al.}{2013}]{Zuntz2013}
{Zuntz} J.,  {Kacprzak} T.,  {Voigt} L.,  {Hirsch} M.,  {Rowe} B.,   {Bridle}
  S.,  2013, \mn@doi [\mnras] {10.1093/mnras/stt1125}, \href
  {https://ui.adsabs.harvard.edu/abs/2013MNRAS.434.1604Z} {434, 1604}

\bibitem[\protect\citeauthoryear{{Zuntz} et~al.,}{{Zuntz}
  et~al.}{2018}]{Zuntz2018}
{Zuntz} J.,  et~al., 2018, \mn@doi [\mnras] {10.1093/mnras/sty2219}, \href
  {https://ui.adsabs.harvard.edu/abs/2018MNRAS.481.1149Z} {481, 1149}

\bibitem[\protect\citeauthoryear{{de Jong} et~al.,}{{de Jong}
  et~al.}{2019}]{deJONG2019}
{de Jong} R.~S.,  et~al., 2019, \mn@doi [The Messenger]
  {10.18727/0722-6691/5117}, \href
  {https://ui.adsabs.harvard.edu/abs/2019Msngr.175....3D} {175, 3}

\bibitem[\protect\citeauthoryear{{von Wietersheim-Kramsta} et~al.,}{{von
  Wietersheim-Kramsta} et~al.}{2021}]{Wietersheim2021}
{von Wietersheim-Kramsta} M.,  et~al., 2021, \mn@doi [\mnras]
  {10.1093/mnras/stab1000}, \href
  {https://ui.adsabs.harvard.edu/abs/2021MNRAS.504.1452V} {504, 1452}

\makeatother
\end{thebibliography}


\appendix

\section{Impact of One Percent Flux Calibration on Variations in $\Delta\Sigma$}\label{onep_flux_calibration}

Modern day surveys control flux calibration to about 1\% or better. This means that $\delta f/f\sim 0.01$. Stellar mass is proportional to flux. Ignoring the fact that we need to use colors to estimate $M^*$ and considering instead only the scaling of $M^*$ with total luminosity, this means the impact of flux uncertainties on $M^*$ is also of order 1\%. Assuming the stellar-to-halo mass relation of \citet[][]{Leauthaud:2012a} and the redshift and mass range of CMASS, this corresponds to a halo mass shift of 0.0043 dex. The lensing observable $\Delta\Sigma$ in the one-halo regime scales as $\Delta\Sigma \propto (M_{\rm halo})^{2/3}$ so this imparts a 0.6\% shift on $\Delta\Sigma$. This level of difference is not detectable with current surveys. Other effects, such as the ability to accurately model and measure the total luminosities of massive galaxies, including their outer envelopes \citep[e.g.,][]{Huang:2018,Huang:2019aa}, or dust corrections, are likely to be more important.

\section{Derivation of \texorpdfstring{f$_{\rm bias}$}{f\_bias} Correction Factor}\label{fbiasappendix}

Using photo-$z$ point source estimates for the source redshift in the calculation of $\Delta\Sigma_{\rm crit}$ can bias estimates of $\Delta\Sigma$. However, this bias, which we call $f_{\rm bias}$, can be estimated from a calibration sample. This calibration sample can be formed from galaxies with known spectroscopic redshifts or from galaxies with higher quality photometric redshifts, such as the COSMOS 30-band catalogue \citep[e.g.][]{Speagle:2019aa}. Here, we derive the estimator for $f_{\rm bias}$ if such a calibration sample is available.

$f_{\rm bias}$ is the expected ratio of the true $\Delta\Sigma_{\rm T}$ versus the estimate using photometric redshifts $\Delta\Sigma_{\rm P}$:
\begin{equation}
    f_{\rm bias}^{-1} = \frac{\Delta\Sigma_{\rm P}}{\Delta\Sigma_{\rm T}} = \frac{1}{\Delta\Sigma_{\rm T}} \frac{\sum_{\rm Ls} w_{\rm sys} w_{\rm Ls} \ \gamma_{\rm t, Ls} \ \Sigma_{{\rm c}, \rm Ls, P}}{\sum_{\rm Ls} w_{\rm sys} w_{\rm Ls}}\,.
\end{equation}
In the above equation $\Delta\Sigma_{\rm T}$ is defined analogously to $\Delta\Sigma_{\rm P}$ but with $\Sigma_{{\rm c}, \rm Ls, T}$ instead of $\Sigma_{{\rm c}, \rm Ls, P}$. Here, $\Sigma_{{\rm c}, \rm Ls, T}$ and $\Sigma_{{\rm c}, \rm Ls, P}$ correspond to the critical surface density derived from high-quality and photometric redshifts, respectively.

In principle, the above equation can only be solved if high-quality redshifts for all sources are known. However, we can get an estimate for $f_{\rm bias}$ using a representative sample of sources with high-quality redshifts. The high-quality sample does not need to be perfectly representative of the full source sample. Instead, we can account for systematic differences with respect to color using calibration weights ($w_{\rm calib, s}$) applied to the high-quality sample.

Furthermore, we note that the expectation value for the tangential shear $\gamma_{\rm t, Ls}$ is $\Delta\Sigma_{\rm T} / \Sigma_{\rm c, Ls, T}$. Thus, we can approximate:
\begin{equation}
    f_{\rm bias}^{-1} \approx \frac{\sum_{\rm Ls} w_{\rm sys} \ w_{\rm calib, s} w_{\rm Ls} \ \Sigma_{\rm c, Ls, P} \Sigma_{\rm c, Ls, T}^{-1}}{\sum_{\rm Ls} w_{\rm sys} w_{\rm calib, s} \ w_{\rm Ls}}
\end{equation}
We note that the above expression does not depend anymore on the actual value of the excess surface density, $\Delta\Sigma$. Thus, unless the systematic lens weights depend systematically on redshift, they can safely be neglected from the above equation. Finally, substituting Eq.~\eqref{eq:weight} for the lens-source weight $w_{\rm Ls}$ into the above expression leads to Eq.~\eqref{fbias}. This form for $f_{\rm bias}$ includes the dilution factor correction. Note that $f_{\rm bias}$ is only meant to correct for photometric redshift errors of sources physically uncorrelated with the lens. Biases in $\Delta\Sigma$ due to sources physically associated with the lens are incorporated in the boost factor.

\section{Spatial maps of homogeneity}\label{app:homog}

Figure \ref{fig:psfinhsc} shows spatial variations of the SDSS PSF in HSC regions. Strong spatial variations in the SDSS PSF mean that errors cannot be estimated from jack-knife when separating the data into separate regions of low and high SDSS PSF.

\begin{figure*}
\centering
\includegraphics[width=12cm]{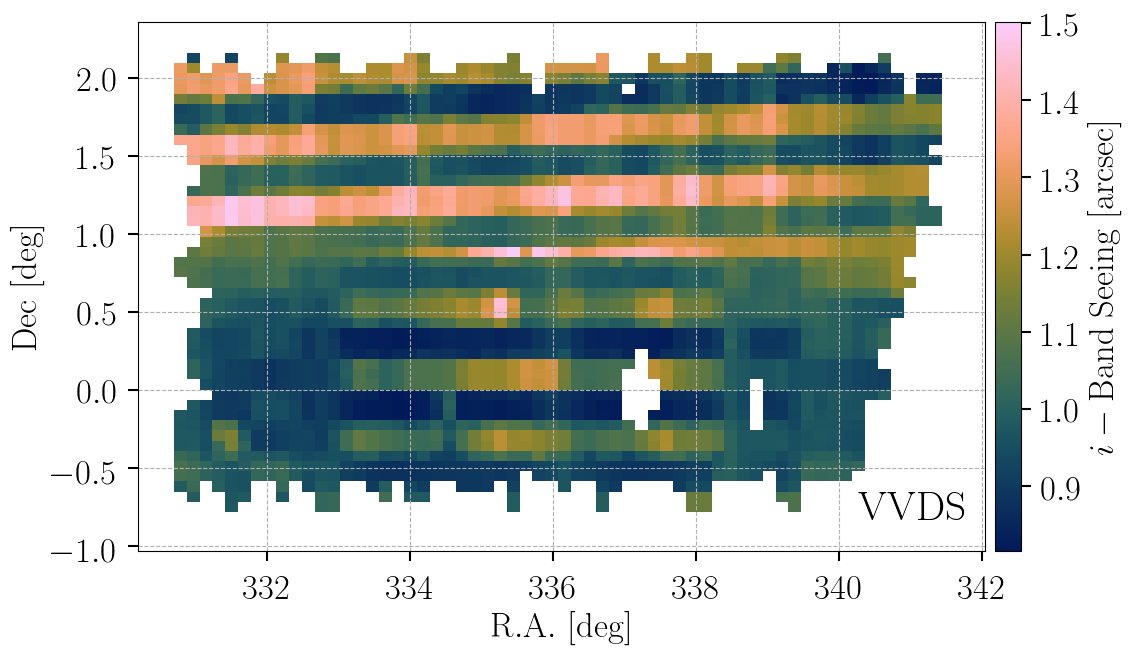}
\caption{SDSS PSF in the HSC regions. Strong spatial variations in the PSF mean that it is difficult to directly estimate inhomogeneity effects from the data because one cannot use jack-knife errors when separating regions with low and high PSF.}
\label{fig:psfinhsc}
\end{figure*}

\begin{figure*}
  \centering
  \begin{tabular}{@{}c@{}}
    \subfloat[]{\includegraphics[width=3.1in]{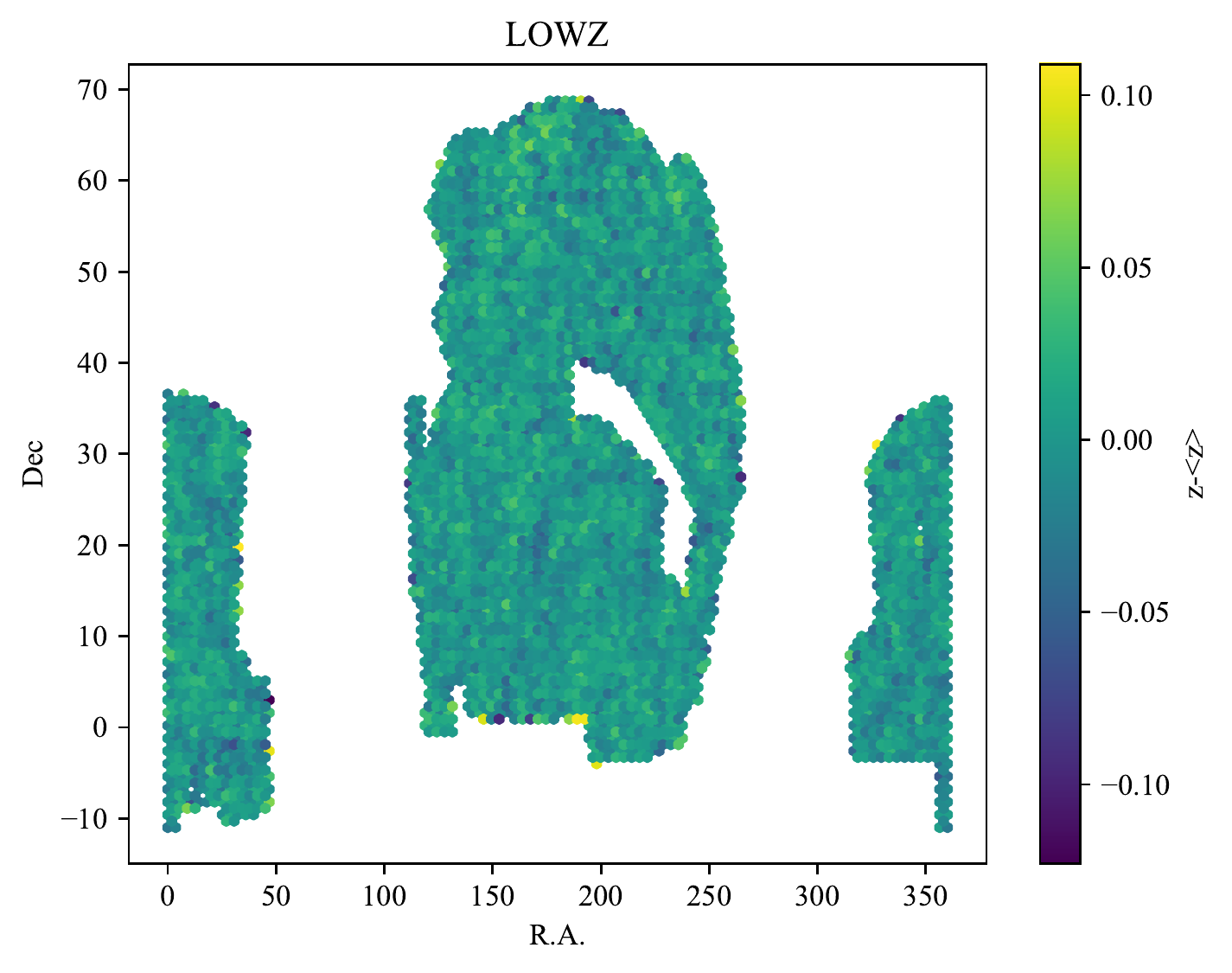}} 
	\subfloat[]{\includegraphics[width=3.1in]{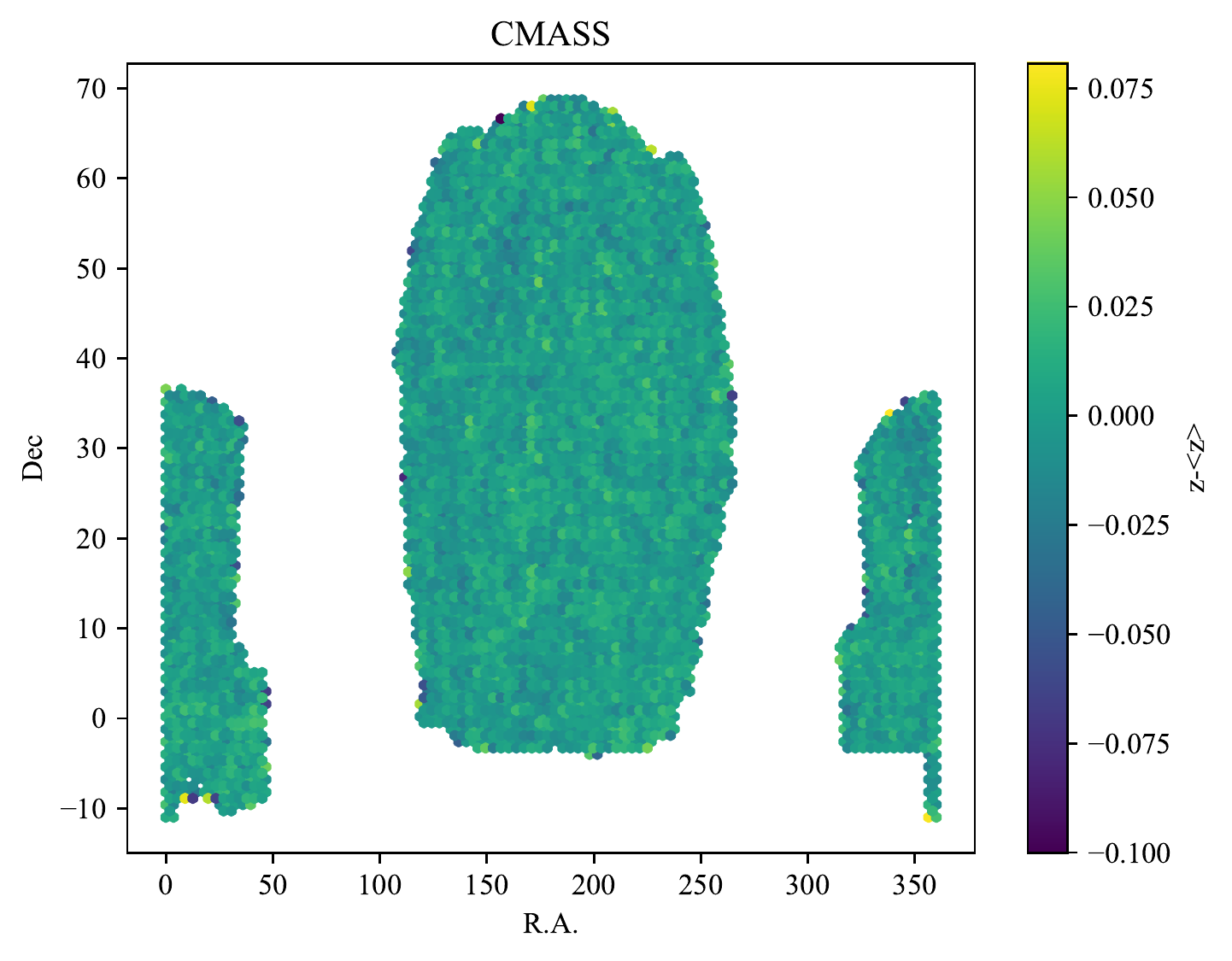}}
  \end{tabular}

  \begin{tabular}{@{}c@{}}
   \subfloat[]{\includegraphics[width=3.1in]{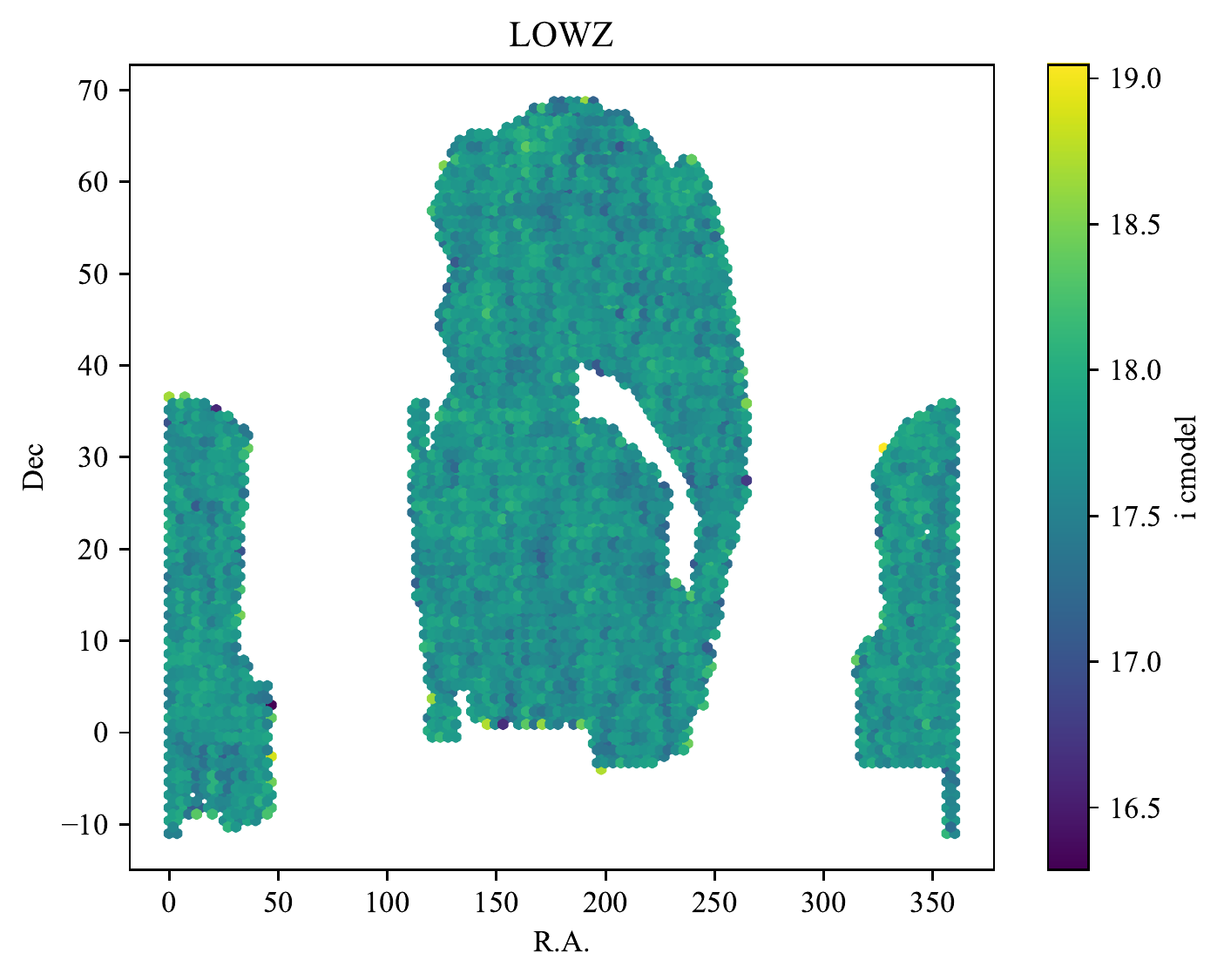}} 
\subfloat[]{\includegraphics[width=3.1in]{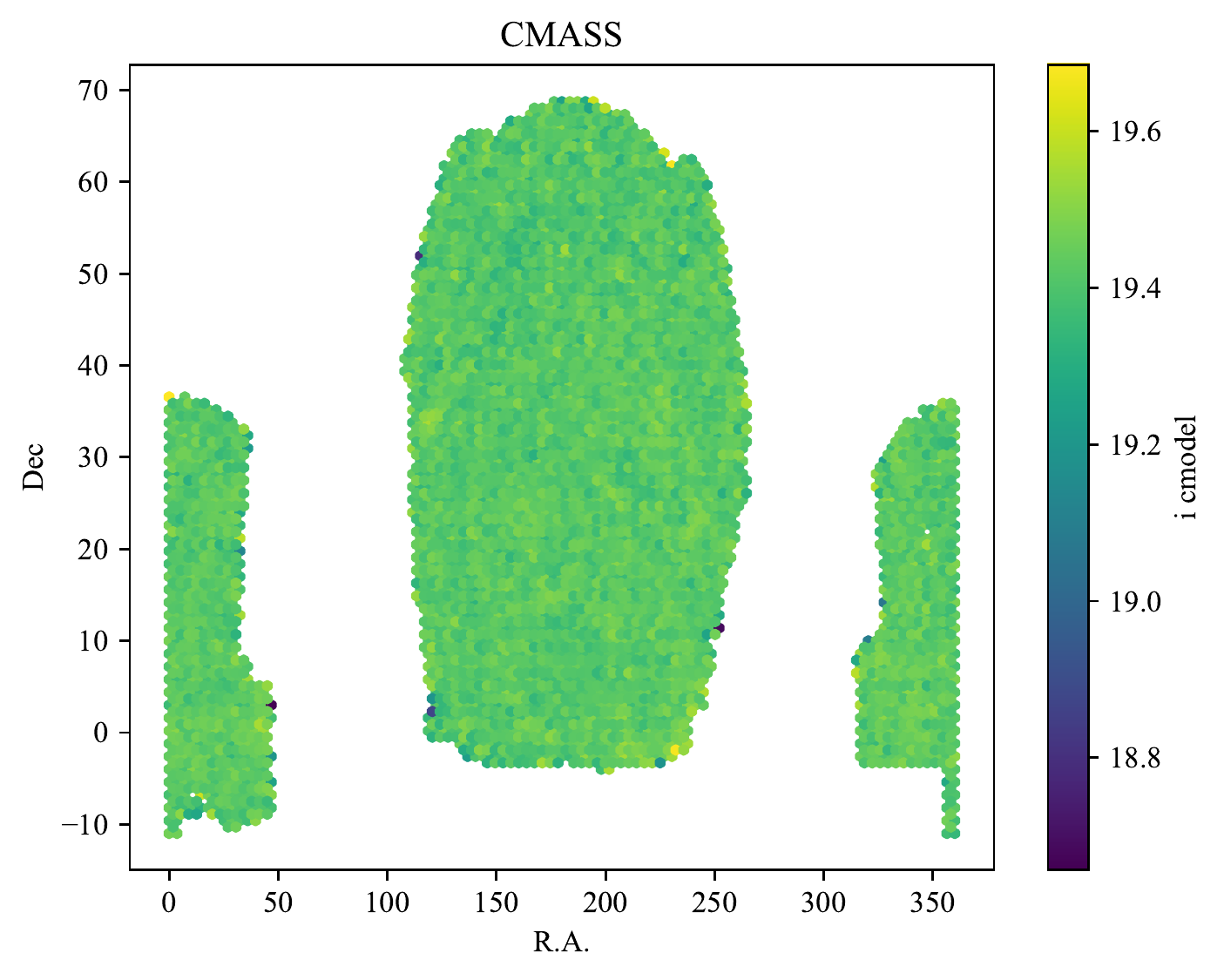}}
  \end{tabular}
  
    \begin{tabular}{@{}c@{}}
\subfloat[]{\includegraphics[width=3.1in]{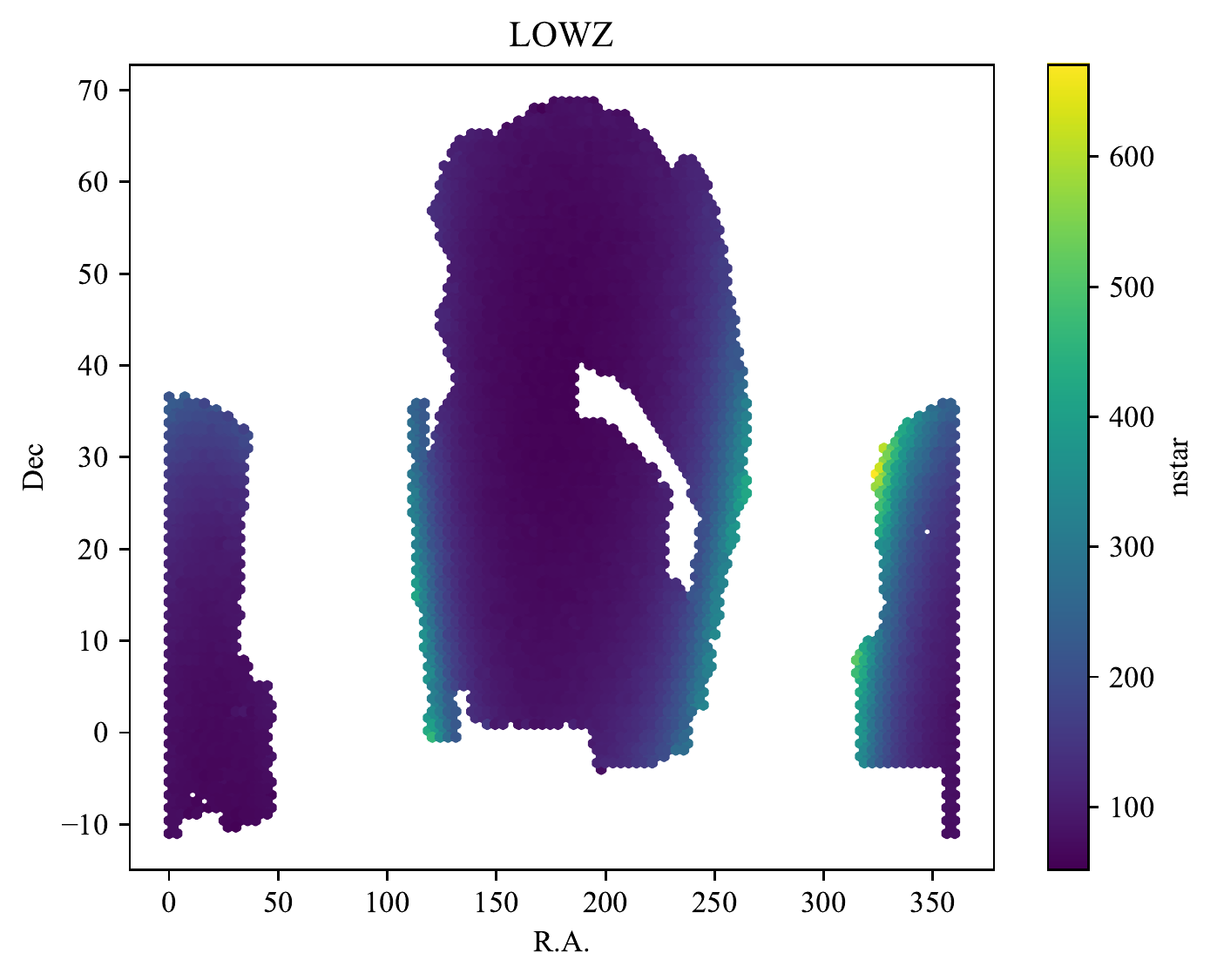}} 
\subfloat[]{\includegraphics[width=3.1in]{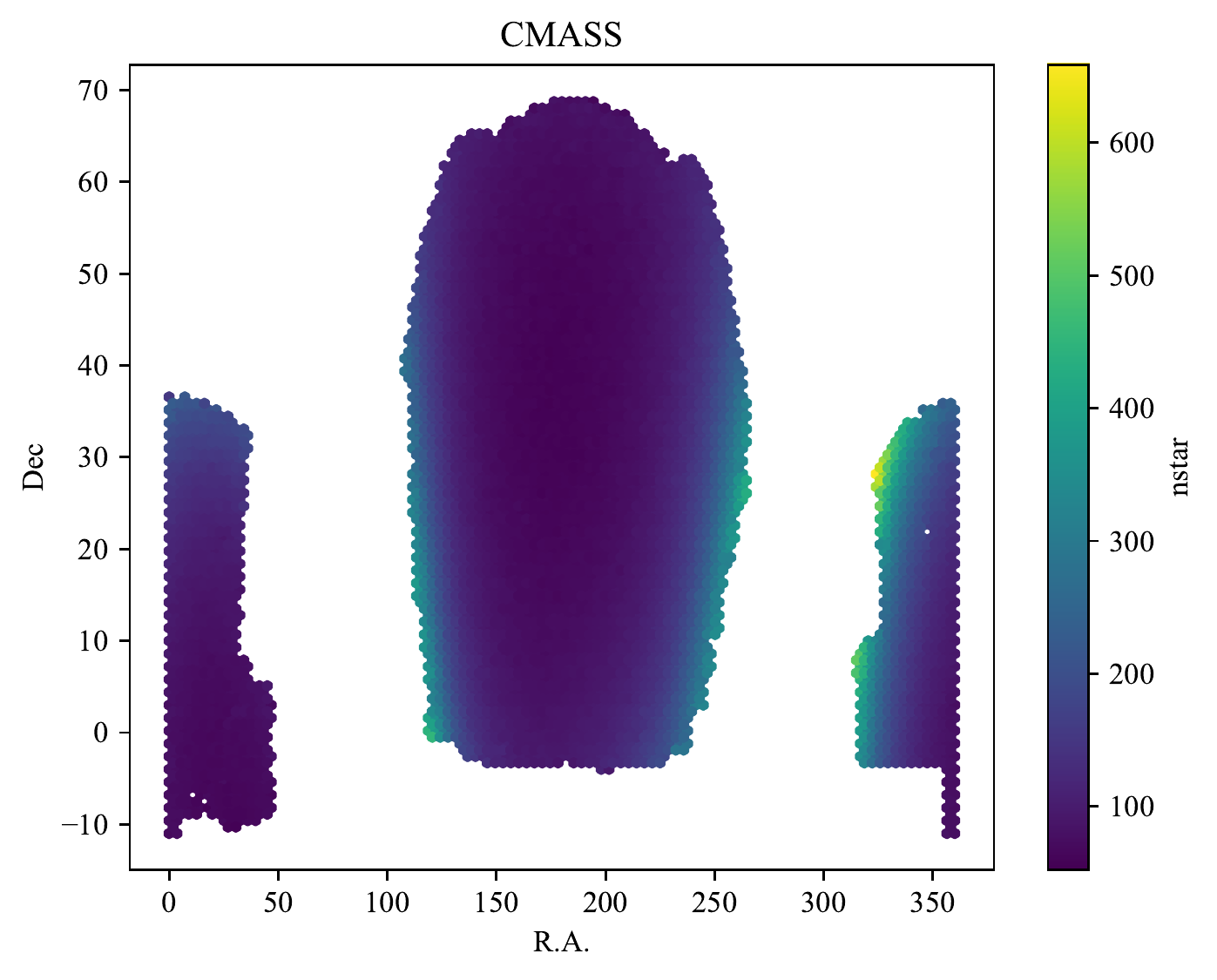}}
  \end{tabular}

  \caption{Spatial variation of $z-\overline{z}$, \ic, and $n_{\rm star}$ for LOWZ and CMASS DR14 data.}\label{fig:a1spatialmaps}
\end{figure*}

\section{Impact of Missing Redshifts}\label{missingredshiftsappendix}

Our fiducial signals include the BOSS weight, $w_{\rm z}$, to account for missing redshifts. Nonetheless, we have also studied the impact of missing redshifts for the LOWZ sample using SDSS. In Figure~\ref{fig:lowz_z_wt} we show the ratio of impact of redshift weights 
(weights accounting for missing redshifts due to close pairs and redshift 
failures). Using the redshift weights increases the $\Delta\Sigma$ by 
$\sim(2\pm1)\%$ percent in the North, and $\sim(5\pm2)\%$ in the South (see Table \ref{missingztable}). For our lensing signals, the missing redshift weight is applied as part of the total systematic lens weight $w_{\rm tot}$. If the $w_{\rm z}$ weight is not applied, we can expect differences related to missing redshift to be small, less than 5\%. 

\begin{figure}
  \centering
  \includegraphics[width=\linewidth]{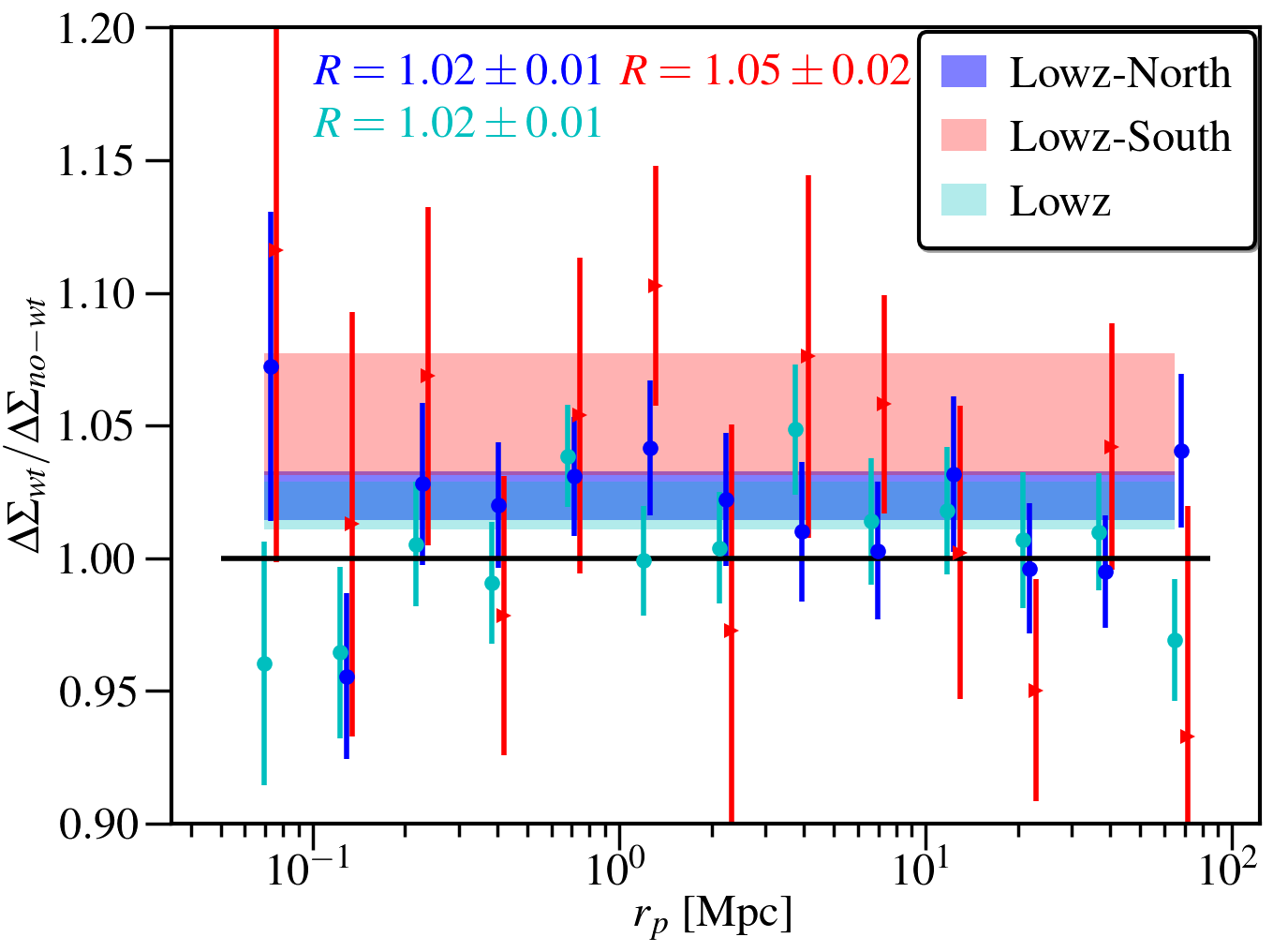}
  \caption{The ratio, $R$, of $\Delta\Sigma$ obtained using $z$ weights and no 
  		weights. Since $z$-weights tend to up weight the regions with more galaxies, $\Delta
        \Sigma$ obtained using weights is slightly larger. The bands and the 
		numbers quoted in the text are the mean values of $R$ obtained by fitting a 
        constant to values between $0.5<r_{\rm p}<15$ Mpc.}
  \label{fig:lowz_z_wt}
\end{figure}

\begin{table}
  \caption{Results for differences between North and South and the impact of missing redshifts. This was computed using SDSS lensing and the LOWZ sample.}
\begin{tabular}{@{}lc}
\hline
Effect & All R \\
\hline
North v.s. South    & 0.95$\pm$0.08 \\
Missing redshifts       &  1.02$\pm$0.01\\
Missing redshifts -  North & 1.02$\pm$0.01 \\
Missing redshifts  -  South & 1.05$\pm$0.02 \\
\hline
\hline
\end{tabular}
\label{missingztable}
\end{table}

\section{Trend with $n_{\rm star}$}\label{nsappendix}

Here we present further discussion on possible origins for the $n_{\rm star}$ trend.

\subsection{Lens inhomogeneity}

One explanation is there may be intrinsic variations in the halo masses of CMASS that correlate with $n_{\rm star}$. This is plausible because we know from \citet[][]{Ross:2017aa} that the number density of CMASS galaxies varies with $n_{\rm star}$ with number density variations that depend on galaxy surface brightness. These effects are strongest in the C2 bin. In addition, \cite{Singh:2020} recently found that higher $n_{\rm star}$ on average leads to lower observed surface brightness of galaxies and that surface brightness was observed to be negatively correlated with the galaxy bias. On the other hand, our internal test using HSC displayed in  Figure \ref{fig:amps_hsc_internal} using HSC data alone does not favour this hypothesis. The question of lens inhomogeneity therefore remains a puzzle.
 
If lens inhomogeneity is indeed the correct explanation, there are a number of interesting consequences to consider. First, the impact of inhomogeneity in lens samples on combined probe analyses of lensing and clustering has typically not been well studied. The trend identified in Figure \ref{fig:ampsys_nstar} could impact combined probe studies because (with the exception of SDSS) most lensing surveys cover smaller areas than the BOSS footprint and have different $n_{\rm star}$ distributions compared to BOSS. Second, it is also interesting to note that the BOSS systematic weights do not correct for the effects. This is because $w_{\rm sys}$ is simply designed to homogenise number densities across the BOSS footprint but knows nothing about correlations between galaxies that are lost near bright stars and halo mass. In order to design a  $w_{\rm sys}$ that would correct for this effect, one would first need to understand and map the physical origin of the $n_{\rm star}$ which will necessitate a) a physical understanding of which galaxies are being lost near bright stars, and b) connecting this with knowledge about the high mass end of the stellar-to-halo mass relation. Third, lens inhomogeneity could be of importance for the derivation of covariance matrices for both clustering and for lensing (halo mass variations across the sky are typically not considered).

In future work, it will be important to devise additional ways of testing whether or not lens inhomogeneity is present in the CMASS data. However, BOSS will also soon be superseded by data from DESI. DESI lens samples are expected to be more homogeneous than those of BOSS \citep[][]{Kitanidis2020} but the precision of joint probe studies with DESI will also be greater compared to BOSS and so the requirement of homogeneous lens samples will also be more stringent for DESI compared to BOSS. 

\subsection{Source systematic that correlates with $n_{\rm star}$}

 The second explanation is that instead of being a variation in the halo masses of the lenses, this could be a systematic in the source sample that correlates with $n_{\rm star}$ and leads to biased estimates of $\Delta\Sigma$. It is possible that the same stars that cause number density variations in CMASS are also responsible for some lensing systematic. For example, $n_{\rm star}$ could also impose a surface brightness selection on source galaxies. The presence of a higher background around bright stars could also impact deblending algorithms. Or the correlation could be due to a different parameter that correlates with $n_{\rm star}$ such as galactic extinction.
 
 If this explanation is correct, then it has important consequences. Indeed, the correlation with $n_{\rm star}$ is the leading cause of scatter in C2. This could therefore be the leading term in the systematic error budget for source galaxies at $z_{\rm s}>0.7$. Investigating the origin of this correlation could therefore help to reduce systematic errors for source galaxies at high redshifts.

\subsection{No trend}\label{nsnotrenddiscuss}

The third possibility is that the trend is a statistical fluctuation and that there is in fact no trend with $n_{\rm star}$ (e.g., see Section \ref{trendmethod}). Indeed, the trend is sensitive to the data point at low $n_{\rm star}$. In this case, the observed spread between the data would be related to source systematics, but these systematics would not correlate with $n_{\rm star}$. It is clear that further work will be required to fully elucidate this question. An interesting avenue to also pursue would be to investigate the correlation between the lensing amplitudes and other quantities such as galactic extinction.

\section{Lens Magnification}\label{lensmagnification}

For CMASS, the $i$-band CMODEL flux is the primary cut that limits the number of CMASS galaxies in the range $z>0.54$. Figure \ref{fig:cmassfluxdist} shows the number of galaxies above a given $i$-band flux divided by the $i$-band flux limit. The flux distribution is fairly steep with slope $\alpha\sim$3 (dotted line) and $\alpha$ is steeper than 1 (dashed line).

Figure \ref{fig:lensmangification} displays an estimate of the lens magnification effect for CMASS assuming $z_{\rm L}=0.6$, $z_{\rm s}=0.8$, and $\alpha=$3 \citep[see also][]{Wietersheim2021}. Figure \ref{fig:lensmangification} shows the spurious additive signal, $\Delta\Sigma_{\rm LM}$, caused by lens magnification. 

\begin{figure}
\centering
\includegraphics[width=7.5cm]{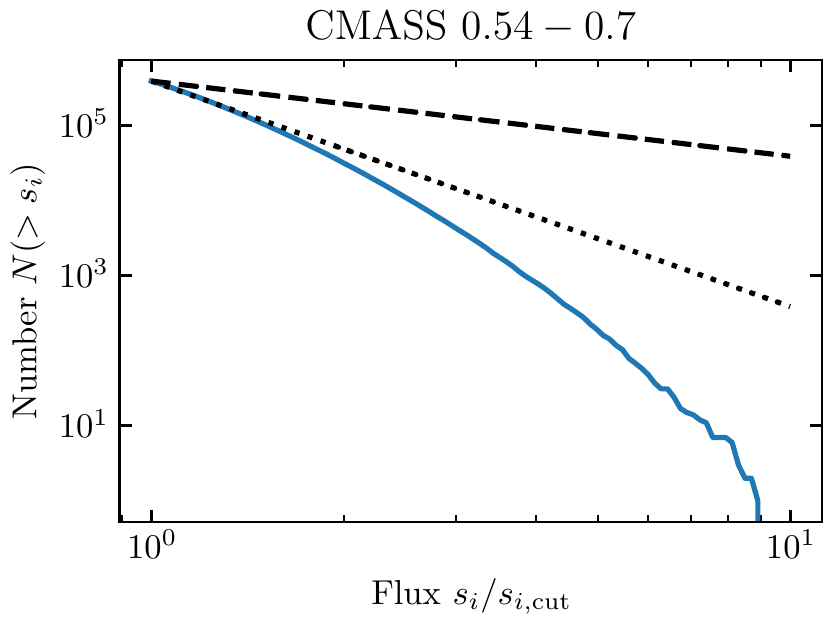}
\caption{Number of galaxies above a given $i$-band flux divided by the $i$-band flux limit for CMASS. CMASS has a slope of $\alpha\sim$3 (dotted line). The slope is steeper than 1 (dashed line).}
\label{fig:cmassfluxdist}
\end{figure}

\begin{figure}
\centering
\includegraphics[width=7.5cm]{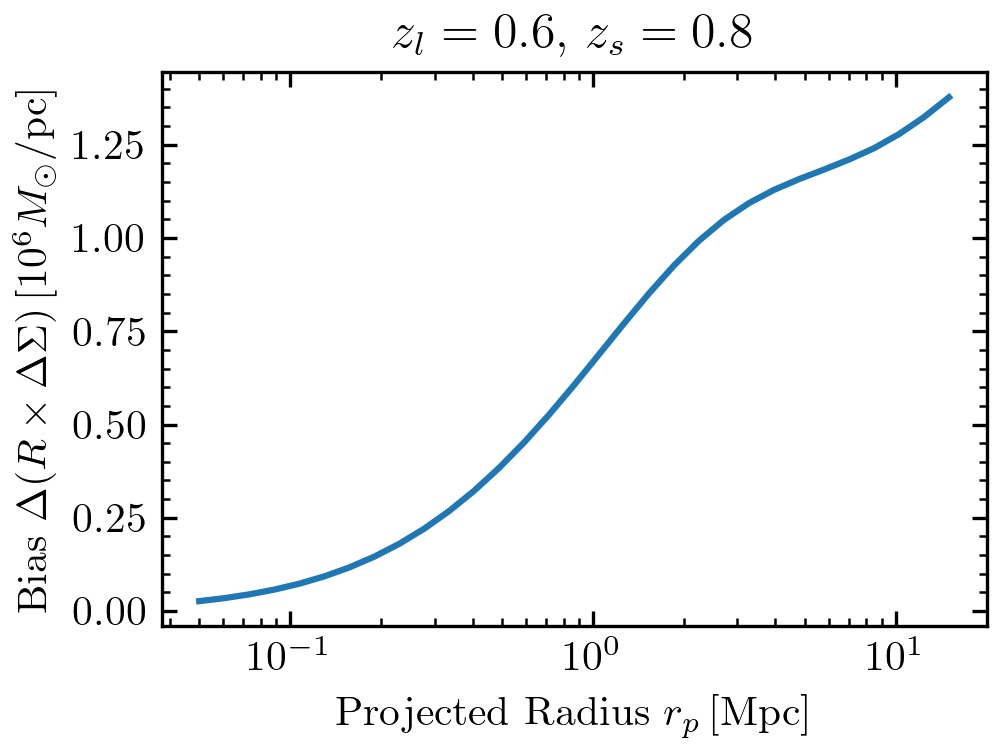}
\caption{Impact of lens magnification on $\Delta\Sigma$. This is the additive spurious signal caused by lens magnificaiton.}
\label{fig:lensmangification}
\end{figure}

\section*{Affiliations}
$^{1}$ Department of Astronomy and Astrophysics, University of California, Santa Cruz, 1156 High Street, Santa Cruz, CA 95064 USA\\
$^{2}$ Kavli Institute for Particle Astrophysics \& Cosmology, P. O. Box 2450, Stanford University, Stanford, CA 94305, USA\\
$^{3}$ McWilliams Center for Cosmology, Department of Physics, Carnegie Mellon University, Pittsburgh, PA 15213, USA\\
$^{4}$ Department of Astronomy, University of California, Berkeley,  501 Campbell Hall, Berkeley, CA 94720, USA\\
$^{5}$ Universit\"ats-Sternwarte, Fakult\"at f\"ur Physik, Ludwig-Maximilians Universit\"at M\"unchen, Scheinerstr. 1, 81679 M\"unchen, Germany\\
$^{6}$ Department of Astrophysical Sciences, Princeton University, Peyton Hall, Princeton, NJ 08544, USA\\
$^{7}$ Institute of Astronomy, University of Cambridge, Madingley Road, Cambridge, CB3 0HA\\
$^{8}$ Max Planck Institute for Extraterrestrial Physics, Giessenbachstrasse, 85748 Garching, Germany\\
$^{9}$ Institute for Astronomy, University of Edinburgh, Royal Observatory, Blackford Hill, Edinburgh EH9 3HJ, UK\\
$^{10}$ Ruhr University Bochum, Faculty of Physics and Astronomy, Astronomical Institute (AIRUB), German Centre for Cosmological Lensing, 44780 Bochum, Germany\\
$^{11}$ Centre for Astrophysics \& Supercomputing, Swinburne University of Technology, P.O.\ Box 218, Hawthorn, VIC 3122, Australia\\
$^{12}$ Laboratoire d'Annecy De Physique Des Particules (LAPP), 9 Chemin de Bellevue, 74940 Annecy, France\\
$^{13}$ Laborat\'orio Interinstitucional de e-Astronomia - LIneA, Rua Gal. Jos\'e Cristino 77, Rio de Janeiro, RJ - 20921-400, Brazil\\
$^{14}$ Fermi National Accelerator Laboratory, P. O. Box 500, Batavia, IL 60510, USA\\
$^{15}$ Instituto de F\'{i}sica Te\'orica, Universidade Estadual Paulista, S\~ao Paulo, Brazil\\
$^{16}$ CNRS, UMR 7095, Institut d'Astrophysique de Paris, F-75014, Paris, France\\
$^{17}$ Sorbonne Universit\'es, UPMC Univ Paris 06, UMR 7095, Institut d'Astrophysique de Paris, F-75014, Paris, France\\
$^{18}$ CEA Irfu, DAp, AIM, Universitè Paris-Saclay, Universitè de Paris, CNRS, F-91191 Gif-sur-Yvette, France\\
$^{19}$ Department of Physics, Northeastern University, Boston, MA 02115, USA\\
$^{20}$ Laboratory of Astrophysics, \'Ecole Polytechnique F\'ed\'erale de Lausanne (EPFL), Observatoire de Sauverny, 1290 Versoix, Switzerland\\
$^{21}$ Jodrell Bank Center for Astrophysics, School of Physics and Astronomy, University of Manchester, Oxford Road, Manchester, M13 9PL, UK\\
$^{22}$ Department of Physics \& Astronomy, University College London, Gower Street, London, WC1E 6BT, UK\\
$^{23}$ SLAC National Accelerator Laboratory, Menlo Park, CA 94025, USA\\
$^{24}$ Instituto de Astrofisica de Canarias, E-38205 La Laguna, Tenerife, Spain\\
$^{25}$ Universidad de La Laguna, Dpto. Astrofísica, E-38206 La Laguna, Tenerife, Spain\\
$^{26}$ Center for Astrophysical Surveys, National Center for Supercomputing Applications, 1205 West Clark St., Urbana, IL 61801, USA\\
$^{27}$ Department of Astronomy, University of Illinois at Urbana-Champaign, 1002 W. Green Street, Urbana, IL 61801, USA\\
$^{28}$ Institut de F\'{\i}sica d'Altes Energies (IFAE), The Barcelona Institute of Science and Technology, Campus UAB, 08193 Bellaterra (Barcelona) Spain\\
$^{29}$ Institut d'Estudis Espacials de Catalunya (IEEC), 08034 Barcelona, Spain\\
$^{30}$ Institute of Space Sciences (ICE, CSIC),  Campus UAB, Carrer de Can Magrans, s/n,  08193 Barcelona, Spain\\
$^{31}$ Physics Department, 2320 Chamberlin Hall, University of Wisconsin-Madison, 1150 University Avenue Madison, WI  53706-1390\\
$^{32}$ Center for Cosmology and Astro-Particle Physics, The Ohio State University, Columbus, OH 43210, USA\\
$^{33}$ Astronomy Unit, Department of Physics, University of Trieste, via Tiepolo 11, I-34131 Trieste, Italy\\
$^{34}$ INAF-Osservatorio Astronomico di Trieste, via G. B. Tiepolo 11, I-34143 Trieste, Italy\\
$^{35}$ Institute for Fundamental Physics of the Universe, Via Beirut 2, 34014 Trieste, Italy\\
$^{36}$ Observat\'orio Nacional, Rua Gal. Jos\'e Cristino 77, Rio de Janeiro, RJ - 20921-400, Brazil\\
$^{37}$ Hamburger Sternwarte, Universit\"{a}t Hamburg, Gojenbergsweg 112, 21029 Hamburg, Germany\\
$^{38}$ Centro de Investigaciones Energ\'eticas, Medioambientales y Tecnol\'ogicas (CIEMAT), Madrid, Spain\\
$^{39}$ Lawrence Berkeley National Laboratory, 1 Cyclotron Road, Berkeley, CA 94720, USA\\
$^{40}$ Santa Cruz Institute for Particle Physics, Santa Cruz, CA 95064, USA\\
$^{41}$ Faculty of Physics, Ludwig-Maximilians-Universit\"at, Scheinerstr. 1, 81679 Munich, Germany\\
$^{42}$ Department of Physics and Astronomy, University of Pennsylvania, Philadelphia, PA 19104, USA\\
$^{43}$ Department of Astronomy, University of Michigan, Ann Arbor, MI 48109, USA\\
$^{44}$ Department of Physics, University of Michigan, Ann Arbor, MI 48109, USA\\
$^{45}$ Institute of Theoretical Astrophysics, University of Oslo. P.O. Box 1029 Blindern, NO-0315 Oslo, Norway\\
$^{46}$ Instituto de Fisica Teorica UAM/CSIC, Universidad Autonoma de Madrid, 28049 Madrid, Spain\\
$^{47}$ Department of Astronomy, University of Geneva, ch. d'\'Ecogia 16, CH-1290 Versoix, Switzerland\\
$^{48}$ Department of Physics, The Ohio State University, Columbus, OH 43210, USA\\
$^{49}$ ASTRAVEO LLC, PO Box 1668, MA 01931\\
$^{50}$ Department of Physics and Astronomy, University College London, Gower Street, London WC1E 6BT, UK\\
$^{51}$ Department of Astrophysical Sciences, Princeton University, 4 Ivy Lane, Princeton, NJ 08544, USA\\
$^{52}$ Department of Astronomy/Steward Observatory, University of Arizona, 933 North Cherry Avenue, Tucson, AZ 85721-0065, USA\\
$^{53}$ Australian Astronomical Optics, Macquarie University, North Ryde, NSW 2113, Australia\\
$^{54}$ Lowell Observatory, 1400 Mars Hill Rd, Flagstaff, AZ 86001, USA\\
$^{55}$ Leiden Observatory, Leiden University, P.O.Box 9513, 2300RA Leiden, The Netherlands\\
$^{56}$ Departamento de F\'isica Matem\'atica, Instituto de F\'isica, Universidade de S\~ao Paulo, CP 66318, S\~ao Paulo, SP, 05314-970, Brazil\\
$^{57}$ Department of Applied Mathematics and Theoretical Physics, University of Cambridge, Cambridge CB3 0WA, UK\\
$^{58}$ International Center for Advanced Studies \& ICIFI, CONICET, ECyT-UNSAM, 1650, Buenos Aires, Argentina\\
$^{59}$ Centro Brasileiro de Pesquisas F\'isicas, Rua Dr. Xavier Sigaud 150, CEP 22290-180, Rio de Janeiro, RJ, Brazil\\
$^{60}$ George P. and Cynthia Woods Mitchell Institute for Fundamental Physics and Astronomy, and Department of Physics and Astronomy, Texas A\&M University, College Station, TX 77843,  USA\\
$^{61}$ Instituci\'o Catalana de Recerca i Estudis Avan\c{c}ats, E-08010 Barcelona, Spain\\
$^{62}$ Kobayashi-Maskawa Institute for the Origin of Particles and the Universe (KMI), Nagoya University, Nagoya, 464-8602, Japan\\
$^{63}$ Kavli Institute for the Physics and Mathematics of the Universe (WPI), University of Tokyo, 5-1-5 Kashiwanoha, Kashiwa, Japan, 2778583\\
$^{64}$ Instituto de Fisica, Universidade Federal do Rio de Janeiro, 21941-972, Rio de Janeiro, RJ, Brazil\\
$^{65}$ The Inter-University Centre for Astronomy and Astrophysics, Pune, India, 411007\\
$^{66}$ Department of Physics, Stanford University, 382 Via Pueblo Mall, Stanford, CA 94305, USA\\
$^{67}$ Kavli Institute for Cosmological Physics, University of Chicago, Chicago, IL 60637, USA\\
$^{68}$ Institute of Astronomy, University of Cambridge, Madingley Road, Cambridge CB3 0HA, UK\\
$^{69}$ Department of Astronomy and Astrophysics, University of Chicago, Chicago, IL 60637, USA\\
$^{70}$ Department of Physics, Carnegie Mellon University, Pittsburgh, Pennsylvania 15312, USA\\
$^{71}$ Jet Propulsion Laboratory, California Institute of Technology,4800 Oak Grove Drive,Pasadena, CA 91109, USA\\
$^{72}$ Lawrence Berkeley National Laboratory, 1 Cyclotron Rd, Berkeley, CA 94720 USA\\
$^{73}$ Instituto de F\'isica, Pontificia Universidad Cat\'olica de Valpara\'iso, Casilla 4059, Valpara\'iso, Chile\\
$^{74}$ School of Physics and Astronomy, University of Southampton,  Southampton, SO17 1BJ, UK\\
$^{75}$ Department of Statistical Sciences, University of Toronto, Toronto, M5S 3G3, Canada\\
$^{76}$ David A. Dunlap Department of Astronomy \& Astrophysics, University of Toronto, Toronto, M5S 3H4, Canada\\
$^{77}$ Dunlap Institute for Astronomy \& Astrophysics, University of Toronto, Toronto, M5S 3H4, Canada\\
$^{78}$ Computer Science and Mathematics Division, Oak Ridge National Laboratory, Oak Ridge, TN 37831\\
$^{79}$ Institute of Cosmology and Gravitation, University of Portsmouth, Portsmouth, PO1 3FX, UK\\
$^{80}$ Center for Cosmology and Particle Physics, Department of Physics, New York University, New York, NY, 10012, USA\\
$^{81}$ Department of Physics, Duke University Durham, NC 27708, USA\\
$^{82}$ Department of Astronomy and Astrophysics, University of British Columbia, 6224 Agricultural Road, V6T 1Z1, Vancouver, Canada\\

\label{lastpage}												


\end{document}